\documentclass[aps,prd,groupedaddress,nofootinbib]{revtex4-2}
\usepackage[a4paper, total={7in, 8in}]{geometry}
\usepackage[utf8]{inputenc}
\usepackage{amsthm}
\usepackage{enumerate}
\usepackage{amssymb}
\usepackage{amsmath}
\usepackage{bbold}
\usepackage{hyperref}
\usepackage{graphicx}
\usepackage{epstopdf}
\usepackage{color} % for inkscape
\graphicspath{{./Figures/}}
\usepackage{caption}
\usepackage{subcaption}
\usepackage{xcolor}

\def\q{{\mathbf q}}
\def\k{{\mathbf k}}
\def\l{{\mathbf l}}
\def\x{{\mathbf x}}  
\def\y{{\mathbf y}}
\def\z{{\mathbf z}}
\def\K{{\mathbf K}}

\newcommand{\slq}{\raise.15ex\hbox{$/$}\kern-.53em\hbox{$q$}}
\newcommand{\slk}{\raise.15ex\hbox{$/$}\kern-.53em\hbox{$k$}}
\newcommand{\sll}{\raise.15ex\hbox{$/$}\kern-.48em\hbox{$l$}}
\newcommand{\sln}{\raise.15ex\hbox{$/$}\kern-.53em\hbox{$n$}}
\newcommand{\slnbar}{\raise.15ex\hbox{$/$}\kern-.53em\hbox{$\bar{n}$}}

\begin{document}

%\begin{frontmatter}

\title{ Renormalization of the gluon distribution function
 in the background field formalism}

\author{Tolga Altinoluk$^{a}$, Guillaume Beuf$\,^{a}$ and Jamal Jalilian-Marian$\,^{a,b,c}$} 
%\author[a,b]{Pedro Agostini}
%%\ead{pedro.agostini@usc.es}
%\author[a]{Tolga Altinoluk}
%%\ead{tolga.altinoluk@ncbj.gov.pl}
%\author[a]{Guillaume Beuf}
%%\ead{guillaume.beuf@ncbj.gov.pl}
%\author[b,c]{Jamal Jalilian-Marian}
%\ead{jamal.jalilian-marian@baruch.cuny.edu}

\affiliation{ $^{a}$Theoretical Physics Division, National Centre for Nuclear Research,
Pasteura 7, Warsaw 02-093, Poland\\
$^{b}$Department of Natural Sciences, Baruch College, CUNY, 17 Lexington Avenue, New York, NY 10010, USA\\
$^{c}$City University of New York Graduate Center, 365 Fifth Avenue, New York, NY 10016, USA}

%\address[a]{Theoretical Physics Division, National Centre for Nuclear Research, Pasteura 7, Warsaw 02-093,
%Poland}
%\address[b]{Department of Natural Sciences, Baruch College, CUNY, 17 Lexington Avenue, New York, NY 10010, USA}
%\address[c]{City University of New York Graduate Center, 365 Fifth Avenue, New York, NY 10016, USA}

%\address[c]{Theoretical Physics Department, Esplanade des Particules 1, CH-1211 Gen\`eve 23, Switzerland}

%\date{\today}

\date{\today}

\begin{abstract}
We derive the Leading Order DGLAP evolution of gluon distribution function in the target light cone gauge starting from its standard operator definition. The derivation is performed using the background 
field formalism also employed in the Color Glass Condensate effective theory of small $x$ QCD. We adopt Mandelstam-Leibbrandt  prescription to regulate in an unambiguous way the spurious singularity appearing in the light-cone gauge Feynman propagator. UV divergences are regulated via dimensional regularization.
The methods introduced in this paper represent the first steps in the construction of a unified framework for QCD evolution, which could address collinear physics as well as small $x$ physics and gluon saturation.
 %The methods introduced in this paper will be used to !!!!! 

\end{abstract}

\maketitle

\tableofcontents

%\end{frontmatter}
%%%%%%%%%
%%%%%%%%%
\section{Introduction}
\label{sec:introduction}

pQCD-based collinear factorization formalism has been extremely successful in describing production of high $p_t$ particles 
in high energy collisions. When it can be proven for a class of processes~\cite{Collins:1989gx},  it guarantees a clean separation of perturbative
from non-perturbative dynamics up to power suppressed corrections. An essential ingredient in this approach is the evolution (scale dependence) of parton 
distribution functions~\cite{Gribov:1972ri,Altarelli:1977zs,Dokshitzer:1977sg} which is calculable perturbatively in powers of $\alpha_s$. This evolution arises from renormalization 
of parton distribution functions, two point correlation functions of quark or gluon field operators which exhibit the usual divergences present in relativistic quantum field theories. 
Nevertheless collinear factorization is expected to break down at very high energy (small $x$) due to 
the large gluon density in a proton or nucleus wave function generated by the fast rise of gluon distribution function. 
At such high gluon densities the concept of a quasi-free parton as envisioned by Feynman is not very useful and it may 
be more appropriate to describe this high occupancy state via semi-classical methods. Color Glass Condensate (CGC) formalism 
is an effective theory of QCD at small $x$ that uses classical color fields to describe such a high occupancy state (see \cite{Gelis:2010nm,Albacete:2014fwa,Blaizot:2016qgz} and references therein). 
In this formalism and in the context of dilute-dense collisions in the so-called hybrid approach appropriate to the 
forward rapidity kinematics, one considers scattering of a projectile parton in a dilute proton on the dense system of 
gluons described as a classical color field. The typical momentum exchanged in such a scattering is of the order of the 
saturation scale $Q_s$ which roughly defines the border between dense and dilute regions of the target wave function.
Such a formalism however can not be used at high transverse momenta since pQCD evolution of the hard scale becomes significant and 
one must use the collinear factorization formalism. 

As the current and the proposed future colliders have a large phase space in $p_t$ and/or large $x$ it is 
imperative to try to combine the two approaches into one unified formalism that has both low $p_t$ (small $x$) 
and high $p_t$ (large $x$) dynamics built in. While there is an obvious need for and significant rewards for deriving 
such a unified formalism it is also a daunting task due to the complexities of the calculations involved as well as
the fact that the underlying approximations and assumptions are vastly different. 
There is already some work done towards this goal \cite{Jalilian-Marian:2017ttv,Jalilian-Marian:2018iui,Jalilian-Marian:2019kaf} where one includes scattering of a projectile parton not only
from small $x$ gluons of the target described by a classical color field, but also from large $x$ gluon field of the target. One loop corrections to this leading order result would then lead to an evolution equation which should reduce to DGLAP \cite{Gribov:1972ri,Altarelli:1977zs,Dokshitzer:1977sg} and 
JIMWLK \cite{Jalilian-Marian:1997qno,Jalilian-Marian:1997jhx,Jalilian-Marian:1997ubg,Kovner:1999bj,Kovner:2000pt,Weigert:2000gi,Iancu:2000hn,Ferreiro:2001qy} evolution equations in the appropriate limits.   

The treatment introduced in \cite{Jalilian-Marian:2017ttv,Jalilian-Marian:2018iui,Jalilian-Marian:2019kaf} goes beyond the standard eikonal approximation that is frequently adopted in the CGC computations. Indeed, there has been a lot of efforts to include the subeikonal corrections in the CGC computations over the last decade. In Refs. \cite{Altinoluk:2014oxa,Altinoluk:2015gia} a subset of subeikonal corrections to the gluon propagator are computed at next-to-next-to-eikonal order. The effects of these corrections on various observables in proton-proton and in proton-nucleus collisions are studied in \cite{Altinoluk:2015xuy,Agostini:2019avp,Agostini:2019hkj,Agostini:2022ctk,Agostini:2022oge}. Subeikonal corrections to the quark propagator and their applications to various DIS observables have been also studied at next-to-eikonal accuracy \cite{Altinoluk:2020oyd,Altinoluk:2021lvu,Altinoluk:2022jkk,Altinoluk:2023qfr}. In \cite{Kovchegov:2015pbl,Kovchegov:2016zex,Kovchegov:2016weo,Kovchegov:2017jxc,Kovchegov:2017lsr,Kovchegov:2018znm,Kovchegov:2018zeq,Cougoulic:2019aja,Kovchegov:2020hgb,Cougoulic:2020tbc,Adamiak:2021ppq,Kovchegov:2021lvz,Kovchegov:2021iyc,Cougoulic:2022gbk,Kovchegov:2022kyy,Borden:2023ugd} quark and gluon helicity evolutions have been computed for at next-to-eikonal accuracy. 
Subeikonal corrections to both quark and gluon propagators have been studied in the high-energy operator product expansion (OPE) formalism, and applied to study the polarized structure functions $g_1$ at low-$x$ in \cite{Chirilli:2018kkw,Chirilli:2021lif}. Apart from the aforementioned direct studies of the subeikonal corrections, in \cite{Balitsky:2015qba,Balitsky:2016dgz,Balitsky:2017flc,Balitsky:2017gis}
rapidity evolution of transverse momentum dependent parton distributions (TMDs) that interpolates between the low and moderate energies have been studied. A similar idea has been pursued in \cite{Boussarie:2020fpb,Boussarie:2021wkn} for the unintegrated gluon distributions. Finally, effects of subeikonal corrections are studied in the context of orbital angular momentum in \cite{Hatta:2016aoc,Boussarie:2019icw}.

The main goal of this paper is take the first steps towards understanding how the scale evolution of the collinear parton distributions can be embedded in the CGC effective theory of a dense target. We start with the standard operator definition of the gluon distribution function in the target light cone gauge and use the background field formalism to calculate the one loop corrections to the tree level result.
 Even though the target light cone gauge is not the standard one used in CGC (but instead the projectile light-cone gauge), it is the gauge in which the parton model picture for the target is the most manifest. It is also a useful first step in order to understand the subtleties of light-cone gauges in general. 
 In particular, we use the Mandelstam-Leibbrandt (ML) prescription since it provides an unambiguous treatment of the spurious singularity appearing in the Feynman propagator in the light cone gauge \cite{Mandelstam:1982cb,Leibbrandt:1983pj,Bassetto:1984dq}. As expected, we encounter UV divergences which are then 
treated with renormalization factors applied to the gluon distribution function, leading to its scale dependence (evolution). Hence we show that the DGLAP 
evolution of gluon distribution function corresponds in this context to the standard UV renormalization of a composite operator constructed 
in the background field formalism. A notable feature of our derivation is that no IR divergence appears at any step in the calculation, thanks to the joint use of the background field method and of the ML prescription for the light-cone gauge. In this study, we focus on the gluon distribution case, in which all the subtleties and complications arise. The case of the quark distribution can then be addressed following the same method, without futher issue. 
Finally, we provide a summary of the results and discuss the outlook. 

%%%%%%%%%%%%%%%%%%%%%%%%%%%%%%%%%%%%%%%%%%%%%%%%%%%%%%%%%%%%%%%%%%%%
%%%%%%%%%%%%%%%%%%%%%%%%%%%%%%%%%%%%%%%%%%%%%%%%%%%%%%%%%%%%%%%%%%%%
%%%%%%%%%%%%%%%%%%%%%%%%%%%%%%%%%%%%%%%%%%%%%%%%%%%%%%%%%%%%%%%%%%%%
%%%%%%%%%%%%%%%%%%%%%%%%%%%%%%%%%%%%%%%%%%%%%%%%%%%%%%%%%%%%%%%%%%%%
%%%%%%%%%%%%%%%%%%%%%%%%%%%%%%%%%%%%%%%%%%%%%%%%%%%%%%%%%%%%%%%%%%%%
%%%%%%%%%%%%%%%%%%%%%%%%%%%%%%%%%%%%%%%%%%%%%%%%%%%%%%%%%%%%%%%%%%%%
	
\section{General definitions and setup}
\label{sec:general}

%%%%%%%%%%%%%%%%%%%%%%%%%%%%%%%%%%%%%%%%%%%%%%%%%%%%%%%%%%%%%%%%%%%%
%%%%%%%%%%%%%%%%%%%%%%%%%%%%%%%%%%%%%%%%%%%%%%%%%%%%%%%%%%%%%%%%%%%%
%%%%%%%%%%%%%%%%%%%%%%%%%%%%%%%%%%%%%%%%%%%%%%%%%%%%%%%%%%%%%%%%%%%%
%%%%%%%%%%%%%%%%%%%%%%%%%%%%%%%%%%%%%%%%%%%%%%%%%%%%%%%%%%%%%%%%%%%%
%%%%%%%%%%%%%%%%%%%%%%%%%%%%%%%%%%%%%%%%%%%%%%%%%%%%%%%%%%%%%%%%%%%%
%%%%%%%%%%%%%%%%%%%%%%%%%%%%%%%%%%%%%%%%%%%%%%%%%%%%%%%%%%%%%%%%%%%%

\subsection{Gluon propagators in light-cone gauge}

In the present study, we adopt the light-cone gauge 
\begin{align}
 n\!\cdot\!A(x)=&\, 0
 \label{eq:def-LCG}
\end{align}
for the gluon field, where $n^{\mu}$ is a fixed light-like vector, meaning that $n^2=0$. By convention, $n^{\mu}$ is oriented towards the future ($n^0>1$).

The Feynman propagator for gluon in vacuum, defined as the time-ordered correlator
\begin{align}
\langle 0| {\cal T} \left\{A_a^{\mu}(x)  A_b^{\nu}(y)\right\} |0\rangle=&\,
\left[ {\mathbb {1}}\right]_{ab}\; { G}_{0,F}^{\mu \nu}(x,y)
=\left[ {\mathbb {1}}\right]_{ab}\; \int \frac{d^D k}{(2\pi)^D}\, e^{-ik\cdot(x-y)}\, {\tilde G}_{0,F}^{\mu \nu}(k)
\label{eq:Feyn-prop_def_pos}
\, ,
\end{align} 
is obtained in the light-cone gauge~\eqref{eq:def-LCG} as
\begin{align}
{\tilde G}_{0,F}^{\mu \nu}(k)=&\,
\frac{i}{\left(k^2+i0^+\right)}\,
\left\{
-g^{\mu \nu}
+\frac{(k^{\mu}n^{\nu}+n^{\mu}k^{\nu})}{[n\!\cdot\!k]}
\right\}
\label{eq:Feyn-prop_def}
\end{align} 
in momentum space. In addition to the usual $k^2$ denominator, regularized by the  $+i0^+$ in the standard way for Feynman propagators, there is an extra denominator $[n\!\cdot\!k]$ which require a regularization as well, in order to fully specify the propagator \eqref{eq:Feyn-prop_def}. This issue is related to the residual gauge freedom, after imposing the light-cone gauge condition~\eqref{eq:def-LCG}.  In early studies of QCD processes in light-cone gauge (in particular in Ref.~\cite{Curci:1980uw}), the extra denominator was regularized with the Cauchy principal 
value prescription. However, such regularization leads to various complications, such as preventing one from performing Wick rotations, and the loss of power counting criterion for the convergence of Feynman integrals. A better regularization for that denominator, compatible with Wick rotations and power counting, is the Mandelstam-Leibbrandt (ML) prescription~\cite{Mandelstam:1982cb,Leibbrandt:1983pj}, defined as  
\begin{align}
\frac{1}{[n\!\cdot\!k]}
\equiv&\,
\frac{(\bar n\!\cdot\!k)}{(n\!\cdot\!k)(\bar n\!\cdot\!k)+i0^+} 
= \frac{1}{\left((n\!\cdot\!k)+i0^+\right)}\, \theta(\bar n\!\cdot\!k) +  \frac{1}{\left((n\!\cdot\!k)-i0^+\right)}\, \theta(-\bar n\!\cdot\!k) 
\label{eq:ML_prescription}
\, ,
\end{align} 
where $\bar n^{\mu}$ is an additional light-like vector, with $\bar n^2=n^2=0$ and $\bar n\cdot n=1$.
In Ref.~\cite{Bassetto:1984dq}, the Hamiltonian quantization of QCD in the instant form (using the ordinary time $x^0$ to formulate the dynamics) has been performed in the light-cone gauge, leading unambiguously to the ML prescription~\eqref{eq:ML_prescription} for the Feynman propagator. 
By contrast, in other approaches to the quantization of QCD in the light-cone gauge, such as path integral quantization, or Hamiltonian quantization in the front form (with a light-cone variable $n\cdot x$ or $\bar n\cdot x$ replacing ordinary time to formulate the dynamics of the theory), no clear guidance could be obtained concerning the handling of the  extra denominator $[n\cdot k]$. In such a way, the ML prescription is the best motivated one, and the only one which can be argued to derive from first principles. 
 Interestingly, the Feynman propagator \eqref{eq:Feyn-prop_def} with the ML prescription can be rewritten as 
\begin{align}
{\tilde G}_{0,F}^{\mu \nu}(k)=&\,
\frac{i}{\left(k^2+i0^+\right)}\,
\left\{
-g^{\mu \nu}
+\frac{2(\bar n\!\cdot\!k)}{\k^2}
(k^{\mu}n^{\nu}+n^{\mu}k^{\nu})
\right\}
-\frac{i}{\big(2(n\!\cdot\!k)(\bar n\!\cdot\!k)+i0^+\big)}\,
 \frac{2(\bar n\!\cdot\!k)}{\k^2}\, 
 (k^{\mu}n^{\nu}+n^{\mu}k^{\nu})
\label{eq:Feyn-prop_2}
\end{align} 
in which the second term can be interpreted as a ghost propagating along a light-like direction, resulting from the residual gauge freedom in the light-cone gauge~\cite{Bassetto:1984dq}.

%%%%%%%%%%%%%%%%%%%%%%%%%%%%%%%%%%%%%%%%%%%%%%%%%%%%%%%%%%%%%%%%%%%%
%%%%%%%%%%%%%%%%%%%%%%%%%%%%%%%%%%%%%%%%%%%%%%%%%%%%%%%%%%%%%%%%%%%%
%%%%%%%%%%%%%%%%%%%%%%%%%%%%%%%%%%%%%%%%%%%%%%%%%%%%%%%%%%%%%%%%%%%%
%%%%%%%%%%%%%%%%%%%%%%%%%%%%%%%%%%%%%%%%%%%%%%%%%%%%%%%%%%%%%%%%%%%%
%%%%%%%%%%%%%%%%%%%%%%%%%%%%%%%%%%%%%%%%%%%%%%%%%%%%%%%%%%%%%%%%%%%%
%%%%%%%%%%%%%%%%%%%%%%%%%%%%%%%%%%%%%%%%%%%%%%%%%%%%%%%%%%%%%%%%%%%%

\subsection{Operator definition of gluon distribution and target light-cone gauge}

Let us consider a boosted hadronic target of momentum $P^{\mu}$, with a large light-cone component $n\!\cdot\!P=P^-$, where $n^2=0$. 
In general, the operator definition of the gluon distribution in that target is (up to the renormalization procedure, described in the next subsection)\footnote{Note that we use lightcone coordinates defined as $x^{\pm}=(x^0\pm x^3)/\sqrt{2}$. Transverse indices are indicated by latin letters $i, j,\dots$. Moreover, transverse vectors are written in bold caracters, and with a euclidian scalar product, so that $x\!\cdot\!y \equiv x^+ y^- + x^- y^+ - \x\!\cdot\!\y$.
In the present study, we choose the two light-like vectors $\bar n^{\mu}$ and $\bar n^{\mu}$ to be related to light-cone coordinates as $n\cdot x =x^-$ and $\bar n\cdot x =x^+$.}
\begin{align}
g(\textrm{x})&=\frac{1}{\textrm{x}P^-}\int \frac{d\Delta x^+}{(2\pi)}\, e^{-i\textrm{x}P^-\Delta x^+}\, \langle P|F^{-i}_b(\Delta x^+,0;0)
\, 
U_{A}(\Delta x^+,0)_{ba} 
\, 
F^{-i}_a(0,0;0)|P\rangle
\label{eq:gluon-def-gengauge}
\, ,
\end{align}
%.
up to UV renormalization, where $\textrm{x}>0$ is the fraction of the momentum $P^-$ of the target carried by the gluon. In Eq.~\eqref{eq:gluon-def-gengauge} the two field strength operators are connected in color by the adjoint gauge link operator
\begin{align}
U_{A}( {x'}^+,x^+) =&\, {\cal P}^{+} \exp \left\{
-ig \int_{x^+}^{{x'}^+}
\!\!\!\!\!\!\!\!dz^+\,  T^c\, A_c^-(z^+,z^-=0;\z=0) 
\right\}
\, ,\label{eq:def-gaugelink}
\end{align}
where ${\cal P}^{+}$ indicates path ordering along the $z^+$ direction.
%Since the mometum $P^{\mu}$ of the target is the same in the bra and in the ket in Eq.~\eqref{eq:gluon-def-gengauge}, the translation of the operator does not generate a phase factor, and one has equivalently
%%
%\begin{align}
%G(\textrm{x})&=\frac{1}{\textrm{x}P^-}\int \frac{d\Delta x^+}{(2\pi)}\, e^{-i\textrm{x}P^-\Delta x^+}\, \langle P|F^{-i}_b(0,0;0)
%\, 
%U_{A}(0,-\Delta x^+)_{ba} 
%\, 
%F^{-i}_a(-\Delta x^+,0;0)|P\rangle
%\label{eq:gluon-def-gengauge_bis}
%\, .
%\end{align}
%%.

In the present study, for simplicity, we will restrict ourselves to the target light-cone gauge
\begin{align}
 n\!\cdot\!A(x)\equiv &\, A^-(x)=0
 \, ,
 \label{eq:def-targetLCG}
\end{align}
 with the same light-like vector $n^{\mu}$ specifying both the gauge choice and the main component of the momentum of the target.\footnote{Alternatively, one could choose the projectile light-cone gauge, with still $n\!\cdot\!A(x)=0$ but now $\bar n\!\cdot\!P$ instead of $n\!\cdot\!P$ as the main component of the boosted target momentum. The projectile light-cone gauge is the most commonly used in the CGC literature, but the gauge link in the definition of the gluon distribution survives in that gauge. We plan to consider the case of projectile light-cone gauge for parton distributions as a future study.} With that gauge choice, the gauge link \eqref{eq:def-gaugelink} reduces to the identity matrix, and Eq.~\eqref{eq:gluon-def-gengauge} becomes
\begin{align}
g(\textrm{x})&=\frac{1}{\textrm{x}P^-}\int \frac{d\Delta x^+}{(2\pi)}\, e^{-i\textrm{x}P^-\Delta x^+}\, \langle P|F^{-i}_a(\Delta x^+,0;0)F^{-i}_a(0,0;0)|P\rangle
\, .\label{eq:gluon-def}
\end{align}
Moreover, in the target light-cone gauge, the field strength tensor components with one upper $-$ index simplify as 
\begin{align}
F^{-\nu}_a(x)=&\, \partial^- A_a^{\nu}(x) = \partial_+ A_a^{\nu}(x)
\, .\label{eq:F_minus_nu_simple}
\end{align}
As a remark, since the target state with the same mometum $P^{\mu}$ is applied on the left and on the right of the operator, the phase factors generated when applying a translation to each field strength compensate each other, and one has
\begin{align}
 \langle P|F^{\rho\sigma}_b(x')F^{\mu\nu}_a(x)|P\rangle 
 =&\,  \langle P|F^{\rho\sigma}_b(x'\!-\!w)F^{\mu\nu}_a(x\!-\!w)|P\rangle
 =  \langle P|F^{\rho\sigma}_b(x'\!-\!x)F^{\mu\nu}_a(0)|P\rangle 
 =  \langle P|F^{\rho\sigma}_b(0)F^{\mu\nu}_a(x\!-\!x')|P\rangle 
 \label{Eq:invar_transl}
 \end{align}
in general, meaning in particular
\begin{align}
 \langle P|F^{-i}_a(\Delta x^+,0;0)F^{-i}_a(0,0;0)|P\rangle =&\,  \langle P|F^{-i}_a(x^+\!+\!\Delta x^+,0;0)F^{-i}_a(x^+,0;0)|P\rangle
 =  \langle P|F^{-i}_a(0,0;0)F^{-i}_a(-\Delta x^+,0;0)|P\rangle 
 \, .\label{Eq:invar_transl_LC}
 \end{align}
Hence, Eq.~\eqref{eq:gluon-def} can be equivalently written as
\begin{align}
g(\textrm{x})&=
\frac{1}{\textrm{x}P^-}
\int \frac{d\Delta x^+}{(2\pi)}\, e^{-i\textrm{x}P^-\Delta x^+}\,
 \langle P|F^{-i}_a(0,0;0)F^{-i}_a(-\Delta x^+,0;0)|P\rangle
\nonumber\\
&=
\frac{1}{(2\pi)\textrm{x}P^-}\, \frac{1}{\left[\int dx^+\right]}\int dx^+\int d x^{\prime+} e^{-i\textrm{x}P^-(x^{\prime+}-x^+)}
\langle P|F^{-i}_a(x^{\prime+},0;0)F^{-i}_a(x^+,0;0)|P\rangle
\, ,\label{eq:gluon-def_alt}
 \end{align}
where the last expression, although formal since it involves infinite numerator and denominator, can be convenient in the course of our calculations.

The expression \eqref{eq:gluon-def} has a clear interpretation if one adopts light-front quantization, with $n\!\cdot\!x =x^-$ replacing $x^0$ as evolution variable. Then, due to the sign in the phases in Eq.~\eqref{eq:gluon-def_alt} (with $\textrm{x}P^->0$), the partial Fourier transform selects the annihilation operator piece of the rightmost field strength and discard its creation operator piece. By contrast, the partial Fourier transform selects creation operator piece of the leftmost field strength, and discard its annihilation operator piece. Hence, the rightmost field strength in Eq.~\eqref{eq:gluon-def_alt} removes a gluon with light-cone momentum $\textrm{x}P^-$ from the target, and the leftmost field strength adds it back. The partial Fourier transforms in Eq.~\eqref{eq:gluon-def} thus automatically selects a purely normal-ordered piece of the operator product. In light-front quantization along $n\!\cdot\!x =x^-$, all states have a positive (or vanishing) light-cone momentum $k^-$, so that it is not possible to remove a gluon with a momentum $\textrm{x}P^-$ larger than the total momentum $P^-$ of the target. For that reason, 
\begin{align}
\textrm{x}>1\; \Rightarrow \; g(\textrm{x})=0
\label{eq:support_pdf}
 \end{align}
so that the gluon distribution $g(\textrm{x})$ has a support $0\leq \textrm{x}\leq 1$. Although that property has been shown within light-front quantization, it should be valid in general, for any quantization procedure. 

Finally, an important feature in the operator definition~\eqref{eq:gluon-def} of the gluon distribution is that the field strength operators are always in the same order, as in a Wightman propagator, instead of being time ordered, like in a Feynman propagator.
This is closely related to the fact than such parton distribution appears at the cross section level for inclusive observables. Indeed, the identity operator can be inserted between the two fields strengths, and decomposed as a sum over states, interpreted as the unobserved final states in the inclusive process. Then, the field strength on the right belongs to the amplitude and the field strength on the left belongs to the complex conjugate amplitude of the scattering process.

Such operator product with a specific order complicates perturbative calculations, and would require the use of the Schwinger-Keldysh formalism for example. Nevertheless, it is possible to bypass this issue by reformulating the operator definition of parton distributions in terms of a time-ordered operator product.
This can be understood in the following way. From the definition of the time-ordered product,
\begin{align}
{\cal T} F^{-i}_a(x')F^{-i}_a(x)  
\equiv &\, 
\theta({x'}^0\!-\! x^0)\, F^{-i}_a(x')F^{-i}_a(x) 
 + \theta(x^0\!-\! {x'}^0)\, F^{-i}_a(x)F^{-i}_a(x') 
\, ,
\end{align}
one finds the relation 
\begin{align}
 F^{-i}_a(x')F^{-i}_a(x)  
= &\, 
{\cal T} F^{-i}_a(x')F^{-i}_a(x)  
- \theta(x^0\!-\! {x'}^0)\, \big[F^{-i}_a(x) ,\, F^{-i}_a(x')\big] 
\nonumber\\
= &\, 
{\cal T} F^{-i}_a(x')F^{-i}_a(x)  
- \theta(x^0\!-\! {x'}^0)\, \partial_{x^+}\partial_{{x'}^+} \big[A^{i}_a(x) ,\, A^{i}_a(x')\big] 
\label{tOrd_fixedOrd_relation}
\, .
\end{align}
In the second term in the right-hand side of Eq.~\eqref{tOrd_fixedOrd_relation}, the commutator is proportional to the identity as a quantum operator. Hence, when inserting the expression \eqref{tOrd_fixedOrd_relation} into the operator definition  \eqref{eq:gluon-def}, the second term in the right-hand side will provide a vacuum contribution, disconnected from the target. By contrast, as we have mentioned earlier, the partial Fourier transforms in Eq.~\eqref{eq:gluon-def} select only a normal-ordered contribution from the fixed order product of field strengths. That contribution is thus fully connected with the target. Hence, if one uses a time ordered product instead of the fixed ordered product in the operator definition  \eqref{eq:gluon-def} of the gluon distribution, one obtains two types of contributions, either connected or disconnected to the target. The first type of contributions correspond to the ones obtained from original definition \eqref{eq:gluon-def}, whereas the second type of contributions was absent from  \eqref{eq:gluon-def} and should thus be discarded.
Hence, the gluon distrubution can alternatively be defined as 
\begin{align}
g(\textrm{x})&=\frac{1}{\textrm{x}P^-}\int \frac{d\Delta x^+}{(2\pi)}\, e^{-i\textrm{x}P^-\Delta x^+}\, \langle P|{\cal T}F^{-i}_a(\Delta x^+,0;0)F^{-i}_a(0,0;0)|P\rangle_c
\, ,\label{eq:gluon-def_T-ord}
\end{align}
with the subscript $c$ indicating that one should discard all diagrams in which the operator is disconnected from the target. 

Similarly, the quark and antiquark parton distributions of flavor $f$ can be defined (again up to renormalization) as
\begin{align}
q_f(\textrm{x})&=\int \frac{d\Delta x^+}{2\pi}\, e^{-i\textrm{x}P^-\Delta x^+}\, \langle P|\overline{\Psi}_f(\Delta x^+,0;0) \frac{\gamma^-}{2}
\, 
U_{F}(\Delta x^+,0)
\, 
\Psi_f(0,0;0)|P\rangle
\label{eq:quark-def-gengauge}
\\
\bar{q}_f(\textrm{x})&=\int \frac{d\Delta x^+}{2\pi}\, e^{-i\textrm{x}P^-\Delta x^+}\, \langle P|
\textrm{Tr}\left[U_{F}(+\infty,\Delta x^+)\Psi_f(\Delta x^+,0;0)
\overline{\Psi}_f(0,0;0) \frac{\gamma^-}{2}
\, 
U_{F}(+\infty,0)^{\dag}
\, 
\right]|P\rangle
\label{eq:antiquark-def-gengauge}
\, ,
\end{align}
%.
with the fundamental gauge links $U_{F}$ analog to $U_A$ defined in Eq.~\eqref{eq:def-gaugelink} but with the fundamental color generators instead of the adjoint ones. In Eq.~\eqref{eq:antiquark-def-gengauge} the trace acts both on the implicit fundamental color indices and Dirac spinor indices.
Like for the gluon distribution case, the gauge links reduce to identity matrices in the target light-cone gauge \eqref{eq:def-targetLCG}. And with the same reasoning as in the gluon case, one can rewrite Eqs.~\eqref{eq:quark-def-gengauge} and \eqref{eq:antiquark-def-gengauge} in terms of time-ordered operator products, with the convention that the contributions disconnected from the target should be discarded. Hence,
\begin{align}
q_f(\textrm{x})&=\int \frac{d\Delta x^+}{2\pi}\, e^{-i\textrm{x}P^-\Delta x^+}\, \langle P|{\cal T}\overline{\Psi}_f(\Delta x^+,0;0) \frac{\gamma^-}{2}
\, 
\Psi_f(0,0;0)|P\rangle_c
\label{eq:quark-def_T-ord}
\\
\bar{q}_f(\textrm{x})&=\int \frac{d\Delta x^+}{2\pi}\, e^{-i\textrm{x}P^-\Delta x^+}\, \langle P|
{\cal T}\textrm{Tr}\left[\Psi_f(\Delta x^+,0;0)
\overline{\Psi}_f(0,0;0) \frac{\gamma^-}{2}
\right]|P\rangle_c
\label{eq:antiquark-def_T-ord}
\, .
\end{align}
%.

%%%%%%%%%%%%%%%%%%%%%%%%%%%%%%%%%%%%%%%%%%%%%%%%%%%%%%%%%%%%%%%%%%%%
%%%%%%%%%%%%%%%%%%%%%%%%%%%%%%%%%%%%%%%%%%%%%%%%%%%%%%%%%%%%%%%%%%%%
%%%%%%%%%%%%%%%%%%%%%%%%%%%%%%%%%%%%%%%%%%%%%%%%%%%%%%%%%%%%%%%%%%%%
%%%%%%%%%%%%%%%%%%%%%%%%%%%%%%%%%%%%%%%%%%%%%%%%%%%%%%%%%%%%%%%%%%%%
%%%%%%%%%%%%%%%%%%%%%%%%%%%%%%%%%%%%%%%%%%%%%%%%%%%%%%%%%%%%%%%%%%%%
%%%%%%%%%%%%%%%%%%%%%%%%%%%%%%%%%%%%%%%%%%%%%%%%%%%%%%%%%%%%%%%%%%%%

\subsection{Renormalization of parton distributions: general form}

When evaluating partons distributions at higher orders in perturbation theory, starting from the definitions provided in the previous subsection, UV divergences arise. Since the parton distributions involve nonlocal composite operators, their renormalization cannot be performed simply via a multiplicative constant, but instead as a convolution, as\footnote{See for example Ref.~\cite{Collins_TMD_book} for a in-depth review of this topic.}
\begin{align}
f_{\alpha}(\textrm{x},\mu^2)
=&\,
\sum_{\beta}\int_{\textrm{x}}^1 \frac{d\textrm{z}}{\textrm{z}}\; Z_{\alpha\beta} (\textrm{z},g,\epsilon) 
f^{(0)}_{\beta}\left(\frac{\textrm{x}}{\textrm{z}}\right)
\, ,\label{eq:pdf_renorm_1}
\end{align}
in dimensional regularization (with $D=4-2\epsilon$), and where the indices $\alpha$ and $\beta$ correspond to the type of parton: gluons, or quark or antiquark of any light quark flavor.   
The parton distribution on the left-hand side of Eq.~\eqref{eq:pdf_renorm_1} corresponds to the renormalized parton distribution, which is then finite but scale dependent.  The parton distribution $f^{(0)}_{\beta}(\textrm{x})$ on the right-hand side of Eq.~\eqref{eq:pdf_renorm_1} is a fully bare distribution, obtained from an operator definition like Eqs.~\eqref{eq:gluon-def_T-ord}, \eqref{eq:quark-def_T-ord} or \eqref{eq:antiquark-def_T-ord} evaluated in bare perturbation theory, with bare field operators and bare coupling.  $f^{(0)}_{\beta}(\textrm{x})$ is thus independent of the dimensional regularization scale $\mu$. The renormalization matrix $Z_{\alpha\beta}$, typically defined in the $\overline{\textrm{MS}}$ scheme, contains the necessary UV counterterms.
Hence, the $\mu$ scale dependence of the renormalized distribution $f_{\alpha}(\textrm{x},\mu^2)$ entirely comes from $Z_{\alpha\beta}$, as
\begin{align}
\mu^2 \frac{d}{d\mu^2} f_{\alpha}(\textrm{x},\mu^2)
=&\,
\sum_{\beta}\int_{\textrm{x}}^1 \frac{d\textrm{z}}{\textrm{z}}\;\left[ \mu^2 \frac{d}{d\mu^2} Z_{\alpha\beta} (\textrm{z},g,\epsilon) \right]
f^{(0)}_{\beta}\left(\frac{\textrm{x}}{\textrm{z}}\right)
\label{eq:pdf_renorm_2}
\, .
\end{align}
Defining the splitting function $P_{\alpha\beta}$ from parton $\beta$ to parton $\alpha$ implicitly via the relation 
\begin{align}
\mu^2 \frac{d}{d\mu^2} Z_{\alpha\beta} (\textrm{z},g,\epsilon)
=&\,
\sum_{\gamma}\int_{\textrm{z}}^1 \frac{d\textrm{z}'}{\textrm{z}'}\;  P_{\alpha\gamma} (\textrm{z}',g,\epsilon)\;   Z_{\gamma\beta} \left(\frac{\textrm{z}}{\textrm{z}'},g,\epsilon\right) 
\label{eq:def_split_funct}
\, , 
\end{align}
one obtains from Eqs.~\eqref{eq:pdf_renorm_2} and \eqref{eq:def_split_funct} the DGLAP equations 
\begin{align}
\mu^2 \frac{d}{d\mu^2} f_{\alpha}(\textrm{x},\mu^2)
=&\,
\sum_{\beta}\int_{\textrm{x}}^1 \frac{d\textrm{z}}{\textrm{z}}\; P_{\alpha\beta} (\textrm{z},g,\epsilon)\;
f_{\beta}\left(\frac{\textrm{x}}{\textrm{z}},\mu^2\right)
\label{eq:DGALP_formal}
\, ,
\end{align}
which involve only the renormalized parton distributions, so that the limit $\epsilon \rightarrow 0$ can be taken safely at this stage.

A first possible method to derive the splitting functions (and thus the DGLAP equations) at a given order in perturbation theory is thus the following
\begin{enumerate}
\item Starting from the operator definition of the parton distribution (like  \eqref{eq:gluon-def_T-ord}, \eqref{eq:quark-def_T-ord} or \eqref{eq:antiquark-def_T-ord}) in terms of bare fields and coupling, perform the expansion in bare perturbation theory to the appropriate order.
\item Deduce the required renormalization matrix $Z_{\alpha\beta}$ to apply as in Eq.~\eqref{eq:pdf_renorm_1} in order to cancel the UV divergences in the parton distributions at that order. 
\item Calculate the splitting functions from $Z_{\alpha\beta}$ thanks to the relation \eqref{eq:def_split_funct}.
\end{enumerate}
  
In practice, it is often more convenient to use renormalized perturbation theory than bare perturbation theory, in particular at higher orders. For that purpose, one should use the operator definitions from the previous subsection expressed in terms of renormalized fields and coupling instead of bare  one. Fields are usually renormalized multiplicatively, as
\begin{align}
\phi_{\alpha}^{(0)}(x) 
=&\, 
\sqrt{Z_{\alpha}(g,\epsilon)}\;
\phi_{\alpha}(x)
\, .
\label{Eq:field_renom_general} 
\end{align}
Focusing for simplicity on the cases in which the gauge link does not play a role, like in target light-cone gauge, the operator definition reduce to a correlator of two fields, so that 
\begin{align}
f^{(0)}_{\alpha}\left({\textrm{x}}\right)
=&\,
\sqrt{Z_{\alpha}(g,\epsilon)}\; \sqrt{Z_{\alpha}(g,\epsilon)}\;
f^{\textrm{n.r.}}_{\alpha}\left({\textrm{x}}, \mu^2\right)
=
Z_{\alpha}(g,\epsilon)\;
f^{\textrm{n.r.}}_{\alpha}\left({\textrm{x}}, \mu^2\right)
\, .
\label{Eq:fully_vs_partially_bare_pdfs}
\end{align}
In Eq.~\eqref{Eq:fully_vs_partially_bare_pdfs}, $f^{\textrm{n.r.}}_{\alpha}\left({\textrm{x}}, \mu^2\right)$ corresponds to the parton distributions obtained in renormalized perturbation theory from the operator definitions like \eqref{eq:gluon-def_T-ord}, \eqref{eq:quark-def_T-ord} or \eqref{eq:antiquark-def_T-ord} written in terms of renormalized fields, but with no further renormalization applied beyond the ones of the elementary fields and coupling. 
Introducing the relation \eqref{Eq:fully_vs_partially_bare_pdfs} in Eq.~\eqref{eq:pdf_renorm_1}, one finds
\begin{align}
f_{\alpha}(\textrm{x},\mu^2)
=&\,
\sum_{\beta}\int_{\textrm{x}}^1 \frac{d\textrm{z}}{\textrm{z}}\; Z_{\alpha\beta} (\textrm{z},g,\epsilon) 
Z_{\beta}(g,\epsilon)\;
f^{\textrm{n.r.}}_{\beta}\left(\frac{\textrm{x}}{\textrm{z}}, \mu^2\right)
\, .\label{eq:pdf_renorm_3}
\end{align}
Hence, a second version of the method to calculate the splitting functions is the following:
\begin{enumerate}
\item Starting from the operator definition of the parton distribution (like \eqref{eq:gluon-def_T-ord}, \eqref{eq:quark-def_T-ord} or \eqref{eq:antiquark-def_T-ord}) in terms of renormalized fields and coupling, perform the expansion in renormalized perturbation theory to the appropriate order.
\item Deduce the required renormalization matrix $[Z_{\alpha\beta}Z_{\beta}]$ to apply as in Eq.~\eqref{eq:pdf_renorm_3} in order to cancel the leftover UV divergences at that order. 
\item Calculate the field renormalization constants $Z_{\beta}$ to the appropriate order.
\item Deduce $Z_{\alpha\beta}$ and calculate the splitting functions from $Z_{\alpha\beta}$ thanks to the relation \eqref{eq:def_split_funct}.
\end{enumerate}
This is the method that we will follow in the present study to derive the evolution of the gluon distribution in target light-cone gauge. The first two steps will be performed in Sec.~\ref{sec:real}, the third step in Sec.~\ref{sec:virtual}, and the fourth step in Sec.~\ref{sec:dglap}.
But first, let us introduce the background field formalism that we will use in this study, in particular in Sec.~\ref{sec:real}.

In this study, we focus on the renormalization and evolution of the gluon distribution. In that case, Eq.~\eqref{eq:pdf_renorm_3} writes
\begin{align}
g(\textrm{x},\mu^2)
=&\,
\int_{\textrm{x}}^1 \frac{d\textrm{z}}{\textrm{z}}\; Z_{gg} (\textrm{z},g,\epsilon) 
Z_{3}(g,\epsilon)\;
g^{\textrm{n.r.}}\left(\frac{\textrm{x}}{\textrm{z}}, \mu^2\right)
+
\sum_{f}\int_{\textrm{x}}^1 \frac{d\textrm{z}}{\textrm{z}}\; Z_{gq} (\textrm{z},g,\epsilon) 
Z_{2}(g,\epsilon)\;
q^{\textrm{n.r.}}_{f}\left(\frac{\textrm{x}}{\textrm{z}}, \mu^2\right)
\nonumber\\
&\,
+
\sum_{f}\int_{\textrm{x}}^1 \frac{d\textrm{z}}{\textrm{z}}\; Z_{g\bar{q}} (\textrm{z},g,\epsilon) 
Z_{2}(g,\epsilon)\;
{\bar q}^{\textrm{n.r.}}_{f}\left(\frac{\textrm{x}}{\textrm{z}}, \mu^2\right)
\, ,\label{eq:q_pdf_renorm_gen}
\end{align}
with $Z_3=1+O(g^2)$ and $Z_2=1+O(g^2)$ the gluon and quark field renormalization constants respectively. Since $Z_{gg}=\delta(\textrm{z}\!-\!1)+O(g^2)$, $Z_{gq}=O(g^2)$ and $Z_{g\bar{q}}=O(g^2)$, Eq.~\eqref{eq:q_pdf_renorm_gen} can be expanded as 
\begin{align}
g(\textrm{x},\mu^2)
=&\,
g^{\textrm{n.r.}}\left({\textrm{x}}, \mu^2\right)
+\int_{\textrm{x}}^1 \frac{d\textrm{z}}{\textrm{z}}\; \Big[Z_{gg} (\textrm{z},g,\epsilon) 
Z_{3}(g,\epsilon)-\delta(\textrm{z}\!-\!1)\Big]\;
g^{\textrm{n.r.}}\left(\frac{\textrm{x}}{\textrm{z}}, \mu^2\right)
\nonumber\\
&\,
+
\sum_{f}\int_{\textrm{x}}^1 \frac{d\textrm{z}}{\textrm{z}}\; Z_{gq} (\textrm{z},g,\epsilon) 
q^{\textrm{n.r.}}_{f}\left(\frac{\textrm{x}}{\textrm{z}}, \mu^2\right)
+
\sum_{f}\int_{\textrm{x}}^1 \frac{d\textrm{z}}{\textrm{z}}\; Z_{g\bar{q}} (\textrm{z},g,\epsilon) 
{\bar q}^{\textrm{n.r.}}_{f}\left(\frac{\textrm{x}}{\textrm{z}}, \mu^2\right)
+O(g^4)
\, ,\label{eq:q_pdf_renorm_1loop}
\end{align}
where the first term is the contribution given by the operator definition, and the last three terms have now the interpretation of one-loop counterterms.

Moreover, at the one-loop accuracy, the relation \eqref{eq:def_split_funct} simplifies into
\begin{align}
\mu^2 \frac{d}{d\mu^2} Z_{\alpha\beta} (\textrm{z},g,\epsilon)
=&\,
 P_{\alpha\beta} (\textrm{z},g,\epsilon) +O(g^4)
\label{eq:def_split_funct_1L}
\, .
\end{align}
% 

%%%%%%%%%%%%%%%%%%%%%%%%%%%%%%%%%%%%%%%%%%%%%%%%%%%%%%%%%%%%%%%%%%%%
%%%%%%%%%%%%%%%%%%%%%%%%%%%%%%%%%%%%%%%%%%%%%%%%%%%%%%%%%%%%%%%%%%%%
%%%%%%%%%%%%%%%%%%%%%%%%%%%%%%%%%%%%%%%%%%%%%%%%%%%%%%%%%%%%%%%%%%%%
%%%%%%%%%%%%%%%%%%%%%%%%%%%%%%%%%%%%%%%%%%%%%%%%%%%%%%%%%%%%%%%%%%%%
%%%%%%%%%%%%%%%%%%%%%%%%%%%%%%%%%%%%%%%%%%%%%%%%%%%%%%%%%%%%%%%%%%%%
%%%%%%%%%%%%%%%%%%%%%%%%%%%%%%%%%%%%%%%%%%%%%%%%%%%%%%%%%%%%%%%%%%%%

\subsection{Expansion around a background field }

In order to calculate the expansion of the gluon distribution in renormalized perturbation theory, starting from the operator definition Eq.~\eqref{eq:gluon-def_T-ord}, we are using the background field method.\footnote{The DGLAP evolution was derived in Ref.~\cite{Bassetto:1998uv} in the target light-cone gauge using the ML prescription as well, but using the formalism developed in Ref.~\cite{Curci:1980uw}, involving on-shell partons in the target instead of background fields. Using on-shell partons produces unphysical IR divergence as artifacts, whereas the use of the background field method allows us to check that no IR divergence arises in the one loop correction to the parton distributions and their UV renormalization.
} 
Namely, we are splitting the renormalized gluon field into a background contribution and a fluctuation contribution, as 
\begin{align}
A^\mu(x)&={\cal A}^\mu(x)+\delta A^\mu(x)
\label{eq:backgd_exp}
\end{align}
at the gauge field level and 
\begin{align}
F^{\mu\nu}(x)&={\cal F}^{\mu\nu}(x)+\delta F^{\mu\nu}(x)
\end{align}
at the field strength level, and then we integrate over the fluctuation contribution to first order in the coupling $\alpha_s$ at the level of the gluon distribution. Roughly speaking, the background field encodes in particular the non-perturbative modes in the target, whereas the fluctuation field represents the perturbative harder and shorter lived modes.
For simplicity, we apply the target light-cone gauge~\eqref{eq:def-targetLCG} not only to the fluctuation field ($\delta A^-(x)=0$) but also to the background field, ${\cal A}^-(x)=0$. Then, one obtains both ${\cal F}^{-\nu}(x)=\partial_+{\cal A}^\nu(x)$ and $\delta{F}^{-\nu}(x)=\partial_+(\delta{A}^\nu)(x)$.
The quark field is also split into a background contribution and a fluctuation as 
\begin{align}
\Psi(x)&=\psi(x)+\delta \Psi(x)
\, .
\label{eq:backgd_exp_q}
\end{align}

First, by neglecting entirely the fluctuation fields and keeping only the background field in the definition~\eqref{eq:gluon-def_T-ord}, one obtains
what we call the background contribution to the gluon distribution
\begin{align}
g^{\textrm{Bckgd}}(\textrm{x},\mu^2)&=\frac{1}{\textrm{x}P^-}\int \frac{d\Delta x^+}{(2\pi)}\, e^{-i\textrm{x}P^-\Delta x^+}\, \langle P|{\cal T}{\cal F}^{-i}_a(\Delta x^+,0;0){\cal F}^{-i}_a(0,0;0)|P\rangle_c
\, .\label{eq:gluon-pdf_Bckgd}
\end{align}
Similarly, one obtains from Eqs.~\eqref{eq:quark-def_T-ord} and \eqref{eq:antiquark-def_T-ord} the background contribution to the quark and antiquark distributions
\begin{align}
q_f^{\textrm{Bckgd}}(\textrm{x},\mu^2)&=\int \frac{d\Delta x^+}{2\pi}\, e^{-i\textrm{x}P^-\Delta x^+}\, \langle P|{\cal T}\overline{\psi}_f(\Delta x^+,0;0) \frac{\gamma^-}{2}
\, 
\psi_f(0,0;0)|P\rangle_c
\label{eq:quark-pdf_Bckgd}
\\
\bar{q}_f^{\textrm{Bckgd}}(\textrm{x},\mu^2)&=\int \frac{d\Delta x^+}{2\pi}\, e^{-i\textrm{x}P^-\Delta x^+}\, \langle P|
{\cal T}\textrm{Tr}\left[\psi_f(\Delta x^+,0;0)
\overline{\psi}_f(0,0;0) \frac{\gamma^-}{2}
\right]|P\rangle_c
\label{eq:antiquark-pdf_Bckgd}
\, .
\end{align}
%.

Corrections beyond that background contribution \eqref{eq:gluon-pdf_Bckgd} for the gluon distribution are obtained by expanding the correlator of two field strengths in the target hadron state as
\begin{align}
&\, \langle P|{\cal T}F_a^{-\mu}(x')F_a^{-\nu}(x)|P\rangle_c
\nonumber\\
&\, =
 \langle P|{\cal T}{\cal F}_a^{-\mu}(x'){\cal F}_a^{-\nu}(x)|P\rangle_c
+\langle P|{\cal T} \delta F_a^{-\mu}(x')\delta F_a^{-\nu}(x)|P\rangle_c
\nonumber\\
&\, =
  \langle P|{\cal T}{\cal F}_a^{-\mu}(x'){\cal F}_a^{-\nu}(x)|P\rangle_c
+\partial_{{x'}^+}\partial_{x^+}\langle P| 
{\cal T} \delta{A}_a^{\mu}(x')\; \delta{A}_a^\nu(x)|P\rangle_c 
\label{eq:op_expansion}
\, .
\end{align}
Note that there is no contribution linear in the fluctuation field. Indeed, in the background field method, the terms linear in the fluctuation field can be made to vanish by restricting ourselves to background fields which obey their equation of motion (including quantum corrections when needed on top of the classical equation of motion).

Inserting the expansion \eqref{eq:op_expansion} of the field strength correlator into the operator definition \eqref{eq:gluon-def_T-ord}, one finds
\begin{align}
g^{\textrm{n.r.}}\left({\textrm{x}}, \mu^2\right)
=&\,
g^{\textrm{Bckgd}}(\textrm{x},\mu^2)
+
 \frac{\textrm{x}P^-}{(2\pi)\,}\int d\Delta x^+e^{-i\textrm{x}P^-\Delta x^+}
\delta^{ii'}\left[ {\mathbb {1}}\right]_{aa'}\, \langle P|\, 
{\mathcal T} 
\delta A^{i'}_{a'}(\Delta x^+, 0; 0) \delta A^i_a(0)|P\rangle_c
\label{eq:G-real-back_and_fluct}
\, .
\end{align}
The second term in Eq.~\eqref{eq:op_expansion} contains the fluctuation field Feynman propagator in the target state. It can be  
 expanded as a series both in the QCD coupling $g$ and in the background field. 

As a reminder, we consider here the connected correlators, meaning that the operators have their vacuum expectation value subtracted. Equivalently, the contribution in which the vacuum propagator for the fluctuation factorizes from the expectation value in the target state is discarded. Only terms involving 
background field insertions in the propagator of the fluctuation are kept.
 Moreover, contributions linear in the background field are absent. Indeed, 
\begin{align}
\langle P|{\cal F}_a^{\mu\nu}(x)|P\rangle &\, = 0 
\end{align}
by color symmetry. The contributions quadratic in the background field are thus the leading ones in the second term of Eq.~\eqref{eq:G-real-back_and_fluct}. 
In this study, we will focus on them, and neglect contributions of higher order in the background field, because, in the case of a dilute target, they are also of higher order in the coupling $g$.
%Hence, in the second term in Eq.~\eqref{eq:G-real-back_and_fluct}, we will consider only the contributions with two powers of the background field inserted in the propagator of the gluon fluctuation field. 

So far, we have shown that all non-zero corrections to the gluon distribution in the expansion~\eqref{eq:G-real-back_and_fluct} around the background field and at any order in $g$ are at least quadratic in the background field. In Sec.~\ref{sec:real}, we calculate the corrections which are exactly quadratic in the background field and of order $g^2$, in order to check in Sec.~\ref{sec:dglap} that we recover the standard LO DGLAP evolution of the gluon distribution from its UV renormalization in light-cone gauge, within the background field formalism.

%%%%%%%%%%%%%%%%%%%%%%%%%%%%%%%%%%%%%%%%%%%%%%%%%%%%%%%%%%%%%%%%%%%%
%%%%%%%%%%%%%%%%%%%%%%%%%%%%%%%%%%%%%%%%%%%%%%%%%%%%%%%%%%%%%%%%%%%%
%%%%%%%%%%%%%%%%%%%%%%%%%%%%%%%%%%%%%%%%%%%%%%%%%%%%%%%%%%%%%%%%%%%%
%%%%%%%%%%%%%%%%%%%%%%%%%%%%%%%%%%%%%%%%%%%%%%%%%%%%%%%%%%%%%%%%%%%%
%%%%%%%%%%%%%%%%%%%%%%%%%%%%%%%%%%%%%%%%%%%%%%%%%%%%%%%%%%%%%%%%%%%%
%%%%%%%%%%%%%%%%%%%%%%%%%%%%%%%%%%%%%%%%%%%%%%%%%%%%%%%%%%%%%%%%%%%%

\section{Real contributions}
\label{sec:real}	
%%%%%%%%%%%%%%%%%%%%%%%%%%%%%%%%%%%%%%%%%%%%%%%%%%%%%%%%%%%%%%%%%%%%
%%%%%%%%%%%%%%%%%%%%%%%%%%%%%%%%%%%%%%%%%%%%%%%%%%%%%%%%%%%%%%%%%%%%
%%%%%%%%%%%%%%%%%%%%%%%%%%%%%%%%%%%%%%%%%%%%%%%%%%%%%%%%%%%%%%%%%%%%
%%%%%%%%%%%%%%%%%%%%%%%%%%%%%%%%%%%%%%%%%%%%%%%%%%%%%%%%%%%%%%%%%%%%
%%%%%%%%%%%%%%%%%%%%%%%%%%%%%%%%%%%%%%%%%%%%%%%%%%%%%%%%%%%%%%%%%%%%
%%%%%%%%%%%%%%%%%%%%%%%%%%%%%%%%%%%%%%%%%%%%%%%%%%%%%%%%%%%%%%%%%%%%

\begin{figure}[ht]
\subfloat[Quark-to-gluon ladder diagram \label{Fig:q2g}]{%
       \includegraphics[width=0.40\textwidth]{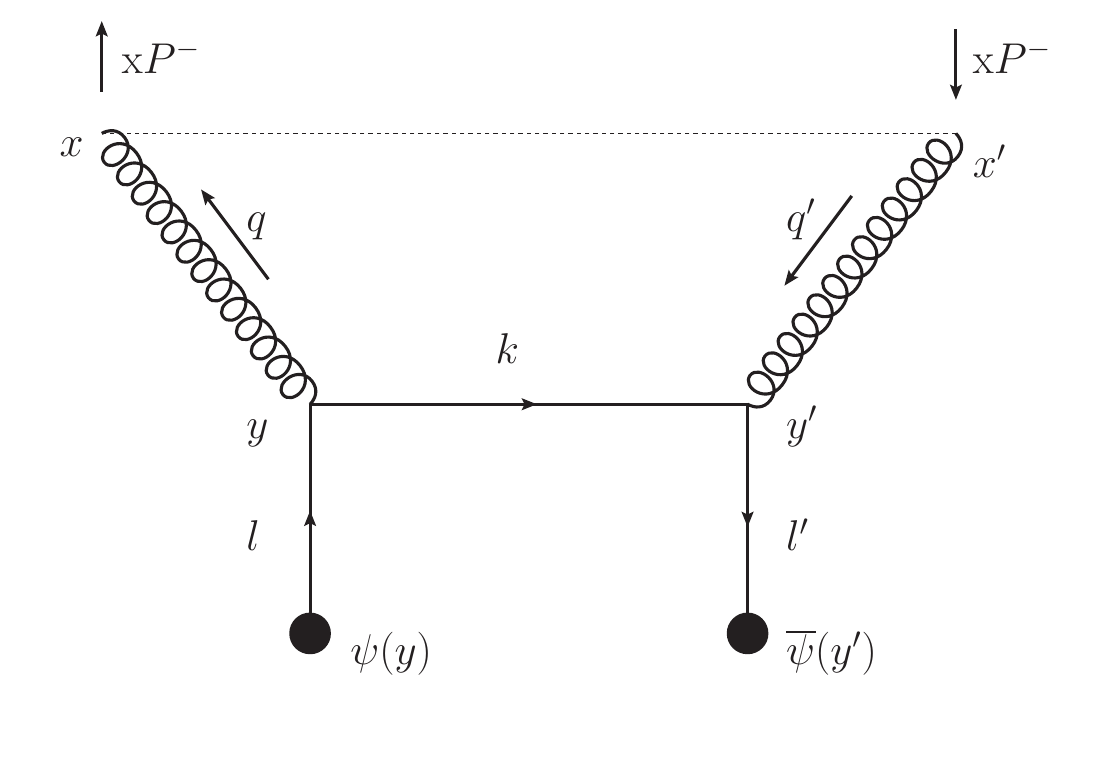}
     }
\hfill
\subfloat[Antiquark-to-gluon ladder diagram \label{Fig:qbar2g}]{%
       \includegraphics[width=0.40\textwidth]{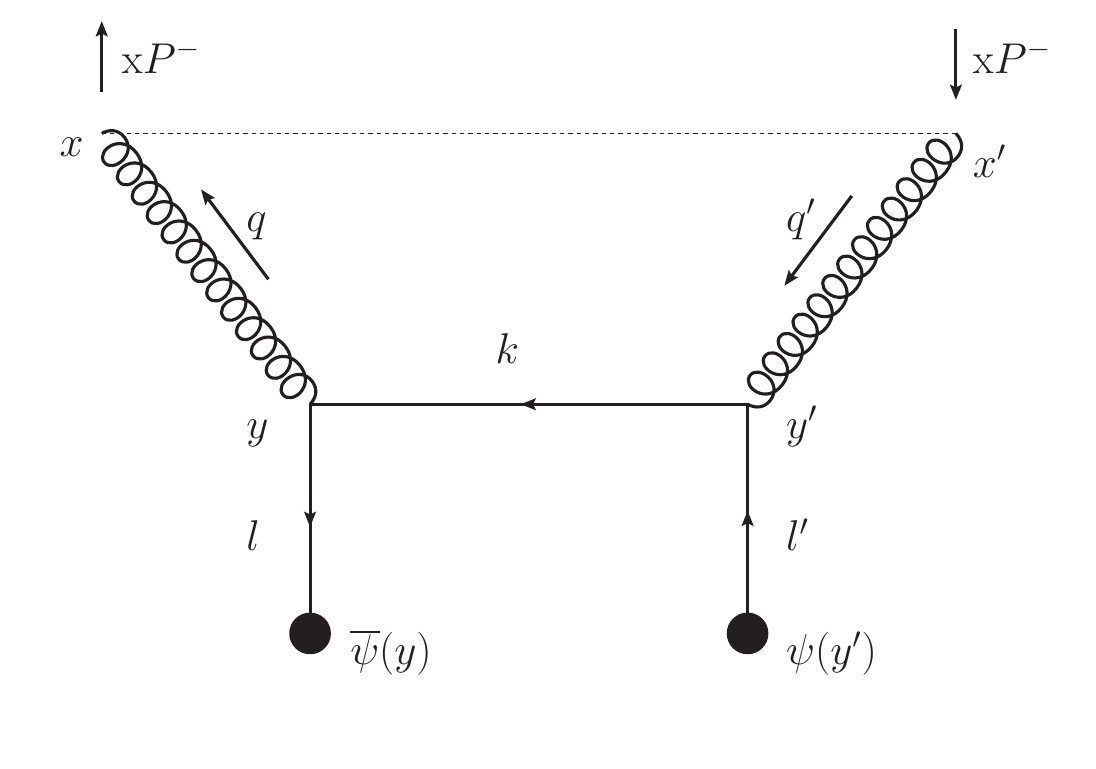}
     }

\subfloat[Four gluon vertex diagram \label{Fig:4gV_real}]{%
       \includegraphics[width=0.40\textwidth]{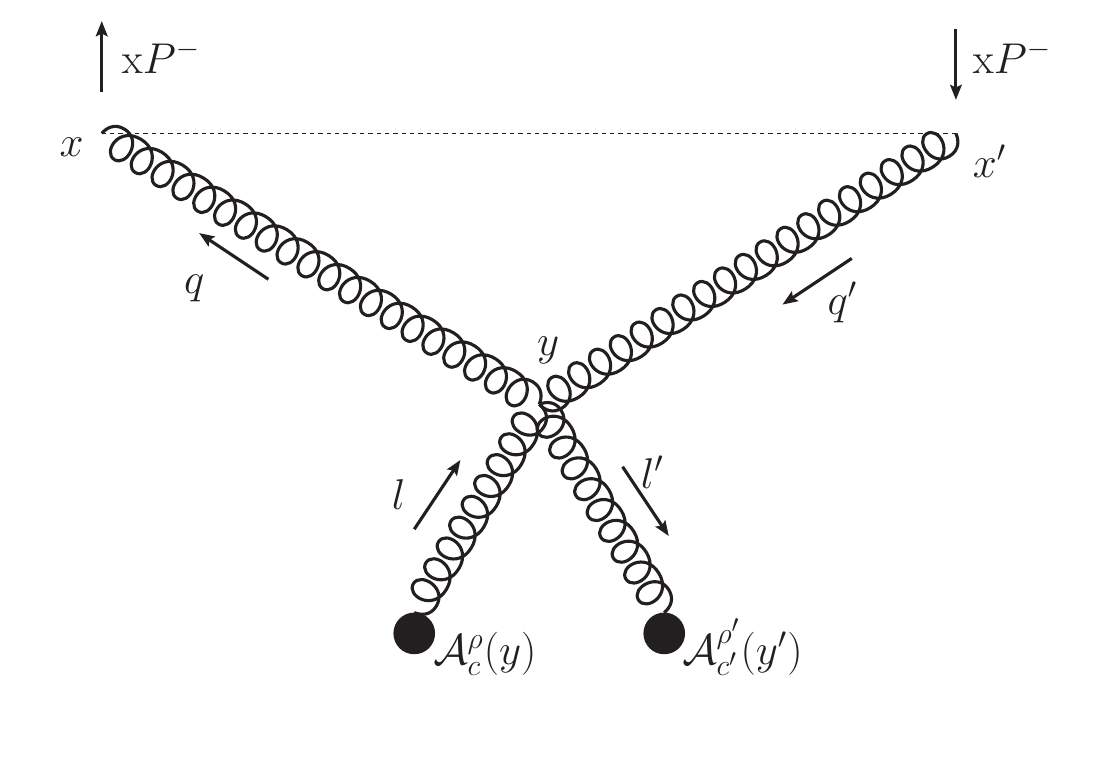}
     }
\hfill
\subfloat[Gluon-to-gluon ladder diagram\label{Fig:g2g_ladder}]{%
       \includegraphics[width=0.40\textwidth]{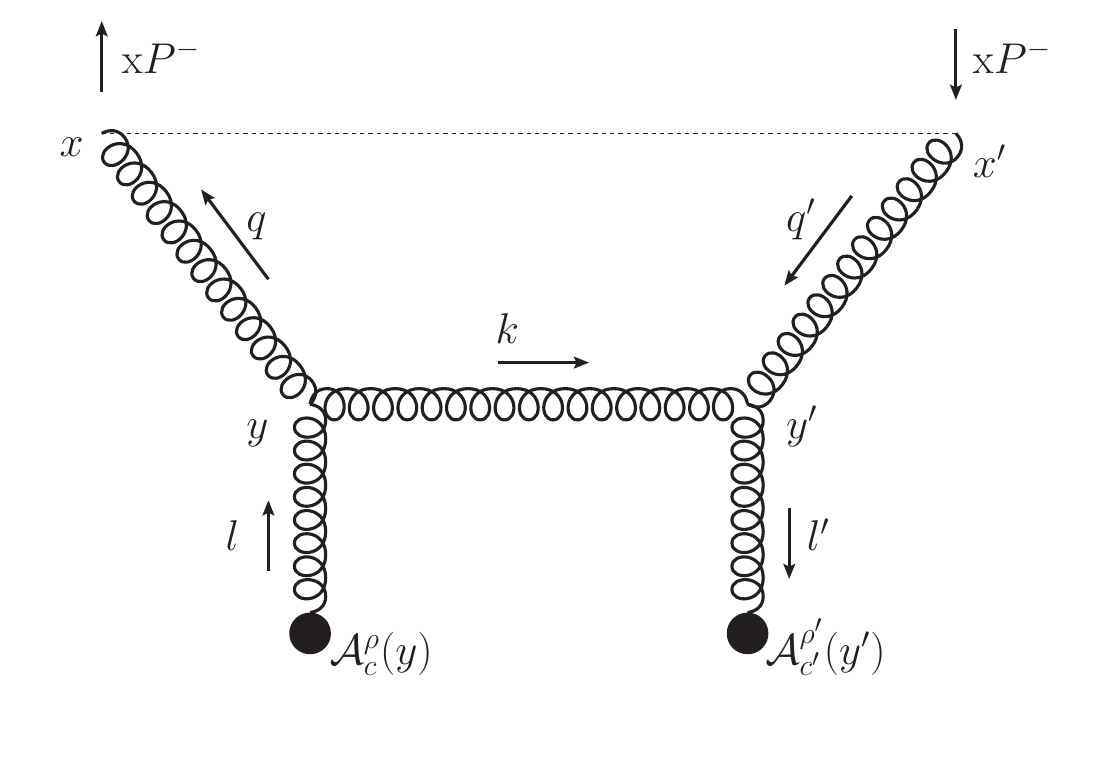}
     }
\caption{\label{Fig:real_diags} Real NLO diagrams contributing to the gluon distribution.}
\end{figure}

In this section, we calculate the real corrections  to the gluon pdf beyond the background contribution, see Eq.~\eqref{eq:G-real-back_and_fluct}. At the considered accuracy, it features the propagator of the gluon fluctuation field with two insertions of background fields of the target, as we have discussed earlier. There are two sets of contributions, depending on the type of background fields inserted: either twice the gluon background field, or twice the (anti)quark background field. All these contributions are represented on Fig.~\ref{Fig:real_diags}.

%%%%%%%%%%%%%%%%%%%%%%%%%%%%%%%%%%%%%%%%%%%%%%%%%%%%%%%%%%%%%%%%%%%%%%%%%%%%%%%%%%%%%%
%%%%%%%%%%%%%%%%%%%%%%%%%%%%%%%%%%%%%%%%%%%%%%%%%%%%%%%%%%%%%%%%%%%%%%%%%%%%%%%%%%%%%%
%%%%%%%%%%%%%%%%%%%%%%%%%%%%%%%%%%%%%%%%%%%%%%%%%%%%%%%%%%%%%%%%%%%%%%%%%%%%%%%%%%%%%%
%%%%%%%%%%%%%%%%%%%%%%%%%%%%%%%%%%%%%%%%%%%%%%%%%%%%%%%%%%%%%%%%%%%%%%%%%%%%%%%%%%%%%%

\subsection{Quark to gluon channel}
\label{sec:q2g}	
%%%%%%%%%%%%%%%%%%%%%%%%%%%%%%%%%%%%%%%%%%%%%%%%%%%%%%%%%%%%%%%%%%%%
%%%%%%%%%%%%%%%%%%%%%%%%%%%%%%%%%%%%%%%%%%%%%%%%%%%%%%%%%%%%%%%%%%%%

Let us first focus on the diagram \ref{Fig:q2g}, which is one of the simpler ones, in order to present the general method of the calculation. That diagram is the only one contributing to the quark to gluon channel of the DGLAP evolution (corresponding to the $P_{gq}$ splitting function), as will be confirmed by the calculation presented here.
The contribution of the diagram \ref{Fig:q2g} to the propagator of the gluon fluctuation field can be written  as
\begin{align}
\langle P|{\mathcal T}\; \delta A^{i'}_{a'}(x')\delta A^i_a(x) |P\rangle_c\bigg|_{\ref{Fig:q2g}} 
=&\, 
\int d^Dy \int d^Dy' \; G_{0, F}^{i\nu}(x,y)\,    G_{0, F}^{\nu'i'}(y',x')\,
\nonumber\\
&\,\times\;
 \langle P|{\mathcal T}\;  \overline{\psi}(y')\, 
  \Big(-ig \mu^{\epsilon} t^{a'} \gamma_{\nu'}\Big)
  S_{0, F}(y',y)\,
    \Big(-ig \mu^{\epsilon} t^{a} \gamma_{\nu}\Big)
 \psi(y)|P\rangle_c 
\nonumber\\
=&\,
-g^2  \mu^{2\epsilon}\int d^Dy \int d^Dy'
\int \frac{d^Dk}{(2\pi)^D}
\int \frac{d^Dq}{(2\pi)^D}  \int \frac{d^Dq'}{(2\pi)^D}
\; 
 e^{-iq\cdot (x-y)}\,   e^{-ik\cdot (y'-y)}\,  e^{-iq'\cdot (y'-x')}\,
\nonumber\\
&\,\times\;
{\tilde G}_{0, F}^{i\nu}(q)\,    {\tilde G}_{0, F}^{\nu'i'}(q')\,
  \langle P|{\mathcal T}\;  \overline{\psi}(y')\, t^{a'} t^{a}\, 
   \gamma_{\nu'}
  {\tilde S}_{0, F}(k)\,
    \gamma_{\nu}
 \psi(y)|P\rangle_c 
\label{eq:fluct_prop_q2g_1}
\, ,
\end{align}
where $D=4-2\epsilon$. For simplicity we keep the quark flavor index $f$ and the summation over it implicit in this section.
Thanks to the analog of the relation \eqref{Eq:invar_transl} in the case of quark fields, the matrix element in Eq.~\eqref{eq:fluct_prop_q2g_1} depends on $y$ and $y'$ only through the difference $\Delta y \equiv y'\!-\!y$. Hence, after the change of variables $(y,y')\mapsto(y, \Delta y)$, one can perform the integration over $y$, which results in a factor $(2\pi)^D\delta^{(D)}(q'\!-\!q)$. In such a way, one obtains
\begin{align}
\langle P|{\mathcal T}\; \delta A^{i'}_{a'}(x')\delta A^i_a(x) |P\rangle_c\bigg|_{\ref{Fig:q2g}} 
=&\,
-g^2  \mu^{2\epsilon}\int d^D\Delta y
\int \frac{d^Dk}{(2\pi)^D}
\int \frac{d^Dq}{(2\pi)^D} 
\; 
 e^{+iq\cdot (x'-x)}\,   e^{-i(k+q)\cdot \Delta y}\,  
 {\tilde G}_{0, F}^{i\nu}(q)\,    {\tilde G}_{0, F}^{\nu'i'}(q)\,
\nonumber\\
&\,\times\;
  \langle P|{\mathcal T}\;  \overline{\psi}(\Delta y)\, t^{a'} t^{a}\, 
   \gamma_{\nu'}
  {\tilde S}_{0, F}(k)\,
    \gamma_{\nu}
 \psi(0)|P\rangle_c 
 \, .
\label{eq:fluct_prop_q2g_2}
\end{align}
Inserting the expression \eqref{eq:fluct_prop_q2g_2} into \eqref{eq:G-real-back_and_fluct} and integrating over $\Delta x^+$, one finds the contribution of the diagram \ref{Fig:q2g} to the gluon distribution
\begin{align}
g^{\textrm{n.r.}}\left({\textrm{x}}, \mu^2\right)\bigg|^{\ref{Fig:q2g}}
=&\, 
-g^2C_F\,  \mu^{2\epsilon}\, \frac{\textrm{x}P^-}{(2\pi)}\,
 \int d^D\Delta y
\int \frac{d^Dk}{(2\pi)^D}
\int \frac{d^Dq}{(2\pi)^D}\,  
  e^{-i(k+q)\cdot \Delta y}\,
  2\pi\delta(q^-\!-\!\textrm{x}P^-)\, 
 \nonumber\\
&\,\times\;
{\tilde G}_{0, F}^{i\nu}(q)\,    {\tilde G}_{0, F}^{\nu'i}(q)\,
 \langle P|{\mathcal T}\;  \overline{\psi}(\Delta y)\,  
   \gamma_{\nu'}
  {\tilde S}_{0, F}(k)\,
    \gamma_{\nu}
 \psi(0)|P\rangle_c 
\nonumber\\
=&\, 
\int \frac{d^D\Delta y}{4\pi}\,    \int \frac{d^Dk}{(2\pi)^D}
\int \frac{d^Dq}{(2\pi)^D}\,   e^{-i(k+q)\cdot \Delta y}\,
\frac{i\,  (2\pi)\delta(q^-\!-\!\textrm{x}P^-)}{\left(q^2+i0^+\right)^2\left(k^2+i0^+\right)}
 \nonumber\\
&\,\times\;
 \langle P|{\mathcal T}\;  \overline{\psi}(\Delta y)\,  
   {\cal N}^{gq}(k;q)\,
 \psi(0)|P\rangle_c 
\label{eq:G-real-NLO_q2g_1}
\end{align}
with the numerator 
\begin{align}
{\cal N}^{gq}(k;q)\equiv &\,
2 g^2 C_F\, \mu^{2\epsilon}\, \textrm{x}P^-\,
\bigg[-g^{i\nu}+\frac{(\q^i n^{\nu}+n^i q^{\nu})}{[n\!\cdot\!q]}\bigg]
\bigg[-g^{\nu' i }+\frac{(\q^i n^{\nu'}+n^i q^{\nu'})}{[n\!\cdot\! q]}\bigg]
\gamma_{\nu'}  \slk  \gamma_{\nu}
\, .
\label{def_num_q2g}
\end{align}
The extra denominators $[n\!\cdot\! q] = [q^-]$ present in Eq.~\eqref{def_num_q2g} should in principle be defined following the ML prescription \eqref{eq:ML_prescription}, since they come from the gluon propagators present in the diagram  \ref{Fig:q2g}. However, due to the constraint $q^-=\textrm{x}P^->0$ present in the expression \eqref{eq:G-real-NLO_q2g_1}, these denominators cannot vanish, so that the ML prescription is irrelevant in that case.

The upper part of the diagram  \ref{Fig:q2g} plays the role of a loop subdiagram, with $l=k+q$ being an external momentum to that loop. It is convenient to take $k$ to parametrize the loop momentum integration, use the constraint  $q^-=\textrm{x}P^-$, and trade the other components of $q$ for the corresponding components of $l$. In such a way, Eq.~\eqref{eq:G-real-NLO_q2g_1} becomes 
\begin{align}
g^{\textrm{n.r.}}\left({\textrm{x}}, \mu^2\right)\bigg|^{\ref{Fig:q2g}}
=&\, 
\int \frac{d^D\Delta y}{4\pi}\,   
\int \frac{d^{D-2}\l}{(2\pi)^{D-2}}\, e^{i\l \cdot \Delta \y}\,
 \int \frac{dl^+}{2\pi}\, e^{-i l^+  \Delta y^-}\,
 \langle P|{\mathcal T}\;  \overline{\psi}(\Delta y)\,  
   {\cal H}^{gq}(l^+,\l;\Delta y^+)\,
 \psi(0)|P\rangle_c 
\label{eq:G-real-NLO_q2g_2}
\, ,
\end{align}
where
\begin{align}
{\cal H}^{gq}(l^+,\l;\Delta y^+)
\equiv &\, 
 \int \frac{d^Dk}{(2\pi)^D}
 e^{-i(k^-+\textrm{x}P^-)\Delta y^+}\,
\frac{i\; {\cal N}^{gq}(k;l^+\!-\!k^+,\textrm{x}P^-,\l\!-\!\k)
}{\Big(2\textrm{x}P^-(l^+\!-\!k^+)-(\l\!-\!\k)^2+i0^+\Big)^2
\left(k^2+i0^+\right)}\,
\,
\label{def_hard_fact_q2g}
\, .
\end{align}

The numerator \eqref{def_num_q2g} can be simplified into\footnote{In that purely algebraic calculation of the numerator, one encounters the trace of the metric over the transverse directions, ${g^i}_i$. Its value depends on the chosen version of dimensional regularization:  ${g^i}_i =D\!-\!2 =2(1\!-\!\epsilon)$ in conventional dimensional regularization (CDR), but ${g^i}_i =2$ in dimensional reduction (DRED) or in the four-dimensional helicity scheme (FDH). Hence, we introduce a regularization-scheme-dependent constant $\delta_s$, with $\delta_s=1$ in CDR and $\delta_s=0$ in DRED and FDH in order to treat all these cases at once.}
\begin{align}
{\cal N}^{gq}(k;q)= &\,
4 g^2 C_F\, \mu^{2\epsilon}\, \textrm{x}P^-\,
\bigg\{(1\!-\! \delta_s \epsilon) \slk
+\left[\k^i \!-\! \frac{k^-}{[q^-]}\, \q^i\right]\left[\gamma^i \!-\! \frac{\q^i}{[q^-]}\, \gamma^-\right]
\bigg\}
\, ,
\label{num_q2g_1}
\end{align}
so that
\begin{align}
{\cal N}^{gq}(k;l^+\!-\!k^+,\textrm{x}P^-,\l\!-\!\k)= &\,
4 g^2 C_F\, \mu^{2\epsilon}\, \textrm{x}P^-\,
\bigg\{(1\!-\! \delta_s \epsilon) \slk
+\frac{(k^-\!+\!\textrm{x}P^-)}{\textrm{x}P^-}
\left[\k^i \!-\! \frac{k^-}{(k^-\!+\!\textrm{x}P^-)}\, \l^i\right]
\left[\gamma^i \!+\! \frac{(\k^i\!-\!\l^i)}{\textrm{x}P^-}\, \gamma^-\right]
\bigg\}
\, .
\label{num_q2g_1}
\end{align}
In the expression \eqref{def_hard_fact_q2g}, the easiest is to perform the integration over $k^+$ first. The numerator \eqref{num_q2g_1} has a term dependent on $k^+$, in $\slk$. In order to avoid the appearance of a potentially problematic $1/k^-$ factor, one should simplify the $k^+$ dependence in the numerator with one of the $q^2$ denominators in Eq.~\eqref {def_hard_fact_q2g} instead of the $k^2$ denominator. For that purpose, the numerator \eqref{num_q2g_1} can be written as
\begin{align}
&\,{\cal N}^{gq}(k;l^+\!-\!k^+,\textrm{x}P^-,\l\!-\!\k)= 
4 g^2 C_F\, \mu^{2\epsilon}\, \textrm{x}P^-\,
\bigg\{
-\frac{(1\!-\! \delta_s \epsilon)}{2\textrm{x}P^-}\, \gamma^-\,
\Big(2\textrm{x}P^-(l^+\!-\!k^+)-(\l\!-\!\k)^2\Big)
\nonumber\\
&\,\times\;
(1\!-\! \delta_s \epsilon)\left(k^- \gamma^+ -\k^i\gamma^i +\left(l^+\!-\!\frac{(\l\!-\!\k)^2}{2\textrm{x}P^-}\right) \gamma^-\right)
+\frac{(k^-\!+\!\textrm{x}P^-)}{\textrm{x}P^-}
\left[\k^i \!-\! \frac{k^-}{(k^-\!+\!\textrm{x}P^-)}\, \l^i\right]
\left[\gamma^i \!+\! \frac{(\k^i\!-\!\l^i)}{\textrm{x}P^-}\, \gamma^-\right]
\bigg\}
\, , 
\label{num_q2g_2}
\end{align}
with the first term proportional to a denominator from Eq.~\eqref{def_hard_fact_q2g} and with terms independent of $k^+$ in the second line. 
The $k^+$ integral in \eqref{def_hard_fact_q2g} is then performed using the residue theorem as 
\begin{align}
&\, \int \frac{dk^+}{2\pi}\,  
 \frac{i}{\Big(2\textrm{x}P^-(l^+\!-\!k^+)-(\l\!-\!\k)^2+i0^+\Big)^n
\left(k^2+i0^+\right)}
=
\frac{\theta(k^-)}{2k^-}\, \frac{1}{\Big(2\textrm{x}P^-l^+ -\frac{\textrm{x}P^-}{k^-}\, \k^2-(\l\!-\!\k)^2+i0^+\Big)^n}
\nonumber\\
=&\,
\frac{\theta(k^-)}{2k^-}\, \frac{1}{\Big( 
-\frac{(k^-\!+\!\textrm{x}P^-)}{k^-}\, \left(\k\!-\!\frac{k^-}{(k^-\!+\!\textrm{x}P^-)}\, \l\right)^2
-\frac{\textrm{x}P^-}{(k^-\!+\!\textrm{x}P^-)}\, \l^2 
+2\textrm{x}P^-l^+ +i0^+\Big)^n}
\, , 
\label{kplus_contour_integ}
\end{align}
valid for $\textrm{x}P^->0$ and $n\geq 1$. 

At this stage, applying the changes of variables
\begin{align}
k^-\mapsto  \textrm{z} = \frac{\textrm{x}P^-}{(k^-\!+\!\textrm{x}P^-)}
\label{change_var_kmin_to_z}
\end{align}
and then
\begin{align}
\k\mapsto  \K = \k - (1\!-\!\textrm{z})\l
\label{change_var_kperp_to_Kperp}
\, ,
\end{align}
one arrives at
\begin{align}
&\, {\cal H}^{gq}(l^+,\l;\Delta y^+)
= 
\frac{4 g^2 C_F\, \textrm{x}P^-\,}{4\pi}
\int_{0}^{1} d\textrm{z}\,  e^{-i\frac{\textrm{x}}{\textrm{z}}P^-\Delta y^+}\,
 \mu^{2\epsilon}
 \int \frac{d^{2-2\epsilon}\K}{(2\pi)^{2-2\epsilon}}\,
 \nonumber\\
 &\, \times\,
 \Bigg\{
 \frac{(1\!-\! \delta_s \epsilon)\, \gamma^-}{2\textrm{z}\textrm{x}P^-}\,
  \frac{1}{\Big[ 
 \K^2 +\textrm{z}(1\!-\!\textrm{z})\left(\l^2\!-\!\frac{2\textrm{x}}{\textrm{z}}P^-l^+\right)
- i0^+\Big]}
 \nonumber\\
 &\,\;\;\;
 + \frac{(1\!-\!\textrm{z})}{\textrm{z}}\,
  \frac{\left[(1\!-\! \delta_s \epsilon)\left(\frac{(1\!-\!\textrm{z})}{\textrm{z}}\textrm{x}P^-\gamma^+
  \!-\!\big(\K^i\!+\!(1\!-\!\textrm{z})\l^i\big)\gamma^i
  +\big(2\textrm{x}P^-l^+\!-\!(\K\!-\!\textrm{z}\l)^2\big)\frac{\gamma^-}{2\textrm{x}P^-}
  \right)
  +\frac{\K^i}{\textrm{z}}\left(\gamma^i+\frac{(\K^i\!-\!\textrm{z}\l^i)}{\textrm{x}P^-}\gamma^-\right)
  \right]}{\Big[ 
 \K^2 +\textrm{z}(1\!-\!\textrm{z})\left(\l^2\!-\!\frac{2\textrm{x}}{\textrm{z}}P^-l^+\right)
- i0^+\Big]^2}
 \Bigg\}
  \nonumber\\
  &\,
= 
2 \alpha_s C_F\, 
\int_{0}^{1} \frac{d\textrm{z}}{\textrm{z}}\,  e^{-i\frac{\textrm{x}}{\textrm{z}}P^-\Delta y^+}\,
 \mu^{2\epsilon}
 \int \frac{d^{2-2\epsilon}\K}{(2\pi)^{2-2\epsilon}}\,
 \Bigg\{
  \frac{(1\!-\! \delta_s \epsilon)\, \gamma^-}{\Big[ 
 \K^2 +\textrm{z}(1\!-\!\textrm{z})\left(-l^2\!-\! i0^+\right)
\Big]}
% \nonumber\\
% &\,\;\;\;
 +\frac{(1\!-\!\textrm{z})
 \left(\frac{2}{\textrm{z}}\!-\!(1\!-\! \delta_s \epsilon)\right)\, 
 \K^2\, \gamma^-}{\Big[ 
 \K^2 +\textrm{z}(1\!-\!\textrm{z})\left(-l^2\!-\! i0^+\right)
\Big]^2}
  \nonumber\\
 &\,\;\;\;
 + \frac{(1\!-\! \delta_s \epsilon)\, (1\!-\!\textrm{z})}{\Big[ 
 \K^2 +\textrm{z}(1\!-\!\textrm{z})\left(-l^2\!-\! i0^+\right)
\Big]^2}\,
\bigg[\textrm{z}^2 l^2\, \gamma^- +2\textrm{x}P^- (1\!-\!\textrm{z})\sll 
\bigg]
 \Bigg\}\Bigg|_{l^-\equiv \frac{\textrm{x}}{\textrm{z}}P^-}
\,
\label{hard_fact_q2g_1}
\, ,
\end{align}
dropping the terms linear in $\K$ in the numerator, by symmetry. Finally, performing the $\K$ integral,
\begin{align}
&\, {\cal H}^{gq}(l^+,\l;\Delta y^+)
= 
\frac{ \alpha_s C_F}{2\pi}\, 
\int_{0}^{1} \frac{d\textrm{z}}{\textrm{z}}\,  e^{-il^-\Delta y^+}\,
\Gamma(1\!+\!\epsilon)\, \left(\frac{-l^2\!-\! i0^+}{4\pi \mu^2}\right)^{-\epsilon}\, \textrm{z}^{-\epsilon}(1\!-\!\textrm{z})^{-\epsilon}
 \Bigg\{
  \frac{(1\!-\! \delta_s \epsilon)\, \gamma^-}{\epsilon}
 \nonumber\\
 &\,\;\;\;
 +\frac{(1\!-\!  \epsilon)}{\epsilon}\,(1\!-\!\textrm{z})
 \left(\frac{2}{\textrm{z}}\!-\!(1\!-\! \delta_s \epsilon)\right)
  \gamma^-
%  \nonumber\\
% &\,\;\;\;
 - (1\!-\! \delta_s \epsilon)\,\textrm{z}\, \gamma^-\,
 -2\textrm{x}P^-(1\!-\! \delta_s \epsilon)\, \frac{ (1\!-\!\textrm{z})}{\textrm{z}}\,
\frac{\sll}{\left(l^2\!+\! i0^+\right)} 
 \Bigg\}\Bigg|_{l^-\equiv \frac{\textrm{x}}{\textrm{z}}P^-}
\,
\label{hard_fact_q2g_2}
\, .
\end{align}
Let us discuss the last term in Eq.~\eqref{hard_fact_q2g_2}, proportional to $\sll$. First, let us note that it regular at $\epsilon=0$ and thus it is a UV finite contribution. Second, when inserting the result \eqref{hard_fact_q2g_2} into Eq.~\eqref{eq:G-real-NLO_q2g_2}, we would get a correction to the gluon distribution proportional to
\begin{align}
\alpha_s C_F\,  \int d^D\Delta y\, 
e^{-i l \cdot \Delta y}\,
   \langle P|{\mathcal T}\;  \overline{\psi}(\Delta y)\,  \sll\,  \psi(0)|P\rangle_c 
=&\, 
\alpha_s C_F\,  \int d^D\Delta y\, 
e^{-i l \cdot \Delta y}\,
  (-i)\, \langle P|{\mathcal T}\;  \left(\partial_{\mu}\overline{\psi}\right)\!\!(\Delta y)\,  \gamma^{\mu}\,  \psi(0)|P\rangle_c 
\label{eq:q2g_eom_like_term_1}
\, .
\end{align}
In the background field formalism, in our order to avoid contributions linear in fluctuation fields, the quark background field is taken to obey its equation of motion, including quantum corrections due to the fluctuation fields:
\begin{align}
i \gamma^{\mu} {\cal D}_{x^{\mu}} \psi(x) \equiv &\, i \gamma^{\mu} {\partial}_{x^{\mu}} \psi(x) - g {\cal A}_{\mu}^{a}(x)\, t^a\, \gamma^{\mu}\,  \psi(x) =O(g^2)
\, ,
\label{EOM_q_back}
\end{align}
and equivalently
\begin{align}
 \overline{\psi}(x) (-i) \gamma^{\mu} \overleftarrow{{\cal D}_{x^{\mu}}} \equiv &\,
 \overline{\psi}(x) (-i) \gamma^{\mu} \overleftarrow{\partial_{x^{\mu}}}
 - \overline{\psi}(x)\, t^a\, \gamma^{\mu}\,  g {\cal A}_{\mu}^{a}(x) =O(g^2)
 \, .
\label{EOM_qbar_back}
\end{align}
Inserting Eq.~\eqref{EOM_qbar_back} into Eq.~\eqref{eq:q2g_eom_like_term_1}, one finds
\begin{align}
 \alpha_s C_F\,  \int d^D\Delta y\, 
e^{-i l \cdot \Delta y}\,
   \langle P|{\mathcal T}\;  \overline{\psi}(\Delta y)\,  \sll\,  \psi(0)|P\rangle_c 
=&\, 
g\, \alpha_s C_F\,  \int d^D\Delta y\,  
e^{-i l \cdot \Delta y}\,
 \langle P|{\mathcal T}\;  \overline{\psi}(\Delta y)\,  \gamma^{\mu}\,  t^a\, {\cal A}_{\mu}^{a}(\Delta y) \psi(0)|P\rangle_c + O(g^4)
\label{eq:q2g_eom_like_term_2}
\, .
\end{align}
In dilute regime for the target, in which the collinear factorization is valid, the power counting is such that $ \langle P|  \overline{\psi}\,  {\cal A} \psi|P\rangle \propto g\,  \langle P|  \overline{\psi}\,  \psi|P\rangle$. Indeed, the extra gluon is predominantly radiated by one of the other fields, in the dilute limit. Hence, a loop correction of the type \eqref{eq:q2g_eom_like_term_2} is actually suppressed by $\alpha_s^2$ instead of $\alpha_s$, which is beyond the accuracy of our calculation in the present study. By contrast, note that in the dense regime of gluon saturation for the target, the power counting is instead $g {\cal A} \propto g^0$, so that a contribution of the type \eqref{eq:q2g_eom_like_term_2} would stay of order $\alpha_s$. Here, we have already focused on the dilute regime, and discarded diagrams with more that two background fields inserted, so that we discard the last term in Eq.~\eqref{hard_fact_q2g_2} by consistency.

Then, inserting the expression~\eqref{hard_fact_q2g_2} into Eq.~\eqref{eq:G-real-NLO_q2g_2}, one obtains
\begin{align}
g^{\textrm{n.r.}}\left({\textrm{x}}, \mu^2\right)\bigg|^{\ref{Fig:q2g}}
=&\, 
\frac{ \alpha_s C_F}{2\pi}\, 
\int_{\textrm{x}}^{1} \frac{d\textrm{z}}{\textrm{z}}\,
 \textrm{z}^{-\epsilon}(1\!-\!\textrm{z})^{-\epsilon}
 \Bigg\{
 \frac{2(1\!-\!\textrm{z})}{\textrm{z}}\, \frac{(1\!-\!  \epsilon)}{\epsilon}
 +(1\!-\! \delta_s \epsilon) \left[ \frac{\textrm{z}}{\epsilon}
 \!+\!(1\!-\!  2\textrm{z}) \right]
 \Bigg\}
 \int \frac{d^{2-2\epsilon}\l}{(2\pi)^{2-2\epsilon}}\,
  \int \frac{dl^+}{2\pi}\,
\nonumber\\
&\, \times\,
\Gamma(1\!+\!\epsilon)\, \left(\frac{-l^2\!-\! i0^+}{4\pi \mu^2}\right)^{-\epsilon}\, 
\int \frac{d^{4-2\epsilon}\Delta y}{4\pi}\,   
 e^{-i l \cdot \Delta y}\,
 \langle P|{\mathcal T}\;  \overline{\psi}(\Delta y)\,  
   \gamma^-\,
 \psi(0)|P\rangle_c \Bigg|_{l^-\equiv \frac{\textrm{x}}{\textrm{z}}P^-}
 +O\left(\alpha_s^2\right)
\label{eq:G-real-NLO_q2g_3}
\, .
\end{align}
Note that we have changed the lower bound of the $\textrm{z}$ integral from $0$ to $\textrm{x}$. Indeed, for $0 <\textrm{z}<\textrm{x}$, the operator matrix element in momentum space would vanish, because the LC momentum $l^-=(\textrm{x}/\textrm{z})P^-$ of the parton extracted from the target would be larger than the one of the whole target $P^-$.

In Eq.~\eqref{hard_fact_q2g_2}, the poles at $\epsilon=0$ correspond to UV divergences which cannot be removed by the standard UV counterterms for fields and coupling (or masses) renormalization in the QCD Lagrangian. They correspond to the extra UV divergences that can occur for composite operators in general, and require an extra renormalization such as presented in Eq.~\eqref{eq:pdf_renorm_3} for parton distributions, or Eq.~\eqref{eq:q_pdf_renorm_1loop} for the gluon distribution at NLO.

Note that the  UV poles at $\epsilon=0$ are the only divergences in the expression \eqref{eq:G-real-NLO_q2g_3}. The integration over $\l$ and $l^+$ could be done analytically without encountering divergences, but the obtained expression would not be more illuminating than Eq.~\eqref{eq:G-real-NLO_q2g_3}. The integration in $\textrm{z}$ is regular as well, provided that the momentum fraction $\textrm{x}$ is non-zero (and positive). 

In the $\overline{\textrm{MS}}$ scheme, the counterterm required to compensate the  UV divergence of the contribution \eqref{eq:G-real-NLO_q2g_3} is then 
\begin{align}
g\left({\textrm{x}}, \mu^2\right)\bigg|^{\ref{Fig:q2g};\textrm{ c.t.}}
=&\, 
-\frac{S_{\epsilon}}{\epsilon}\,\frac{ \alpha_s C_F}{2\pi}\, 
\int_{\textrm{x}}^{1} \frac{d\textrm{z}}{\textrm{z}}\,
 \bigg[
 \frac{2(1\!-\!\textrm{z})}{\textrm{z}}
 +  \textrm{z}
 \bigg]
 \int \frac{d^{2-2\epsilon}\l}{(2\pi)^{2-2\epsilon}}\,
  \int \frac{dl^+}{2\pi}\,
\nonumber\\
&\, \times\,
\int \frac{d^{4-2\epsilon}\Delta y}{4\pi}\,   
 e^{-i l \cdot \Delta y}\,
 \langle P|{\mathcal T}\;  \overline{\psi}(\Delta y)\,  
   \gamma^-\,
 \psi(0)|P\rangle_c \Bigg|_{l^-\equiv \frac{\textrm{x}}{\textrm{z}}P^-}
\nonumber\\
=&\, 
-\frac{S_{\epsilon}}{\epsilon}\,\frac{ \alpha_s C_F}{2\pi}\, 
\int_{\textrm{x}}^{1} \frac{d\textrm{z}}{\textrm{z}}\,
 \bigg[
 \frac{2(1\!-\!\textrm{z})}{\textrm{z}}
 +  \textrm{z}
 \bigg]
%\nonumber\\
%&\, \times\,
\int \frac{d\Delta y^+}{4\pi}\,   
 e^{-i\frac{\textrm{x}}{\textrm{z}}P^-\Delta y^+}\,
 \langle P|{\mathcal T}\;  \overline{\psi}(\Delta y^+,0;0)\,  
   \gamma^-\,
 \psi(0)|P\rangle_c
 \nonumber\\
=&\, 
-\frac{S_{\epsilon}}{\epsilon}\,\frac{ \alpha_s C_F}{2\pi}\, 
\int_{\textrm{x}}^{1} \frac{d\textrm{z}}{\textrm{z}}\,
 \bigg[
 \frac{2(1\!-\!\textrm{z})}{\textrm{z}}
 +  \textrm{z}
 \bigg]
%\nonumber\\
%&\, \times\,
q^{\textrm{Bckgd}}\left(\frac{\textrm{x}}{\textrm{z}}, \mu^2\right)
\label{eq:G-real-NLO_q2g_ct}
\, ,
\end{align}
with 
\begin{align}
S_{\epsilon}\equiv &\, \Big(4\pi\, e^{-\gamma_E} \Big)^{\epsilon}
\, ,
\end{align}
containing the universal constants,
$\gamma_E$ being the Euler-Mascheroni constant.

By comparison of the counterterm \eqref{eq:G-real-NLO_q2g_ct} with Eq.~\eqref{eq:q_pdf_renorm_1loop}, we can read off  
\begin{align}
Z_{gq} (\textrm{z},g,\epsilon)
=&\, 
-\frac{S_{\epsilon}}{\epsilon}\,\frac{ \alpha_s C_F}{2\pi}\,  
\bigg[ \frac{2(1\!-\!\textrm{z})}{\textrm{z}} +  \textrm{z} \bigg]
+O\left(\alpha_s^2\right)
=-\frac{S_{\epsilon}}{\epsilon}\,\frac{ \alpha_s C_F}{2\pi}\,  
\frac{\big[1+(1\!-\!\textrm{z})^2\big]}{\textrm{z}}
+O\left(\alpha_s^2\right)
\label{Eq:Zgq_1loop}
\, ,
\end{align}
since the diagram \ref{Fig:q2g} is the only one contributing to the $q\rightarrow g$ channel at order $\alpha_s$.

%%%%%%%%%%%%%%%%%%%%%%%%%%%%%%%%%%%%%%%%%%%%%%%%%%%%%%%%%%%%%%%%%%%%
%%%%%%%%%%%%%%%%%%%%%%%%%%%%%%%%%%%%%%%%%%%%%%%%%%%%%%%%%%%%%%%%%%%%
%%%%%%%%%%%%%%%%%%%%%%%%%%%%%%%%%%%%%%%%%%%%%%%%%%%%%%%%%%%%%%%%%%%%
%%%%%%%%%%%%%%%%%%%%%%%%%%%%%%%%%%%%%%%%%%%%%%%%%%%%%%%%%%%%%%%%%%%%
%%%%%%%%%%%%%%%%%%%%%%%%%%%%%%%%%%%%%%%%%%%%%%%%%%%%%%%%%%%%%%%%%%%%
%%%%%%%%%%%%%%%%%%%%%%%%%%%%%%%%%%%%%%%%%%%%%%%%%%%%%%%%%%%%%%%%%%%%

\subsection{Antiquark to gluon channel}
\label{sec:qbar2g}	
%%%%%%%%%%%%%%%%%%%%%%%%%%%%%%%%%%%%%%%%%%%%%%%%%%%%%%%%%%%%%%%%%%%%
%%%%%%%%%%%%%%%%%%%%%%%%%%%%%%%%%%%%%%%%%%%%%%%%%%%%%%%%%%%%%%%%%%%%

The diagram \ref{Fig:qbar2g}, differing from the diagram \ref{Fig:q2g} only by the orientation of the quark line, corresponds to the contribution of the antiquark to gluon channel in the DGLAP evolution. Its calculation can be done in  a very similar way, and leads to the result
\begin{align}
g^{\textrm{n.r.}}\left({\textrm{x}}, \mu^2\right)\bigg|^{\ref{Fig:qbar2g}}
=&\, 
\frac{ \alpha_s C_F}{2\pi}\, 
\int_{\textrm{x}}^{1} \frac{d\textrm{z}}{\textrm{z}}\,
 \textrm{z}^{-\epsilon}(1\!-\!\textrm{z})^{-\epsilon}
 \Bigg\{
 \frac{2(1\!-\!\textrm{z})}{\textrm{z}}\, \frac{(1\!-\!  \epsilon)}{\epsilon}
 +(1\!-\! \delta_s \epsilon) \left[ \frac{\textrm{z}}{\epsilon}
 \!+\!(1\!-\!  2\textrm{z}) \right]
 \Bigg\}
 \int \frac{d^{2-2\epsilon}\l}{(2\pi)^{2-2\epsilon}}\,
  \int \frac{dl^+}{2\pi}\,
\nonumber\\
&\, \times\,
\Gamma(1\!+\!\epsilon)\, \left(\frac{-l^2\!-\! i0^+}{4\pi \mu^2}\right)^{-\epsilon}\, 
\int \frac{d^{4-2\epsilon}\Delta y}{4\pi}\,   
 e^{+i l \cdot \Delta y}\,
 \langle P|{\mathcal T}\; \textrm{Tr}\left[
 \psi(\Delta y) \overline{\psi}(0)\,  
   \gamma^-\right]|P\rangle_c \Bigg|_{l^-\equiv -\frac{\textrm{x}}{\textrm{z}}P^-}
 +O\left(\alpha_s^2\right)
\label{eq:G-real-NLO_qbar2g}
\, , 
\end{align}
analog to Eq.~\eqref{eq:G-real-NLO_q2g_3}.
The $\overline{\textrm{MS}}$ counterterm required to cancel the UV divergence from Eq.~\eqref{eq:G-real-NLO_qbar2g} is then found to be
\begin{align}
g\left({\textrm{x}}, \mu^2\right)\bigg|^{\ref{Fig:qbar2g};\textrm{ c.t.}}
=&\, 
-\frac{S_{\epsilon}}{\epsilon}\,\frac{ \alpha_s C_F}{2\pi}\, 
\int_{\textrm{x}}^{1} \frac{d\textrm{z}}{\textrm{z}}\,
 \bigg[
 \frac{2(1\!-\!\textrm{z})}{\textrm{z}}
 +  \textrm{z}
 \bigg]
%\nonumber\\
%&\, \times\,
{\bar q}^{\textrm{Bckgd}}\left(\frac{\textrm{x}}{\textrm{z}}, \mu^2\right)
\label{eq:G-real-NLO_qbar2g_ct}
\, ,
\end{align}
from which we can read off
\begin{align}
Z_{g{\bar q}} (\textrm{z},g,\epsilon)
=&\, 
Z_{gq} (\textrm{z},g,\epsilon)
=-\frac{S_{\epsilon}}{\epsilon}\,\frac{ \alpha_s C_F}{2\pi}\,  
\frac{\big[1+(1\!-\!\textrm{z})^2\big]}{\textrm{z}}
+O\left(\alpha_s^2\right)
\label{Eq:Zgbarq_1loop}
\, ,
\end{align}
since diagram \ref{Fig:qbar2g} is the only one contributing to the ${\bar q}\rightarrow g$ channel at order $\alpha_s$.

%%%%%%%%%%%%%%%%%%%%%%%%%%%%%%%%%%%%%%%%%%%%%%%%%%%%%%%%%%%%%%%%%%%%
%%%%%%%%%%%%%%%%%%%%%%%%%%%%%%%%%%%%%%%%%%%%%%%%%%%%%%%%%%%%%%%%%%%%
%%%%%%%%%%%%%%%%%%%%%%%%%%%%%%%%%%%%%%%%%%%%%%%%%%%%%%%%%%%%%%%%%%%%
%%%%%%%%%%%%%%%%%%%%%%%%%%%%%%%%%%%%%%%%%%%%%%%%%%%%%%%%%%%%%%%%%%%%
%%%%%%%%%%%%%%%%%%%%%%%%%%%%%%%%%%%%%%%%%%%%%%%%%%%%%%%%%%%%%%%%%%%%
%%%%%%%%%%%%%%%%%%%%%%%%%%%%%%%%%%%%%%%%%%%%%%%%%%%%%%%%%%%%%%%%%%%%

\subsection{Gluon to gluon channel}
\label{sec:g2g}	
%%%%%%%%%%%%%%%%%%%%%%%%%%%%%%%%%%%%%%%%%%%%%%%%%%%%%%%%%%%%%%%%%%%%
%%%%%%%%%%%%%%%%%%%%%%%%%%%%%%%%%%%%%%%%%%%%%%%%%%%%%%%%%%%%%%%%%%%%

\subsubsection{Four gluon vertex diagram}

The contribution of the diagram \ref{Fig:4gV_real} to the propagator of the gluon fluctuation field is
\begin{align}
\langle P|{\mathcal T}\;\delta A^{i'}_{a'}(x')\delta A^i_a(x) |P\rangle_c\bigg|_{\ref{Fig:4gV_real}} 
=&\, 
\int d^Dy  \; G_{0, F}^{i\nu}(x,y)\,    G_{0, F}^{\nu'i'}(y,x')\,
{V_{4g}}_{\nu \rho\nu' \rho'}^{a c a' c'}\,
 \langle P| {\cal A}^{\rho'}_{c'}(y)\, {\cal A}^{\rho}_{c}(y)|P\rangle_c 
\nonumber\\
=&\,
\int d^Dy \int \frac{d^Dq}{(2\pi)^D}  \int \frac{d^Dq'}{(2\pi)^D}
 e^{-iq\cdot (x-y)}\,   e^{-iq'\cdot (y-x')}\,
 \;{\tilde G}_{0, F}^{i\nu}(q)\,    {\tilde G}_{0, F}^{\nu'i'}(q')\,
  \nonumber\\
 &\,\times\; 
{V_{4g}}_{\nu \rho\nu' \rho'}^{a c a' c'}\,
 \langle P| {\cal A}^{\rho'}_{c'}(0)\, {\cal A}^{\rho}_{c}(0)|P\rangle_c 
\nonumber\\
=&\,
{V_{4g}}_{\nu \rho\nu' \rho'}^{a c a' c'}\,
 \langle P| {\cal A}^{\rho'}_{c'}(0)\, {\cal A}^{\rho}_{c}(0)|P\rangle_c 
 \int \frac{d^Dq}{(2\pi)^D} 
 e^{-iq\cdot (x-x')}\,  
 \;{\tilde G}_{0, F}^{i\nu}(q)\,    {\tilde G}_{0, F}^{\nu'i'}(q)\,
\label{eq:fluct_prop_4gV}
\, ,
\end{align}
since the four gluon vertex factor ${V_{4g}}_{\nu \rho\nu' \rho'}^{a c a' c'}$ is independent of the vertex position, and of the momenta involved. Note that the two background field insertion happen at the same vertex, and thus at the same point $y$. For that reason, they can be both shifted to the origin, thanks to the relation~\eqref{Eq:invar_transl}. Inserting the result \eqref{eq:fluct_prop_4gV} into \eqref{eq:G-real-back_and_fluct} and performing the integration over $\Delta x^+$, one finds
\begin{align}
g^{\textrm{n.r.}}\left({\textrm{x}}, \mu^2\right)\bigg|^{\ref{Fig:4gV_real}}
=&\, 
 \frac{\textrm{x}P^-}{(2\pi)}\,
 \delta^{aa'}\,
{V_{4g}}_{\nu \rho\nu' \rho'}^{a c a' c'}\,
 \langle P| {\cal A}^{\rho'}_{c'}(0)\, {\cal A}^{\rho}_{c}(0)|P\rangle_c  
\int \frac{d^Dq}{(2\pi)^D}\,  
  2\pi\delta(q^-\!-\!\textrm{x}P^-)\, \left(\frac{i}{\left(q^2+i0^+\right)}\right)^2
  \nonumber\\
 &\,\times\; 
  \bigg[-g^{i\nu}+\frac{(\q^i n^{\nu}+n^i q^{\nu})}{[n\!\cdot\!q]}\bigg]
\bigg[-g^{\nu' i }+\frac{(\q^i n^{\nu'}+n^i q^{\nu'})}{[n\!\cdot\! q]}\bigg]
\nonumber\\
=&\, 
i\, \frac{\textrm{x}P^-}{(2\pi)}\,
 \delta^{aa'}\,
{V_{4g}}_{\nu \rho\nu' \rho'}^{a c a' c'}\,
 \langle P| {\cal A}^{\rho'}_{c'}(0)\, {\cal A}^{\rho}_{c}(0)|P\rangle_c  
\int \frac{d^{D-2}\q}{(2\pi)^{D-2}}\,  
 \bigg[-g^{i\nu}+\frac{\q^i n^{\nu}}{\textrm{x}P^-}\bigg]
\bigg[-g^{\nu' i }+\frac{\q^i n^{\nu'}}{\textrm{x}P^-}\bigg]
 \nonumber\\
&\,\times\;
\int \frac{dq^+}{2\pi}\,
  \frac{i}{\left(2\textrm{x}P^- q^+-\q^2+i0^+\right)^2}
\label{eq:G-real-NLO_4gV_1}
\, .
\end{align}
The integrand is a pure double pole in $q^+$. Hence, that integral over $q^+$ vanishes, by application of the residue theorem. The diagram \ref{Fig:4gV_real}, involving a four gluon vertex, is therefore not contributing to the gluon pdf at NLO:
\begin{align}
g^{\textrm{n.r.}}\left({\textrm{x}}, \mu^2\right)\bigg|^{\ref{Fig:4gV_real}}
=&\, 
 0
\label{eq:G-real-NLO_4gV_2}
\, .
\end{align}
%

%%%%%%%%%%%%%%%%%%%%%%%%%%%%%%%%%%%%%%%%%%%%%%%%%%%%%%%%%%%%%%%%%%%%
%%%%%%%%%%%%%%%%%%%%%%%%%%%%%%%%%%%%%%%%%%%%%%%%%%%%%%%%%%%%%%%%%%%%

\subsubsection{Gluon ladder diagram}

Let us now consider the diagram \ref{Fig:g2g_ladder}. 
Its main building block is the Feynman propagator for the gluon fluctuation field, with two insertions of the gluon background field of the target at two different vertices. It can be written as
\begin{align}
&\, \langle P|{\mathcal T}\; \delta A^{i'}_{a'}(x')\delta A^i_a(x) |P\rangle\bigg|_{\ref{Fig:g2g_ladder}} 
=
\int d^Dy \int d^Dy' \; \int \frac{d^Dq}{(2\pi)^D} \int \frac{d^Dk}{(2\pi)^D} \int \frac{d^Dq'}{(2\pi)^D}\,
 e^{-iq\cdot (x\!-\!y)}\, e^{-iq'\cdot (y'\!-\!x')}\, e^{-ik\cdot (y'\!-\!y)}\,
\nonumber\\
&\,\times\;
{\tilde G}_{0,F}^{i \nu}(q)\;
{\tilde G}_{0,F}^{\sigma'\sigma}(k)\,
{\tilde G}_{0, F}^{\nu' i'}(q')\,
{V_{3g}}_{\nu \sigma \rho}^{abc}(-q,-k,k\!+\!q)\;
{V_{3g}}_{\nu' \sigma' \rho'}^{a'bc'}(q',k,-q'\!-\!k)\;
 \langle P|{\mathcal T}\; {\cal A}^{\rho'}_{c'}(y')\, {\cal A}^{\rho}_c(y)|P\rangle_c 
 \, .
\label{eq:fluct_prop_g2g_1}
\end{align}
In order to express the integrand in terms of the background gluon distribution \eqref{eq:gluon-pdf_Bckgd}, one should write the background gauge fields in terms of background field strength components, inverting their relation  \eqref{eq:F_minus_nu_simple} in the target light-cone gauge as
\begin{align}
\int d^Dy \, e^{iy\cdot (q\!+\!k)}\,  {\cal A}^{\rho}_c(y) 
=&\,  
\int d^Dy \, e^{iy\cdot (q\!+\!k)}\, \frac{(-i) \overleftarrow{\partial_{y^+}}}{(q^-\!+\!k^-)}\, {\cal A}^{\rho}_c(y) 
=
\int d^Dy \, e^{iy\cdot (q\!+\!k)}\, \frac{(+i)}{(q^-\!+\!k^-)}\,  {\cal F}^{-\rho}_c(y) 
\, ,
\label{gauge_field_to_field_strength}
\end{align}
so that
\begin{align}
 \langle P|{\mathcal T}\; \delta A^{i'}_{a'}(x')\delta A^i_a(x) |P\rangle\bigg|_{\ref{Fig:g2g_ladder}} 
=&\,
\int d^Dy \int d^Dy' \; \int \frac{d^Dq}{(2\pi)^D} \int \frac{d^Dk}{(2\pi)^D} \int \frac{d^Dq'}{(2\pi)^D}\,
\frac{e^{-iq\cdot x}\, e^{iq'\cdot x'}\,  e^{iy\cdot (q\!+\!k)}\,  e^{-iy'\cdot (q'\!+\!k)}\,}{(q^-\!+\!k^-)({q'}^-\!+\!k^-)}
\nonumber\\
\times\;
{\tilde G}_{0,F}^{i \nu}(q)\;
{\tilde G}_{0,F}^{\sigma'\sigma}(k)\,
{\tilde G}_{0, F}^{\nu' i'}(q')\,
&\,
{V_{3g}}_{\nu \sigma \rho}^{abc}(-q,-k,k\!+\!q)\;
{V_{3g}}_{\nu' \sigma' \rho'}^{a'bc'}(q',k,-q'\!-\!k)\;
 \langle P|{\mathcal T}\; {\cal F}^{-\rho'}_{c'}(y')\, {\cal F}^{-\rho}_c(y)|P\rangle_c 
 \nonumber\\
 =&\,
 \int d^D \Delta y\; \int \frac{d^Dq}{(2\pi)^D} \int \frac{d^Dk}{(2\pi)^D}\,
\frac{e^{iq\cdot (x'\!-\!x)}\,    e^{-i(q\!+\!k)\cdot \Delta y}\, }{(q^-\!+\!k^-)^2}\,
{\tilde G}_{0,F}^{i \nu}(q)\;
{\tilde G}_{0,F}^{\sigma'\sigma}(k)\,
{\tilde G}_{0, F}^{\nu' i'}(q)\,
\nonumber\\
&\,\times\;
{V_{3g}}_{\nu \sigma \rho}^{abc}(-q,-k,k\!+\!q)\;
{V_{3g}}_{\nu' \sigma' \rho'}^{a'bc'}(q,k,-q\!-\!k)\;
 \langle P|{\mathcal T}\; {\cal F}^{-\rho'}_{c'}(\Delta y)\, {\cal F}^{-\rho}_c(0)|P\rangle_c 
 \, ,
\label{eq:fluct_prop_g2g_2}
\end{align}
following the same steps as from Eqs.~\eqref{eq:fluct_prop_q2g_1} to \eqref{eq:fluct_prop_q2g_2}. Inserting the expression \eqref{eq:fluct_prop_g2g_2} into \eqref{eq:G-real-back_and_fluct} and integrating over $\Delta x^+$ leads to 
\begin{align}
g^{\textrm{n.r.}}\left({\textrm{x}}, \mu^2\right)\bigg|^{\ref{Fig:g2g_ladder}}
=&\, 
  \frac{\textrm{x}P^-}{(2\pi)}\,
 \int d^D\Delta y\,  \langle P|{\mathcal T}\; {\cal F}^{-\rho'}_{c'}(\Delta y)\, {\cal F}^{-\rho}_c(0)|P\rangle_c\;  
\int \frac{d^Dk}{(2\pi)^D}
\int \frac{d^Dq}{(2\pi)^D}\,  
  \frac{e^{-i(k+q)\cdot \Delta y}}{(q^-\!+\!k^-)^2}\,
  2\pi\delta(q^-\!-\!\textrm{x}P^-)\, 
 \nonumber\\
&\,\times\;
{\tilde G}_{0, F}^{i\nu}(q)\,    {\tilde G}_{0, F}^{\nu'i}(q)\, {\tilde G}_{0,F}^{\sigma'\sigma}(k)\,
 {V_{3g}}_{\nu \sigma \rho}^{abc}(-q,-k,k\!+\!q)\;
{V_{3g}}_{\nu' \sigma' \rho'}^{a'bc'}(q,k,-q\!-\!k)\;
\nonumber\\
=&\, 
\int \frac{d^D\Delta y}{2\pi\, \textrm{x}P^-}\, \langle P|{\mathcal T}\; {\cal F}^{-\rho'}_{c'}(\Delta y)\, {\cal F}^{-\rho}_c(0)|P\rangle_c\;  
  \int \frac{d^Dk}{(2\pi)^D}
\int \frac{d^Dq}{(2\pi)^D}\,   e^{-i(k+q)\cdot \Delta y}\,
 \nonumber\\
&\,\times\;
\frac{i\,  (2\pi)\delta(q^-\!-\!\textrm{x}P^-)}{\left(q^2+i0^+\right)^2\left(k^2+i0^+\right)}\, \delta^{c'c}\, 
   {\cal N}^{gg}_{\rho'\rho}(k;q)\,
   \, ,
\label{eq:G-real-NLO_g2g_1}
\end{align}
with the numerator defined as
\begin{align}
 \delta^{c'c}\, {{\cal N}}^{gg}_{\rho'\rho}(k;q)
 \equiv &\,
-  \frac{(q^-)^2}{(q^-\!+\!k^-)^2}\,
\bigg[-g^{i\nu}+\frac{(\q^i n^{\nu}+n^i q^{\nu})}{[n\!\cdot\!q]}\bigg]
\bigg[-g^{\nu' i }+\frac{(\q^i n^{\nu'}+n^i q^{\nu'})}{[n\!\cdot\! q]}\bigg]
\nonumber\\
&\,\times\;
\bigg[-g^{\sigma'\sigma }+\frac{(k^{\sigma'} n^{\sigma}+n^{\sigma'} k^{\sigma})}{[n\!\cdot\! k]}\bigg]
 {V_{3g}}_{\nu \sigma \rho}^{abc}(-q,-k,k\!+\!q)\;
{V_{3g}}_{\nu' \sigma' \rho'}^{abc'}(q,k,-q\!-\!k)\;
\, .
\label{def_num_g2g_ladder}
\end{align}

Like in the case of the quark-to-gluon contribution, in Sec.~\ref{sec:q2g}, using the constraint  $q^-=\textrm{x}P^-$ and changing variables from $q$ to $l=k+q$ for the other components of $q^-$, leads to a result of the form
\begin{align}
g^{\textrm{n.r.}}\left({\textrm{x}}, \mu^2\right)\bigg|^{\ref{Fig:g2g_ladder}}
=&\, 
\int \frac{d^D\Delta y}{2\pi\, \textrm{x}P^-}\, \langle P|{\mathcal T}\; {\cal F}^{-\rho'}_{c}(\Delta y)\, {\cal F}^{-\rho}_c(0)|P\rangle_c\; 
\int \frac{d^{D-2}\l}{(2\pi)^{D-2}}\, e^{i\l \cdot \Delta \y}\,
 \int \frac{dl^+}{2\pi}\, e^{-i l^+  \Delta y^-}\,
   {{\cal H}}^{gg}_{\rho'\rho}(l^+,\l;\Delta y^+)\,
\label{eq:G-real-NLO_g2g_2}
\, ,
\end{align}
where
\begin{align}
 {{\cal H}}^{gg}_{\rho'\rho}(l^+,\l;\Delta y^+)
\equiv &\, 
 \int \frac{d^Dk}{(2\pi)^D}
 e^{-i(k^-+\textrm{x}P^-)\Delta y^+}\,
\frac{i\; {{\cal N}}^{gg}_{\rho'\rho}(k;l^+\!-\!k^+,\textrm{x}P^-,\l\!-\!\k)
}{\Big(2\textrm{x}P^-(l^+\!-\!k^+)-(\l\!-\!\k)^2+i0^+\Big)^2
\left(k^2+i0^+\right)}\,
\,
\label{def_hard_fact_g2g}
\, .
\end{align}

Using the standard expression for the three gluon vertices in momentum space, \emph{i.e.}
\begin{align}
 {V_{3g}}_{\nu \sigma \rho}^{abc}(-q,-k,k\!+\!q)
 =&\,
 -g\, \mu^{\epsilon}\, f^{abc} \Big\{
 (k_{\rho}\!-\!q_{\rho})\, g_{\nu \sigma} - (2k_{\nu}\!+\!q_{\nu})\, g_{\rho \sigma}+(2q_{\sigma}\!+\!k_{\sigma})\, g_{\nu \rho}
 \Big\}
\label{def_3g_vertex}
\, ,
\end{align}
the numerator \eqref{def_num_g2g_ladder} can be written as
\begin{align}
{{\cal N}}^{gg}_{\rho'\rho}(k;q)
= &\,
g^2 C_A\,  \mu^{2\epsilon}\, \frac{(q^-)^2}{(q^-\!+\!k^-)^2}\,
\bigg[-g^{i\nu}+\frac{\q^i n^{\nu}}{[n\!\cdot\!q]}\bigg]
\bigg[-g^{\nu' i }+\frac{\q^i n^{\nu'}}{[n\!\cdot\! q]}\bigg]
\bigg[-g^{\sigma'\sigma }+\frac{(k^{\sigma'} n^{\sigma}+n^{\sigma'} k^{\sigma})}{[n\!\cdot\! k]}\bigg]\;
\nonumber\\
&\,\times\;
\Big\{
 (k_{\rho}\!-\!q_{\rho})\, g_{\nu \sigma} - (2k_{\nu}\!+\!q_{\nu})\, g_{\rho \sigma}+(2q_{\sigma}\!+\!k_{\sigma})\, g_{\nu \rho}
 \Big\}
\nonumber\\
&\,\times\;
\Big\{
 (k_{\rho'}\!-\!q_{\rho'})\, g_{\nu' \sigma'} - (2k_{\nu'}\!+\!q_{\nu'})\, g_{\rho' \sigma'}+(2q_{\sigma'}\!+\!k_{\sigma'})\, g_{\nu' \rho'}
 \Big\}
\, ,
\label{num_g2g_ladder_1}
\end{align}
or, after some algebra,
\begin{align}
{{\cal N}}^{gg}_{\rho'\rho}(k;q)
= &\,
g^2 C_A\,  \mu^{2\epsilon}\, \frac{(q^-)^2}{(q^-\!+\!k^-)^2}\,
\Bigg\{{g_{\rho}}^i {g_{\rho'}}^i
\left[ 4\frac{q^-}{[k^-]} \big(k^2+2k\!\cdot\! q\big) +k^2-4q^2 + 
4 \left(\k\!-\!\frac{k^-}{[q^-]}\q\right)^2
\right]
\nonumber\\
&\,\;
+2 \left(1\!-\! \frac{q^-}{[k^-]}\right)  \left[\k^i\!-\! \frac{k^-}{[q^-]}\q^i\right]
 \left[(k_{\rho}\!+\!q_{\rho}){g_{\rho'}}^i +{g_{\rho}}^i(k_{\rho'}\!+\!q_{\rho'})\right]\,
%\nonumber\\
%&\,\;
 + 2(1\!-\! \delta_s \epsilon)\, (k_{\rho}\!-\!q_{\rho}) (k_{\rho'}\!-\!q_{\rho'})
 \Bigg\}+\cdots
\, ,
\label{num_g2g_ladder_2}
\end{align}
where the dots stand for terms proportional to either $n_{\rho}$ or $n_{\rho'}$, which will eventually vanish by contraction with the field strengths  in Eq.~\eqref{eq:G-real-NLO_g2g_1}. Moreover, with the same notation, one has
\begin{align}
k_{\rho}\!\pm\!q_{\rho} 
=&\,
{g_{\rho}}^+ (k^-\!\pm\!q^-) - {g_{\rho}}^j (k^j\!\pm\!q^j)+\cdots
\, .
\label{k_pm_q_components}
\end{align}

In Sec.~\ref{sec:q2g}, for the quark-to-gluon channel contribution, at such stage of the calculation, we have performed the integration over $k^+$ first, using the residue theorem. For contributions including a factor $1/[k^-]$ with the ML prescription \eqref{eq:ML_prescription}, it is difficult to perform the $k^+$ integral first without spoiling the ML prescription. Instead, it is more appropriate to perform the $k^-$ integral first. However, it would be very cumbersome to perform the $k^-$ integral for the whole expression \eqref{def_hard_fact_g2g}, due to the complicated dependence on $k^-$ in the integrand.
For these reasons, we will adopt the following strategy: we will split the integrand of Eq.~\eqref{def_hard_fact_g2g} into two types of contributions: on the one hand, terms proportional to $1/[k^-]$ and further depending on $k^-$ only through the phase factor and the scalar propagator denominators, for which we will perform the $k^-$ integration first, and on the other hand terms which are regular at $k^-=0$, for which we will perform the $k^+$ integration first, in the same way as in Sec.~\ref{sec:q2g}.

For convenience let us first split the numerator \eqref{num_g2g_ladder_2} into three parts as 
\begin{align}
{{\cal N}}^{gg}_{\rho'\rho}(k;q)\bigg|_{A}
= &\,
4g^2 C_A\,  {g_{\rho}}^i {g_{\rho'}}^i\, \mu^{2\epsilon}\, \frac{(q^-)^2}{(q^-\!+\!k^-)^2}\,
\frac{q^-}{[k^-]} \big(k^2+2k\!\cdot\! q\big)
\nonumber\\
{{\cal N}}^{gg}_{\rho'\rho}(k;q)\bigg|_{B}
= &\,
2 g^2 C_A\,   \mu^{2\epsilon}\, \frac{(q^-)^2}{(q^-\!+\!k^-)^2}\,
 \left(1\!-\! \frac{q^-}{[k^-]}\right)  \left[\k^i\!-\! \frac{k^-}{[q^-]}\q^i\right]
 \left[(k_{\rho}\!+\!q_{\rho}){g_{\rho'}}^i +{g_{\rho}}^i(k_{\rho'}\!+\!q_{\rho'})\right]
\nonumber\\
{{\cal N}}^{gg}_{\rho'\rho}(k;q)\bigg|_{C}
= &\,
g^2 C_A\,  \mu^{2\epsilon}\, \frac{(q^-)^2}{(q^-\!+\!k^-)^2}\,
\Bigg\{
{g_{\rho}}^i {g_{\rho'}}^i
\left[  k^2-4q^2+4 \left(\k\!-\!\frac{k^-}{[q^-]}\q\right)^2\right]
+2(1\!-\! \delta_s \epsilon)\, (k_{\rho}\!-\!q_{\rho}) (k_{\rho'}\!-\!q_{\rho'})
 \Bigg\}+\cdots
\nonumber\\
{{\cal N}}^{gg}_{\rho'\rho}(k;q)
= &\,
{{\cal N}}^{gg}_{\rho'\rho}(k;q)\bigg|_{A}
+{{\cal N}}^{gg}_{\rho'\rho}(k;q)\bigg|_{B}
+{{\cal N}}^{gg}_{\rho'\rho}(k;q)\bigg|_{C}
\, ,
\label{num_g2g_ladder_3}
\end{align}
isolating in the parts $A$ and $B$ the two structures containing a $1/[k^-]$.

For the part $A$, we have
\begin{align}
{{\cal N}}^{gg}_{\rho'\rho}(k;l^+\!-\!k^+,\textrm{x}P^-,\l\!-\!\k)\bigg|_{A}
= &\,
4g^2 C_A\,   {g_{\rho}}^i {g_{\rho'}}^i\, \mu^{2\epsilon}\, \frac{(\textrm{x}P^-)^2}{(\textrm{x}P^-\!+\!k^-)^2}\,
\frac{\textrm{x}P^-}{[k^-]} 
\Big[2l^+k^- \!+\!2\textrm{x}P^-k^+ \!+\!\k^2 \!-\!2\k\!\cdot\! \l\Big]
\, .
\label{num_g2g_ladder_A1}
\end{align}
Its contribution proportional to $1/[k^-]$ with no further $k^-$ dependence is
\begin{align}
{{\cal N}}^{gg}_{\rho'\rho}(k;l^+\!-\!k^+,\textrm{x}P^-,\l\!-\!\k)\bigg|_{A,\textrm{ ML term}}
= &\,
4g^2 C_A\,   {g_{\rho}}^i {g_{\rho'}}^i\, \mu^{2\epsilon}\, 
\frac{\textrm{x}P^-}{[k^-]} 
\Big[2\textrm{x}P^-k^+ \!+\!\k^2 \!-\!2\k\!\cdot\! \l\Big]
\, ,
\label{num_g2g_ladder_A_ML}
\end{align}
and the leftover which is regular at $k^-=0$ is found to be
\begin{align}
&\,{{\cal N}}^{gg}_{\rho'\rho}(k;l^+\!-\!k^+,\textrm{x}P^-,\l\!-\!\k)\bigg|_{A,\, k^-\textrm{ reg.}}
= 
{{\cal N}}^{gg}_{\rho'\rho}(k;l^+\!-\!k^+,\textrm{x}P^-,\l\!-\!\k)\bigg|_{A}
-{{\cal N}}^{gg}_{\rho'\rho}(k;l^+\!-\!k^+,\textrm{x}P^-,\l\!-\!\k)\bigg|_{A,\textrm{ ML term}}
\nonumber\\
= &\,
4g^2 C_A\,   {g_{\rho}}^i {g_{\rho'}}^i\, \mu^{2\epsilon}\, \frac{\textrm{x}P^-}{(\textrm{x}P^-\!+\!k^-)^2}\,
\Bigg\{2(\textrm{x}P^-)^2 l^+ \!-\! (k^-\!+\!2\textrm{x}P^-)\Big[2\textrm{x}P^-k^+ \!+\!\k^2 \!-\!2\k\!\cdot\! \l\Big]\Bigg\}
\nonumber\\
= &\,
4g^2 C_A\,   {g_{\rho}}^i {g_{\rho'}}^i\, \mu^{2\epsilon}\, \frac{\textrm{x}P^-}{(\textrm{x}P^-\!+\!k^-)^2}\,
\Bigg\{2(\textrm{x}P^-)^2 l^+ \!-\! (k^-\!+\!2\textrm{x}P^-)\Big[2\textrm{x}P^-l^+ \!-\!\l^2 \Big]
\nonumber\\
&\, \hspace{5.5cm}
+(k^-\!+\!2\textrm{x}P^-)\Big[2\textrm{x}P^-(l^+\!-\!k^+)-(\l\!-\!\k)^2\Big]
\Bigg\}
\, .
\label{num_g2g_ladder_A_reg}
\end{align}

\begin{figure}[ht]
    \includegraphics[width=0.40\textwidth]{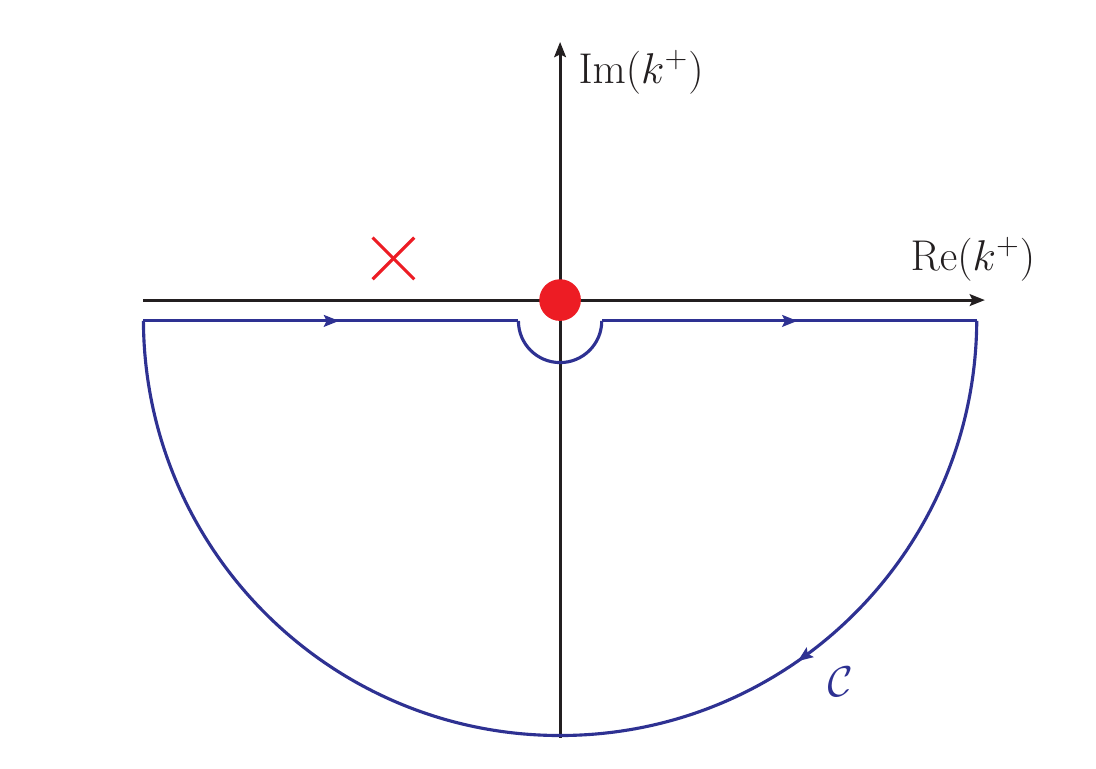}
\caption{\label{Fig:int_contour_kplus_flip} Integration contour in $k^+$ used to demonstrate the relation \eqref{eq:kplus_flip}, for $\Delta y^+<0$. The red dot at the origin represents the essential singularity, whereas the red cross represents the double pole.}
\end{figure}

The contribution of the term \eqref{num_g2g_ladder_A_ML} to the expression \eqref{def_hard_fact_g2g} is then
\begin{align}
 {{\cal H}}^{gg}_{\rho'\rho}(l^+,\l;\Delta y^+)\bigg|_{A,\textrm{ ML term}}
= &\, 
4g^2 C_A\,  {g_{\rho}}^i {g_{\rho'}}^i\, \mu^{2\epsilon}\, 
 \int \frac{d^{D-2}\k}{(2\pi)^{D-2}}    \int \frac{dk^+}{2\pi}
 \frac{\textrm{x}P^-\Big[2\textrm{x}P^-k^+ \!+\!\k^2 \!-\!2\k\!\cdot\! \l\Big]}{\Big(2\textrm{x}P^-(l^+\!-\!k^+)-(\l\!-\!\k)^2+i0^+\Big)^2}\,
\nonumber\\
&\, \times\,
 \int \frac{dk^-}{2\pi}
 e^{-i(k^-+\textrm{x}P^-)\Delta y^+}\,
\frac{i (2k^+)}{\left(k^2+i0^+\right)\left(2k^+k^-+i0^+\right)}\,
\,
\label{hard_fact_g2g_A_ML_1}
\, .
\end{align}
Applying the residue theorem, one finds
\begin{align}
&\, \int \frac{dk^-}{2\pi}
\frac{i (2k^+)\: e^{-i(k^-+\textrm{x}P^-)\Delta y^+}\,}{\left(2k^+k^-\!-\!\k^2+i0^+\right)\left(2k^+k^-+i0^+\right)}\,
=
\frac{1}{\k^2}\, 
 \int \frac{dk^-}{2\pi}
 e^{-i(k^-+\textrm{x}P^-)\Delta y^+}\,
 \left[
\frac{i (2k^+)}{\left(2k^+k^-\!-\!\k^2+i0^+\right)}\,
-
\frac{i (2k^+)}{\left(2k^+k^-+i0^+\right)}\,
\right]
\nonumber\\
= &\, 
\frac{1}{\k^2}\, \Big[\theta(\Delta y^+)\theta(k^+)-\theta(-\Delta y^+)\theta(-k^+)\Big]
\left[e^{-i \left(\textrm{x}P^-+\frac{\k^2}{2k^+}\right)\Delta y^+} -e^{-i \textrm{x}P^-\, \Delta y^+}\right]
\,
\label{kmin_int_first}
\, .
\end{align}
Inserting the result \eqref{kmin_int_first} into Eq.~\eqref{hard_fact_g2g_A_ML_1} thus leads to an integral over $k^+$ of the form
\begin{align}
&\int \frac{dk^+}{2\pi}\, \Big[\theta(\Delta y^+)\theta(k^+)-\theta(-\Delta y^+)\theta(-k^+)\Big]\, f(k^+)
\, ,
\end{align}
with 
\begin{align}
f(k^+)
= &\,
\frac{\textrm{x}P^-\Big[2\textrm{x}P^-k^+ \!+\!\k^2 \!-\!2\k\!\cdot\! \l\Big]}{\Big(2\textrm{x}P^-(l^+\!-\!k^+)-(\l\!-\!\k)^2+i0^+\Big)^2}\,
\left[e^{-i \left(\textrm{x}P^-+\frac{\k^2}{2k^+}\right)\Delta y^+} -e^{-i \textrm{x}P^-\, \Delta y^+}\right]
\, .
\end{align}
That integrand behaves as
\begin{align}
f(k^+)
= &\, O\left((k^+)^{-2}\right) \quad \textrm{for } |k^+|\rightarrow +\infty 
\, .
\end{align}
It has a double pole  in $k^+$ just above the real axis. Its only other singularity is an essential singularity at $k^+=0$.
Noting that 
\begin{align}
e^{-i \frac{\k^2}{2k^+}\Delta y^+} =&\, e^{- \frac{\k^2}{2|k^+|^2}\, \textrm{Im}(k^+)\,\Delta y^+}\,
e^{-i \frac{\k^2}{2|k^+|^2}\, \textrm{Re}(k^+)\,\Delta y^+}
\, ,
\end{align}
it is clear that $f(k^+)$ converges to zero when one approaches the essential singularity from above the real axis if $\Delta y^+>0$, and  when one approaches the essential singularity from below the real axis if $\Delta y^+<0$.  
Thanks to these properties of $f(k^+)$, applying the residue theorem on the contour ${\cal C}$ defined on Fig.~\ref{Fig:int_contour_kplus_flip}, one obtains 
%%%%%%%%%%%%%%%%%%%%%%%%%%%%%%%%%%%%%%%%%%%%%%%%%%%%%%%%%%%%%%%%%%%%%%%%%%%%%%%%%%%%%%
%%%%%%%%%%%%%%%%%%%%%%%%%%%%%%%%%%%%%%%%%%%%%%%%%%%%%%%%%%%%%%%%%%%%%%%%%%%%%%%%%%%%%%
%%%%%%%%%%%%%%%%%%%%%%%%%%%%%%%%%%%%%%%%%%%%%%%%%%%%%%%%%%%%%%%%%%%%%%%%%%%%%%%%%%%%%%
%%%%%%%%%%%%%%%%%%%%%%%%%%%%%%%%%%%%%%%%%%%%%%%%%%%%%%%%%%%%%%%%%%%%%%%%%%%%%%%%%%%%%%
%
\begin{align}
0 =&\, \theta(-\Delta y^+)\int_{{\cal C}} \frac{dk^+}{2\pi}\,  f(k^+)
= \theta(-\Delta y^+)\int_{-\infty}^{0} \frac{dk^+}{2\pi}\,  f(k^+)
+\theta(-\Delta y^+)\int_{0}^{+\infty} \frac{dk^+}{2\pi}\,  f(k^+)
\, ,
\label{eq:kplus_flip}
\end{align}
and thus
\begin{align}
\int \frac{dk^+}{2\pi}\, \Big[\theta(\Delta y^+)\theta(k^+)-\theta(-\Delta y^+)\theta(-k^+)\Big]\, f(k^+)
=&\,
\int_{0}^{+\infty} \frac{dk^+}{2\pi}\,  f(k^+)
\, .
\label{contour_deform_kplus_int}
\end{align}
All in all, the expression \eqref{hard_fact_g2g_A_ML_1} becomes
\begin{align}
 {{\cal H}}^{gg}_{\rho'\rho}(l^+,\l;\Delta y^+)\bigg|_{A,\textrm{ ML term}}
= &\, 
4g^2 C_A\,   {g_{\rho}}^i {g_{\rho'}}^i\, \mu^{2\epsilon}\, 
 \int \frac{d^{D-2}\k}{(2\pi)^{D-2}}\,    \frac{1}{\k^2}\, 
 \int_{0}^{+\infty} \frac{dk^+}{2\pi}\,
 \left[e^{-i \left(\textrm{x}P^-+\frac{\k^2}{2k^+}\right)\Delta y^+} -e^{-i \textrm{x}P^-\, \Delta y^+}\right]
 \nonumber\\
&\, \times\,
 \frac{\textrm{x}P^-\Big[2\textrm{x}P^-k^+ \!+\!\k^2 \!-\!2\k\!\cdot\! \l\Big]}{\Big(2\textrm{x}P^-(l^+\!-\!k^+)-(\l\!-\!\k)^2+i0^+\Big)^2}\,
\,
\label{hard_fact_g2g_A_ML_2}
\, .
\end{align}
Applying the changes of variables
\begin{align}
k^+\mapsto  \textrm{z} = \frac{2\textrm{x}P^-k^+}{(2\textrm{x}P^-k^+\!+\!\k^2)}
\label{change_var_kplus_to_z}
\end{align}
and then
\begin{align}
\k\mapsto  \K = \k - (1\!-\!\textrm{z})\l
\label{change_var_kperp_to_Kperp_bis}
\, ,
\end{align}
one finds
\begin{align}
 {{\cal H}}^{gg}_{\rho'\rho}(l^+,\l;\Delta y^+)\bigg|_{A,\textrm{ ML term}}
= &\, 
4\alpha_s C_A\,   {g_{\rho}}^i {g_{\rho'}}^i\, \int_{0}^{1}\frac{d \textrm{z}}{(1\!-\! \textrm{z})}\,
\left[e^{-i \frac{\textrm{x}}{\textrm{z}}P^-\, \Delta y^+} -e^{-i \textrm{x}P^-\, \Delta y^+}\right]
\nonumber\\
&\, \times\,
\mu^{2\epsilon}\, 
 \int \frac{d^{2-2\epsilon}\K}{(2\pi)^{2-2\epsilon}}\,   
 \frac{\Big[\K^2  \!-\!(1\!-\! \textrm{z})^2 \l^2\Big]}{\Big(\K^2 +  \textrm{z}(1\!-\! \textrm{z}) (-l^2-i0^+)\Big)^2}\Bigg|_{l^-\equiv \frac{\textrm{x}}{\textrm{z}}P^-}
\nonumber\\
=&\,
\frac{\alpha_s C_A}{\pi}\,  {g_{\rho}}^i {g_{\rho'}}^i\, \int_{0}^{1}d \textrm{z}\;
\Gamma(1\!+\!\epsilon)\, \left(\frac{-l^2\!-\! i0^+}{4\pi \mu^2}\right)^{-\epsilon}\, \textrm{z}^{-\epsilon}(1\!-\!\textrm{z})^{-\epsilon}
\nonumber\\
&\, \times\,
\left[e^{-i \frac{\textrm{x}}{\textrm{z}}P^-\, \Delta y^+} -e^{-i \textrm{x}P^-\, \Delta y^+}\right]
\bigg[
\frac{(1\!-\! \epsilon)}{\epsilon\, (1\!-\! \textrm{z})}
+\frac{\l^2}{\textrm{z}(l^2\!+\! i0^+)}
\bigg]\Bigg|_{l^-\equiv \frac{\textrm{x}}{\textrm{z}}P^-}
\,
\label{hard_fact_g2g_A_ML_3}
\, .
\end{align}
Note that the $1/(1\!-\! \textrm{z})$ in this result originates from the $1/[k^-]$ in the starting expression. Due the the ML prescription, the subtraction term $-e^{-i \textrm{x}P^-\, \Delta y^+}$ is generated when performing the integration over $k^-$. The difference between the two phases factors vanishes at $\textrm{z}=1$, which convert the potential pole at $\textrm{z}=1$ into a removable singularity. In lightcone gauge, the ML prescription thus automatically leads to a result which is regular at $\textrm{z}=1$, even before summing over diagrams.

Let us now consider the regular leftover \eqref{num_g2g_ladder_A_reg} from the part $A$ of the numerator. When inserting it in Eq.~\eqref{def_hard_fact_g2g}, the integration over $k^+$ can be performed thanks to the relation \eqref{kplus_contour_integ}. In such a way, one obtains\footnote{A discussion is in order concerning the denominators  $(\textrm{x}P^-\!+\!k^-)$, produced when expressing the background gauge field in terms of the background field strength, see Eq.~\eqref{gauge_field_to_field_strength}. On the one hand, they are reminiscent of the extra denominator apprearing in the gluon propagator in light-cone gauge (Eq.~\eqref{eq:Feyn-prop_def}), that we have defined with the ML prescription \eqref{eq:ML_prescription}. The ML prescription is a consequence of the Hamiltonian quantization in the light-cone gauge   \cite{Bassetto:1984dq}, and as such is applied to the propagator of the gluon fluctuations. By contrast, it is not entirely clear if the ML prescription should be applied as well to Eq.~\eqref{gauge_field_to_field_strength}, involving only background fields. Other prescriptions have been considered in the literature for such denominators, for example the Cauchy principal value, or the retarded prescription.
However, in the present study, it turns out that this choice of prescription does not matter. 
First, there is no such denominator surviving in the ML contributions, like \eqref{num_g2g_ladder_A_ML}, due to the replacement $(\textrm{x}P^-\!+\!k^-)\mapsto \textrm{x}P^- >0$. 
Second, in the regular leftover contributions, like \eqref{num_g2g_ladder_A_reg}, we perform the integration over $k^+$ first, using the residue theorem, but that denominator is independent of $k^+$ in all of the motivated prescriptions. In particular, in our notations for this diagram, the ML prescription for this denominator would be
\begin{align}
\frac{1}{[\textrm{x}P^-\!+\!k^-]}
=&\,
\frac{l^+}{\left(l^+(\textrm{x}P^-\!+\!k^-)+i0^+\right)} 
\, .
\end{align}
Moreover, the result of the integration over $k^+$ is proportional to $\theta(k^-)$, see Eq.~\eqref{kplus_contour_integ}. The denominator  $(\textrm{x}P^-\!+\!k^-)$ is then always strictly positive in this calculation, so that the choice of prescription to define it is irrelevant.}
\begin{align}
 {{\cal H}}^{gg}_{\rho'\rho}(l^+,\l;\Delta y^+)\bigg|_{A,\, k^-\textrm{ reg.}}
= &\, 
4g^2 C_A\,   {g_{\rho}}^i {g_{\rho'}}^i\, \mu^{2\epsilon}\, 
 \int \frac{d^{D-2}\k}{(2\pi)^{D-2}}    \int \frac{dk^-}{2\pi}\,  e^{-i(k^-+\textrm{x}P^-)\Delta y^+}\, 
 \frac{\textrm{x}P^-}{(\textrm{x}P^-\!+\!k^-)^2}\, 
 \frac{\theta(k^-)}{2k^-}
 \nonumber\\
&\, \times\,
\Bigg\{ 
\frac{\Big[
2(\textrm{x}P^-)^2 l^+ \!-\! (k^-\!+\!2\textrm{x}P^-)\Big(2\textrm{x}P^-l^+ \!-\!\l^2 \Big)
\Big]}{\Big[ 
-\frac{(k^-\!+\!\textrm{x}P^-)}{k^-}\, \left(\k\!-\!\frac{k^-}{(k^-\!+\!\textrm{x}P^-)}\, \l\right)^2-\frac{\textrm{x}P^-}{(k^-\!+\!\textrm{x}P^-)}\, \l^2 +2\textrm{x}P^-l^+ +i0^+\Big]^2}
\nonumber\\
&\, \hspace{1cm}
+\frac{(k^-\!+\!2\textrm{x}P^-)}{\Big[ 
-\frac{(k^-\!+\!\textrm{x}P^-)}{k^-}\, \left(\k\!-\!\frac{k^-}{(k^-\!+\!\textrm{x}P^-)}\, \l\right)^2-\frac{\textrm{x}P^-}{(k^-\!+\!\textrm{x}P^-)}\, \l^2 +2\textrm{x}P^-l^+ +i0^+\Big]}
\Bigg\}
\,
\label{hard_fact_g2g_A_reg_1}
\, .
\end{align}
Applying the changes of variables \eqref{change_var_kmin_to_z} and \eqref{change_var_kperp_to_Kperp} like in the quark-to-gluon channel, one arrives at
\begin{align}
 {{\cal H}}^{gg}_{\rho'\rho}(l^+,\l;\Delta y^+)\bigg|_{A,\, k^-\textrm{ reg.}}
= &\, 
4\alpha_s C_A\, , {g_{\rho}}^i {g_{\rho'}}^i\, 
 \int_{0}^{1}d \textrm{z}\,
e^{-i \frac{\textrm{x}}{\textrm{z}}P^-\, \Delta y^+} 
\mu^{2\epsilon}\, 
 \int \frac{d^{2-2\epsilon}\K}{(2\pi)^{2-2\epsilon}}
 \nonumber\\
&\, \times\,
\Bigg\{ 
\frac{\textrm{z}(1\!-\!\textrm{z})\Big[
2\textrm{x}P^- l^+ \!-\! \frac{(1\!+\!\textrm{z})}{\textrm{z}}\big(2\textrm{x}P^-l^+ \!-\!\l^2 \big)
\Big]}{\Big[\K^2 +  \textrm{z}(1\!-\! \textrm{z}) (-l^2-i0^+)\Big]^2}
%\nonumber\\
%&\, \hspace{1cm}
-\frac{(1\!+\!\textrm{z})}{\Big[\K^2 +  \textrm{z}(1\!-\! \textrm{z}) (-l^2-i0^+)\Big]}
\Bigg\}\Bigg|_{l^-\equiv \frac{\textrm{x}}{\textrm{z}}P^-}
\nonumber\\
= &\,
\frac{\alpha_s C_A}{\pi}\,   {g_{\rho}}^i {g_{\rho'}}^i\, \int_{0}^{1}d \textrm{z}\;
\Gamma(1\!+\!\epsilon)\, \left(\frac{-l^2\!-\! i0^+}{4\pi \mu^2}\right)^{-\epsilon}\, \textrm{z}^{-\epsilon}(1\!-\!\textrm{z})^{-\epsilon}\;
e^{-i \frac{\textrm{x}}{\textrm{z}}P^-\, \Delta y^+} 
\nonumber\\
&\, \times\,
\bigg\{
-\frac{1}{(l^2\!+\! i0^+)}\left[2\textrm{x}P^- l^+ 
    -\frac{(1\!+\! \textrm{z})}{ \textrm{z}} \big(2\textrm{x}P^-l^+ \!-\!\l^2 \big)\right]
    -\frac{(1\!+\! \textrm{z})}{\epsilon}
\bigg\}\Bigg|_{l^-\equiv \frac{\textrm{x}}{\textrm{z}}P^-}
\nonumber\\
= &\,
\frac{\alpha_s C_A}{\pi}\,   {g_{\rho}}^i {g_{\rho'}}^i\, \int_{0}^{1}d \textrm{z}\;
\Gamma(1\!+\!\epsilon)\, \left(\frac{-l^2\!-\! i0^+}{4\pi \mu^2}\right)^{-\epsilon}\, \textrm{z}^{-\epsilon}(1\!-\!\textrm{z})^{-\epsilon}\;
e^{-i \frac{\textrm{x}}{\textrm{z}}P^-\, \Delta y^+} 
\nonumber\\
&\, \times\,
\bigg\{
-\frac{(1\!-\!\epsilon\!+\! \textrm{z})}{\epsilon}
-\frac{\l^2}{\textrm{z}(l^2\!+\! i0^+)}
\bigg\}\Bigg|_{l^-\equiv \frac{\textrm{x}}{\textrm{z}}P^-}
\,
\label{hard_fact_g2g_A_reg_2}
\, .
\end{align}

The total contribution $A$ is then obtained by summing the results \eqref{hard_fact_g2g_A_ML_3} and \eqref{hard_fact_g2g_A_reg_2}, as 
\begin{align}
 {{\cal H}}^{gg}_{\rho'\rho}(l^+,\l;\Delta y^+)\bigg|_{A}
=&\,
\frac{\alpha_s C_A}{\pi}\,  {g_{\rho}}^i {g_{\rho'}}^i\, \int_{0}^{1}d \textrm{z}\;
\Gamma(1\!+\!\epsilon)\, \left(\frac{-l^2\!-\! i0^+}{4\pi \mu^2}\right)^{-\epsilon}\, \textrm{z}^{-\epsilon}(1\!-\!\textrm{z})^{-\epsilon}\, 
\bigg\{-\frac{ \textrm{z}}{\epsilon}\, e^{-i \frac{\textrm{x}}{\textrm{z}}P^-\, \Delta y^+}
\nonumber\\
&\, 
+
\frac{(1\!-\!\epsilon)}{\epsilon}\, \frac{1}{(1\!-\! \textrm{z})}
\left[\textrm{z}\, e^{-i \frac{\textrm{x}}{\textrm{z}}P^-\, \Delta y^+} -e^{-i \textrm{x}P^-\, \Delta y^+}\right]
- \frac{\l^2}{\textrm{z}(l^2\!+\! i0^+)}\, e^{-i \textrm{x}P^-\, \Delta y^+}
\bigg\}\Bigg|_{l^-\equiv \frac{\textrm{x}}{\textrm{z}}P^-}
\,
\label{hard_fact_g2g_A_res}
\, .
\end{align}

For the part $B$ of the numerator defined in Eq.~\eqref{num_g2g_ladder_3} , one has
\begin{align}
{{\cal N}}^{gg}_{\rho'\rho}(k;l^+\!-\!k^+,\textrm{x}P^-,\l\!-\!\k)\bigg|_{B}
= &\,
2g^2 C_A\,   \mu^{2\epsilon}\, \frac{\textrm{x}P^-}{[k^-]} \, \frac{(k^-\!-\!\textrm{x}P^-)}{(k^-\!+\!\textrm{x}P^-)}\,
\left[\k^i-\frac{k^-}{(k^-\!+\!\textrm{x}P^-)}\l^i\right]
\nonumber\\
&\, \times\,
\Big[({g_{\rho}}^{+} {g_{\rho'}}^i\!+\!{g_{\rho}}^i {g_{\rho'}}^{+})(k^-\!+\!\textrm{x}P^-)
-({g_{\rho}}^{j} {g_{\rho'}}^i\!+\!{g_{\rho}}^i {g_{\rho'}}^{j})\l^j\Big]
\, .
\label{num_g2g_ladder_B1}
\end{align}
Its contribution proportional to $1/[k^-]$ with no further $k^-$ dependence is thus
\begin{align}
{{\cal N}}^{gg}_{\rho'\rho}(k;l^+\!-\!k^+,\textrm{x}P^-,\l\!-\!\k)\bigg|_{B,\textrm{ ML term}}
= &\,
2g^2 C_A\,   \mu^{2\epsilon}\, \frac{(-1)\textrm{x}P^-}{[k^-]} \, \k^i\,
%\nonumber\\
%&\, \times\,
\Big[({g_{\rho}}^{+} {g_{\rho'}}^i\!+\!{g_{\rho}}^i {g_{\rho'}}^{+})\textrm{x}P^-
-({g_{\rho}}^{j} {g_{\rho'}}^i\!+\!{g_{\rho}}^i {g_{\rho'}}^{j})\l^j\Big]
\, ,
\label{num_g2g_ladder_B_ML}
\end{align}
whereas the regular leftover at $k^-=0$ is 
\begin{align}
&\,{{\cal N}}^{gg}_{\rho'\rho}(k;l^+\!-\!k^+,\textrm{x}P^-,\l\!-\!\k)\bigg|_{B,\, k^-\textrm{ reg.}}
= 
{{\cal N}}^{gg}_{\rho'\rho}(k;l^+\!-\!k^+,\textrm{x}P^-,\l\!-\!\k)\bigg|_{B}
-{{\cal N}}^{gg}_{\rho'\rho}(k;l^+\!-\!k^+,\textrm{x}P^-,\l\!-\!\k)\bigg|_{B,\textrm{ ML term}}
\nonumber\\
= &\,
2g^2 C_A\,  \mu^{2\epsilon}\, \frac{\textrm{x}P^-}{(k^-\!+\!\textrm{x}P^-)}\,
\Bigg\{\l^i
\Big[({g_{\rho}}^{+} {g_{\rho'}}^i\!+\!{g_{\rho}}^i {g_{\rho'}}^{+})\textrm{x}P^-
-({g_{\rho}}^{j} {g_{\rho'}}^i\!+\!{g_{\rho}}^i {g_{\rho'}}^{j})\l^j\Big]
\nonumber\\
&\, 
+\left[\k^i-\frac{k^-}{(k^-\!+\!\textrm{x}P^-)}\l^i\right]
\Big[({g_{\rho}}^{+} {g_{\rho'}}^i\!+\!{g_{\rho}}^i {g_{\rho'}}^{+})(k^-\!+\!\textrm{x}P^-)
-({g_{\rho}}^{j} {g_{\rho'}}^i\!+\!{g_{\rho}}^i {g_{\rho'}}^{j})2\l^j\Big]
\Bigg\}
\, .
\label{num_g2g_ladder_B_reg}
\end{align}

For the former contribution, one finds
\begin{align}
 {{\cal H}}^{gg}_{\rho'\rho}(l^+,\l;\Delta y^+)\bigg|_{B,\textrm{ ML term}}
= &\, 
-2 g^2 C_A\,  \mu^{2\epsilon}\, 
\textrm{x}P^-\, \Big[({g_{\rho}}^{+} {g_{\rho'}}^i\!+\!{g_{\rho}}^i {g_{\rho'}}^{+})\textrm{x}P^-
-({g_{\rho}}^{j} {g_{\rho'}}^i\!+\!{g_{\rho}}^i {g_{\rho'}}^{j})\l^j\Big]
\nonumber\\
&\, \times\,
 \int \frac{d^{D-2}\k}{(2\pi)^{D-2}}\,    \k^i\,  
 \int \frac{dk^+}{2\pi}
 \frac{1}{\Big(2\textrm{x}P^-(l^+\!-\!k^+)-(\l\!-\!\k)^2+i0^+\Big)^2}\,
\nonumber\\
&\, \times\,
 \int \frac{dk^-}{2\pi}
 e^{-i(k^-+\textrm{x}P^-)\Delta y^+}\,
\frac{i (2k^+)}{\left(k^2+i0^+\right)\left(2k^+k^-+i0^+\right)}\,
\,
\nonumber\\
=&\,
-2 g^2 C_A\,  \mu^{2\epsilon}\, 
\textrm{x}P^-\, \Big[({g_{\rho}}^{+} {g_{\rho'}}^i\!+\!{g_{\rho}}^i {g_{\rho'}}^{+})\textrm{x}P^-
-({g_{\rho}}^{j} {g_{\rho'}}^i\!+\!{g_{\rho}}^i {g_{\rho'}}^{j})\l^j\Big]
\nonumber\\
&\, \times\,
 \int \frac{d^{D-2}\k}{(2\pi)^{D-2}}\,    \frac{\k^i}{\k^2}\,  
 \int_{0}^{+\infty} \frac{dk^+}{2\pi}
 \frac{\left[e^{-i \left(\textrm{x}P^-+\frac{\k^2}{2k^+}\right)\Delta y^+} -e^{-i \textrm{x}P^-\, \Delta y^+}\right]}{\Big(2\textrm{x}P^-(l^+\!-\!k^+)-(\l\!-\!\k)^2+i0^+\Big)^2}\,
\label{hard_fact_g2g_B_ML_1}
\, ,
\end{align}
using the identity \eqref{kmin_int_first} as well as the demonstration leading to Eq.~\eqref{contour_deform_kplus_int}. Then, changing variables from $k^+$ to $\textrm{z}$  and from $\k$ to $\K$ (see Eqs.~\eqref{change_var_kplus_to_z} and \eqref{change_var_kperp_to_Kperp_bis}) leads to
\begin{align}
 {{\cal H}}^{gg}_{\rho'\rho}(l^+,\l;\Delta y^+)\bigg|_{B,\textrm{ ML term}}
= &\, 
-\frac{2 g^2 C_A}{4\pi}\,  
 \Big[({g_{\rho}}^{+} {g_{\rho'}}^i\!+\!{g_{\rho}}^i {g_{\rho'}}^{+})\textrm{x}P^-
-({g_{\rho}}^{j} {g_{\rho'}}^i\!+\!{g_{\rho}}^i {g_{\rho'}}^{j})\l^j\Big]
\int_{0}^{1}d \textrm{z}\,
\bigg[e^{-i \frac{\textrm{x}}{\textrm{z}}P^-\, \Delta y^+}  
\nonumber\\
&\, %\times\,
-e^{-i \textrm{x}P^-\, \Delta y^+}\bigg]\,
\mu^{2\epsilon}\,  \int \frac{d^{D-2}\K}{(2\pi)^{D-2}}\,   
 \frac{\left[\K^i +(1\!-\! \textrm{z})\l^i \right]}{\left[\K^2+\textrm{z}(1\!-\! \textrm{z})\big(\l^2\!-\!2\frac{\textrm{x}}{\textrm{z}}P^- l^+\!-\!i0^+\big)\right]^2}\,
 \nonumber\\
 = &\, 
 \frac{\alpha_s C_A}{\pi}\,
 \int_{0}^{1}d \textrm{z}\,
\bigg[e^{-i \frac{\textrm{x}}{\textrm{z}}P^-\, \Delta y^+}  -e^{-i \textrm{x}P^-\, \Delta y^+}\bigg]\,
\Gamma(1\!+\!\epsilon)\, \left(\frac{-l^2\!-\! i0^+}{4\pi \mu^2}\right)^{-\epsilon}\, \textrm{z}^{-\epsilon}(1\!-\!\textrm{z})^{-\epsilon}\;
 \nonumber\\
 &\, \times\,
\Big[({g_{\rho}}^{+} {g_{\rho'}}^i\!+\!{g_{\rho}}^i {g_{\rho'}}^{+})\textrm{x}P^-\, \l^i
-2{g_{\rho}}^i {g_{\rho'}}^{j}\l^i\l^j\Big]\:
%\frac{1}{2\textrm{z}}\, 
\frac{1}{2\textrm{z}\big(l^2\!+\!i0^+\big)}\Bigg|_{l^-\equiv \frac{\textrm{x}}{\textrm{z}}P^-}
\label{hard_fact_g2g_B_ML_2}
\, .
\end{align}

By contrast, for the contribution  \eqref{num_g2g_ladder_B_reg}, regular at $k^-=0$, one obtains 
\begin{align}
 {{\cal H}}^{gg}_{\rho'\rho}(l^+,\l;\Delta y^+)\bigg|_{B,\, k^-\textrm{ reg.}}
= &\, 
2g^2 C_A\,   \mu^{2\epsilon}\, 
 \int \frac{d^{D-2}\k}{(2\pi)^{D-2}}    \int \frac{dk^-}{2\pi}\,  e^{-i(k^-+\textrm{x}P^-)\Delta y^+}\, 
 \frac{\textrm{x}P^-}{(\textrm{x}P^-\!+\!k^-)}\, 
 \nonumber\\
&\, \times\,
\Bigg\{\l^i
\Big[({g_{\rho}}^{+} {g_{\rho'}}^i\!+\!{g_{\rho}}^i {g_{\rho'}}^{+})\textrm{x}P^-
-({g_{\rho}}^{j} {g_{\rho'}}^i\!+\!{g_{\rho}}^i {g_{\rho'}}^{j})\l^j\Big]
\nonumber\\
&\, 
+\left[\k^i-\frac{k^-}{(k^-\!+\!\textrm{x}P^-)}\l^i\right]
\Big[({g_{\rho}}^{+} {g_{\rho'}}^i\!+\!{g_{\rho}}^i {g_{\rho'}}^{+})(k^-\!+\!\textrm{x}P^-)
-({g_{\rho}}^{j} {g_{\rho'}}^i\!+\!{g_{\rho}}^i {g_{\rho'}}^{j})2\l^j\Big]
\Bigg\}
 \nonumber\\
&\, \times\,
%\Bigg\{ 
 \frac{\theta(k^-)}{2k^-}\,
 \frac{1}{\Big[ 
-\frac{(k^-\!+\!\textrm{x}P^-)}{k^-}\, \left(\k\!-\!\frac{k^-}{(k^-\!+\!\textrm{x}P^-)}\, \l\right)^2-\frac{\textrm{x}P^-}{(k^-\!+\!\textrm{x}P^-)}\, \l^2 +2\textrm{x}P^-l^+ +i0^+\Big]^2}
\,
\label{hard_fact_g2g_B_reg_1}
\, ,
\end{align}
by integrating over $k^+$ first, thanks to the identity \eqref{kplus_contour_integ}. Performing the changes of variables from $k^-$ to $\textrm{z}$ (see Eq.~\eqref{change_var_kmin_to_z}) and then from $\k$ to $\K$ (see Eq.~\eqref{change_var_kperp_to_Kperp}), one arrives at
\begin{align}
 {{\cal H}}^{gg}_{\rho'\rho}(l^+,\l;\Delta y^+)\bigg|_{B,\, k^-\textrm{ reg.}}
= &\, 
\frac{2 g^2 C_A}{4\pi}\,  
\int_{0}^{1}d \textrm{z}\,
e^{-i \frac{\textrm{x}}{\textrm{z}}P^-\, \Delta y^+}\,   (1\!-\! \textrm{z})\,
\mu^{2\epsilon}  \int \frac{d^{D-2}\K}{(2\pi)^{D-2}}\,   
\frac{1}{\left[\K^2+\textrm{z}(1\!-\! \textrm{z})\big(\l^2\!-\!2\frac{\textrm{x}}{\textrm{z}}P^- l^+\!-\!i0^+\big)\right]^2}\,
\nonumber\\
&\, \times\,
\bigg\{ \Big[({g_{\rho}}^{+} {g_{\rho'}}^i\!+\!{g_{\rho}}^i {g_{\rho'}}^{+})\textrm{x}P^-\l^i
-2{g_{\rho}}^i {g_{\rho'}}^{j}\l^i\l^j\Big]
\nonumber\\
&\, 
+\K^i \Big[({g_{\rho}}^{+} {g_{\rho'}}^i\!+\!{g_{\rho}}^i {g_{\rho'}}^{+})\, \frac{\textrm{x}}{\textrm{z}}P^-
-({g_{\rho}}^{j} {g_{\rho'}}^i\!+\!{g_{\rho}}^i {g_{\rho'}}^{j})2\l^j\Big]
\bigg\}
\nonumber\\
= &\, 
 -\frac{\alpha_s C_A}{\pi}\,
 \int_{0}^{1}d \textrm{z}\,
e^{-i \frac{\textrm{x}}{\textrm{z}}P^-\, \Delta y^+}  \,
\Gamma(1\!+\!\epsilon)\, \left(\frac{-l^2\!-\! i0^+}{4\pi \mu^2}\right)^{-\epsilon}\, \textrm{z}^{-\epsilon}(1\!-\!\textrm{z})^{-\epsilon}\;
 \nonumber\\
 &\, \times\,
\Big[({g_{\rho}}^{+} {g_{\rho'}}^i\!+\!{g_{\rho}}^i {g_{\rho'}}^{+})\textrm{x}P^-\, \l^i
-2{g_{\rho}}^i {g_{\rho'}}^{j}\l^i\l^j\Big]\:
%\frac{1}{2\textrm{z}}\, 
\frac{1}{2\textrm{z}\big(l^2\!+\!i0^+\big)}\Bigg|_{l^-\equiv \frac{\textrm{x}}{\textrm{z}}P^-}
\,
\label{hard_fact_g2g_B_reg_2}
\, .
\end{align}
Summing the two contributions \eqref{hard_fact_g2g_B_ML_2} and \eqref{hard_fact_g2g_B_reg_2} to the part $B$ then leads to
\begin{align}
 {{\cal H}}^{gg}_{\rho'\rho}(l^+,\l;\Delta y^+)\bigg|_{B}
= &\, 
 -\frac{\alpha_s C_A}{\pi}\,
 \int_{0}^{1}d \textrm{z}\,
e^{-i \textrm{x}P^-\, \Delta y^+}  \,
\Gamma(1\!+\!\epsilon)\, \left(\frac{-l^2\!-\! i0^+}{4\pi \mu^2}\right)^{-\epsilon}\, \textrm{z}^{-\epsilon}(1\!-\!\textrm{z})^{-\epsilon}\;
 \nonumber\\
 &\, \times\,
\Big[({g_{\rho}}^{+} {g_{\rho'}}^i\!+\!{g_{\rho}}^i {g_{\rho'}}^{+})\textrm{x}P^-\, \l^i
-2{g_{\rho}}^i {g_{\rho'}}^{j}\l^i\l^j\Big]\:
%\frac{1}{2\textrm{z}}\, 
\frac{1}{2\textrm{z}\big(l^2\!+\!i0^+\big)}\Bigg|_{l^-\equiv \frac{\textrm{x}}{\textrm{z}}P^-}
\,
\label{hard_fact_g2g_B_res_prelim}
\, .
\end{align}
In that contribution, at this stage, it is convenient to use the notation $l^-\equiv \textrm{x}P^-$ instead of $l^-\equiv \frac{\textrm{x}}{\textrm{z}}P^-$, in order to simplifty the integrand as
\begin{align}
 {{\cal H}}^{gg}_{\rho'\rho}(l^+,\l;\Delta y^+)\bigg|_{B}
= &\, 
\frac{\alpha_s C_A}{\pi}\,
 \int_{0}^{1}d \textrm{z}\,
e^{-i \textrm{x}P^-\, \Delta y^+}  \,
\Gamma(1\!+\!\epsilon)\, \left({4\pi \mu^2}\right)^{\epsilon}\,
\left({\l^2\!-\! \frac{2\textrm{x}P^- l^+}{\textrm{z}}\!-\! i0^+}\right)^{-1-\epsilon}\, 
\textrm{z}^{-\epsilon}(1\!-\!\textrm{z})^{-\epsilon}\;
 \nonumber\\
 &\, \times\,
\frac{1}{2\textrm{z}}\Big[l_{\rho}\, {g_{\rho'}}^i\, \l^i \!+\! {g_{\rho}}^i\, \l^i\,  l_{\rho'}\Big]\
\Bigg|_{l^-\equiv \textrm{x}P^-}
\,
\label{hard_fact_g2g_B_res}
\, ,
\end{align}
where we have dropped contributions proportional to $n_{\rho}$ or $n_{\rho'}$, which would vanish by contraction with the field strengths  in Eq.~\eqref{eq:G-real-NLO_g2g_1}.

Finally, it remains to calculate the contribution of the part $C$ of the numerator (see Eq.~\eqref{num_g2g_ladder_3}), 
\begin{align}
{{\cal N}}^{gg}_{\rho'\rho}(k;l^+\!-\!k^+,\textrm{x}P^-,\l\!-\!\k)\bigg|_{C}
= &\,
g^2 C_A\,   \mu^{2\epsilon}\, \frac{(\textrm{x}P^-)^2}{(k^-\!+\!\textrm{x}P^-)^2} \, 
\Bigg\{
{g_{\rho}}^i {g_{\rho'}}^i
\bigg[  k^2-4\Big(2\textrm{x}P^-(l^+\!-\!k^+)-(\l\!-\!\k)^2\Big)
\nonumber\\
&\,
+4 \frac{(k^-\!+\!\textrm{x}P^-)^2}{(\textrm{x}P^-)^2}\,\left(\k-\frac{k^-}{(k^-\!+\!\textrm{x}P^-)}\l\right)^2\bigg]
\nonumber\\
&\, 
\hspace{-2cm}+2(1\!-\! \delta_s \epsilon)\, \left[(k^-\!-\!\textrm{x}P^-){g_{\rho}}^{+} - (2\k^i\!-\!\l^i) {g_{\rho}}^i \right]\,
\left[(k^-\!-\!\textrm{x}P^-){g_{\rho'}}^{+} - (2\k^j\!-\!\l^j) {g_{\rho'}}^j \right]
 \Bigg\}
\, .
\label{num_g2g_ladder_C1}
\end{align}
Since it does not contain any piece in $1/[k^-]$, the integration over $k^+$ can be performed first for the entire part $C$, when inserting the expression \eqref{num_g2g_ladder_C1} into Eq.~\eqref{def_hard_fact_g2g}. The first term in Eq.~\eqref{num_g2g_ladder_C1} is proportional to $k^2$, compensating the corresponding denominator in Eq.~\eqref{def_hard_fact_g2g}, so that the integrand is a pure double pole in $k^+$, with a vanishing integral, like in Eq.~\eqref{eq:G-real-NLO_q2g_1}.
For the other terms in Eq.~\eqref{num_g2g_ladder_C1}, the integral over $k^+$ can be obtained from the identity \eqref{kplus_contour_integ}, leading to
\begin{align}
 {{\cal H}}^{gg}_{\rho'\rho}(l^+,\l;\Delta y^+)\bigg|_{C}
= &\, 
g^2 C_A\,   \mu^{2\epsilon}\, 
 \int \frac{d^{D-2}\k}{(2\pi)^{D-2}}    \int \frac{dk^-}{2\pi}\,  e^{-i(k^-+\textrm{x}P^-)\Delta y^+}\, 
 \frac{(\textrm{x}P^-)^2}{(k^-\!+\!\textrm{x}P^-)^2}\,  \frac{\theta(k^-)}{2k^-}\,
 \nonumber\\
&\, \times\,
\Bigg\{
- \frac{4{g_{\rho}}^i{g_{\rho'}}^i}{\Big[ 
-\frac{(k^-\!+\!\textrm{x}P^-)}{k^-}\, \left(\k\!-\!\frac{k^-}{(k^-\!+\!\textrm{x}P^-)}\, \l\right)^2-\frac{\textrm{x}P^-}{(k^-\!+\!\textrm{x}P^-)}\, \l^2 +2\textrm{x}P^-l^+ +i0^+\Big]}
\nonumber\\
&\,
+ \frac{4{g_{\rho}}^i{g_{\rho'}}^i\,  \left(\frac{(k^-\!+\!\textrm{x}P^-)}{\textrm{x}P^-}\right)^2 \, \left(\k\!-\!\frac{k^-}{(k^-\!+\!\textrm{x}P^-)}\, \l\right)^2}{\Big[ 
-\frac{(k^-\!+\!\textrm{x}P^-)}{k^-}\, \left(\k\!-\!\frac{k^-}{(k^-\!+\!\textrm{x}P^-)}\, \l\right)^2-\frac{\textrm{x}P^-}{(k^-\!+\!\textrm{x}P^-)}\, \l^2 +2\textrm{x}P^-l^+ +i0^+\Big]^2}
\nonumber\\
&\,
+ \frac{2(1\!-\! \delta_s \epsilon)\,  \left[(k^-\!-\!\textrm{x}P^-){g_{\rho}}^{+} - (2\k^i\!-\!\l^i) {g_{\rho}}^i \right]\,
\left[(k^-\!-\!\textrm{x}P^-){g_{\rho'}}^{+} - (2\k^j\!-\!\l^j) {g_{\rho'}}^j \right]}{\Big[ 
-\frac{(k^-\!+\!\textrm{x}P^-)}{k^-}\, \left(\k\!-\!\frac{k^-}{(k^-\!+\!\textrm{x}P^-)}\, \l\right)^2-\frac{\textrm{x}P^-}{(k^-\!+\!\textrm{x}P^-)}\, \l^2 +2\textrm{x}P^-l^+ +i0^+\Big]^2}
\Bigg\}
\,
\label{hard_fact_g2g_C_reg_1}
\, .
\end{align}
Changing variables from $k^-$ to $\textrm{z}$ (see Eq.~\eqref{change_var_kmin_to_z}) and from $\k$ to $\K$ (see Eq.~\eqref{change_var_kperp_to_Kperp}), as well as using symmetry under rotations of $\K$, one finds
\begin{align}
&\,  {{\cal H}}^{gg}_{\rho'\rho}(l^+,\l;\Delta y^+)\bigg|_{C}
= 
\frac{g^2 C_A}{4\pi}\,  
\int_{0}^{1}d \textrm{z}\,
e^{-i \frac{\textrm{x}}{\textrm{z}}P^-\, \Delta y^+}\,   
\mu^{2\epsilon}  \int \frac{d^{D-2}\K}{(2\pi)^{D-2}}\,   
\nonumber\\
&\, \times\,
\bigg\{
\frac{4\textrm{z}\, {g_{\rho}}^i{g_{\rho'}}^i}{\left[\K^2+\textrm{z}(1\!-\! \textrm{z})\big(\l^2\!-\!2\frac{\textrm{x}}{\textrm{z}}P^- l^+\!-\!i0^+\big)\right]}\,
+\frac{4\frac{(1\!-\! \textrm{z})}{\textrm{z}}\, {g_{\rho}}^i{g_{\rho'}}^i\, \K^2}{\left[\K^2+\textrm{z}(1\!-\! \textrm{z})\big(\l^2\!-\!2\frac{\textrm{x}}{\textrm{z}}P^- l^+\!-\!i0^+\big)\right]^2}\,
\nonumber\\
&\, 
+\frac{2(1\!-\! \delta_s \epsilon){\textrm{z}}{(1\!-\! \textrm{z})}}{\left[\K^2+\textrm{z}(1\!-\! \textrm{z})\big(\l^2\!-\!2\frac{\textrm{x}}{\textrm{z}}P^- l^+\!-\!i0^+\big)\right]^2}
\left[\frac{4\K^2}{(2\!-\! 2 \epsilon)}\,  {g_{\rho}}^i{g_{\rho'}}^i
+(1\!-\!2 \textrm{z})^2\,   
 \left[ \frac{\textrm{x}}{\textrm{z}}P^-\,{g_{\rho}}^{+}\!-\!\l^i {g_{\rho}}^i \right]\,
 \left[ \frac{\textrm{x}}{\textrm{z}}P^-\,{g_{\rho'}}^{+}\!-\!\l^j {g_{\rho'}}^j \right]\,
\right]\,
\bigg\}
\,
\label{hard_fact_g2g_C_reg_2}
%\, .
\end{align}
and thus
\begin{align}
 {{\cal H}}^{gg}_{\rho'\rho}(l^+,\l;\Delta y^+)\bigg|_{C}
= &\, 
 \frac{\alpha_s C_A}{\pi}\,
 \int_{0}^{1}d \textrm{z}\,
e^{-i  \frac{\textrm{x}}{\textrm{z}}P^-\, \Delta y^+}  \,
\Gamma(1\!+\!\epsilon)\, \left(\frac{-l^2\!-\! i0^+}{4\pi \mu^2}\right)^{-\epsilon}\, \textrm{z}^{-\epsilon}(1\!-\!\textrm{z})^{-\epsilon}\;
 \nonumber\\
 &\, \times\,
\bigg\{{g_{\rho}}^{i} {g_{\rho'}}^i
\Big[\frac{\textrm{z}}{\epsilon}
+\frac{(1\!-\!\epsilon)}{\epsilon}\, \frac{(1\!-\!\textrm{z})}{\textrm{z}}
+\frac{(1\!-\!\delta_s \epsilon)}{\epsilon}\, {\textrm{z}}{(1\!-\!\textrm{z})}\Big]\:
-\frac{(1\!-\!\delta_s \epsilon)(1\!-\!2 \textrm{z})^2\, l_{\rho}l_{\rho'}}{2\big(l^2\!+\!i0^+\big)}
\bigg\}\Bigg|_{l^-\equiv \frac{\textrm{x}}{\textrm{z}}P^-}
\,
\label{hard_fact_g2g_C_res}
\, .
\end{align}

Collecting the results for the three parts $A$, $B$, and $C$, given in Eqs.~\eqref{hard_fact_g2g_A_res}, \eqref{hard_fact_g2g_B_res} and \eqref{hard_fact_g2g_C_res} respectively, one gets   
\begin{align}
 {{\cal H}}^{gg}_{\rho'\rho}(l^+,\l;\Delta y^+)
= &\, 
\frac{\alpha_s C_A}{\pi}\,  \int_{0}^{1}d \textrm{z}\;
\Gamma(1\!+\!\epsilon)\, \left(\frac{{\l^2\!-\! \frac{2\textrm{x}P^- l^+}{\textrm{z}}\!-\! i0^+}}{4\pi \mu^2}\right)^{-\epsilon}\, \textrm{z}^{-\epsilon}(1\!-\!\textrm{z})^{-\epsilon}\, 
\nonumber\\
&\, \times\,
\Bigg\{
{g_{\rho}}^i {g_{\rho'}}^i\,\frac{(1\!-\!\epsilon)}{\epsilon}\, \frac{1}{(1\!-\! \textrm{z})}
\left[\textrm{z}\, e^{-i \frac{\textrm{x}}{\textrm{z}}P^-\, \Delta y^+} -e^{-i \textrm{x}P^-\, \Delta y^+}\right]
\nonumber\\
&\, 
+{g_{\rho}}^{i} {g_{\rho'}}^i
\Big[
\frac{(1\!-\!\epsilon)}{\epsilon}\, \frac{(1\!-\!\textrm{z})}{\textrm{z}}
+\frac{(1\!-\!\delta_s \epsilon)}{\epsilon}\, {\textrm{z}}{(1\!-\!\textrm{z})}
\Big]\,
e^{-i \frac{\textrm{x}}{\textrm{z}}P^-\, \Delta y^+}  \,
\nonumber\\
&\, 
+\frac{(1\!-\!\delta_s \epsilon)(1\!-\!2 \textrm{z})^2\, l_{\rho}l_{\rho'}}{2\big({\l^2\!-\! \frac{2\textrm{x}P^- l^+}{\textrm{z}}\!-\! i0^+}\big)}\, 
e^{-i l^-\, \Delta y^+} \Bigg|_{l^-\equiv \frac{\textrm{x}}{\textrm{z}}P^-}
\nonumber\\
&\, 
+\Big[2\l^2 {g_{\rho}}^i {g_{\rho'}}^{i}
+l_{\rho}\, {g_{\rho'}}^i\, \l^i \!+\! {g_{\rho}}^i\, \l^i\,  l_{\rho'}
\Big]\:
\frac{e^{-i l^-\, \Delta y^+}}{2\textrm{z}\big({\l^2\!-\! \frac{2\textrm{x}P^- l^+}{\textrm{z}}\!-\! i0^+}\big)}
\Bigg|_{l^-\equiv \textrm{x}P^-}
\Bigg\}
\,
\label{hard_fact_g2g_total_res}
\, .
\end{align}

When inserting the result \eqref{hard_fact_g2g_total_res} into the expression \eqref{eq:G-real-NLO_g2g_2}, the terms proportional to $l_{\rho'}$ and/or $l_{\rho}$ can be rewritten thanks to the relations
\begin{align}
\alpha_s C_A  \int d^D\Delta y\, 
e^{-i l \cdot \Delta y}\,
 l_{\rho'}\, \langle P|{\mathcal T} {\cal F}^{-{\rho'}}_{a}(\Delta y)\, {\cal F}^{-\rho}_a(0)|P\rangle_c 
=&\, 
-i \alpha_s C_A  \int d^D\Delta y\, 
e^{-i l \cdot \Delta y}\,
   \langle P|{\mathcal T}   \left(\partial_{\rho'}{\cal F}^{-{\rho'}}_{a}\right)\!\!(\Delta y)\, {\cal F}^{-\rho}_a(0)|P\rangle_c 
\label{eq:g2g_eom_like_term_1}
\end{align}
and
\begin{align}
\alpha_s C_A  \int d^D\Delta y\, 
e^{-i l \cdot \Delta y}\,
 l_{\rho}\, \langle P|{\mathcal T}\; {\cal F}^{-{\rho'}}_{a}(\Delta y)\, {\cal F}^{-\rho}_a(0)|P\rangle_c\; 
=&\, 
i \alpha_s C_A  \int d^D\Delta y\, 
e^{-i l \cdot \Delta y}\,
 \langle P|{\mathcal T}   {\cal F}^{-\rho'}_a(0)   \left(\partial_{\rho}{\cal F}^{-{\rho}}_{a}\right)\!\!(-\Delta y)|P\rangle_c 
\nonumber\\
=&\, 
i \alpha_s C_A  \int d^D\Delta y\, 
e^{-i l \cdot \Delta y}\,
   \langle P|{\mathcal T}  {\cal F}^{-\rho'}_a(\Delta y)   \left(\partial_{\rho}{\cal F}^{-{\rho}}_{a}\right)\!\!(0)|P\rangle_c 
\label{eq:g2g_eom_like_term_2}
\, .
\end{align}
Following the backgroubd field method, the gluon background field is taken to obey its equation of motion  (including quantum corrections due the the fluctuation fields)
\begin{align}
 {\cal D}_{{\mu}} {\cal F}^{\mu \nu}(x)_a 
 \equiv &\, 
{\partial}_{{\mu}} {\cal F}^{\mu \nu}_a(x) - g f^{abc}  {\cal A}_{\mu}^{b}(x)\, {\cal F}^{\mu \nu}_c(x) 
= 
g  \overline{\Psi}(x) \gamma^{\nu} t^a  \Psi(x) +O(g^2)
\, ,
\label{EOM_g_back}
\end{align}
so that 
\begin{align}
{\partial}_{\rho} {\cal F}^{-\rho}_a(x)
= &\, 
 g f^{abc}  {\cal A}_{\rho}^{b}(x)\, {\cal F}^{-\rho}_c(x) 
- g  \overline{\Psi}(x) \gamma^{-} t^a  \Psi(x)+O(g^2)
\, .
\label{EOM_g_back}
\end{align}
Hence, on the right hand side of Eqs.~\eqref{eq:g2g_eom_like_term_1} and \eqref{eq:g2g_eom_like_term_2}, one obtains correlators of three background fields (three gluon fields, or a quark-antiquark-gluon correlator), with an overall factor of  $g \alpha_s C_A$. Following our discussion in section \ref{sec:q2g}, such contributions are of higher order in the dilute regime for the target, and should be discarded at the accuracy of our calculations. 

Then, discarding the terms proportional to $l_{\rho'}$ and/or $l_{\rho}$ in the fourth and fifth lines in Eq.~\eqref{hard_fact_g2g_total_res} and inserting the leftover in the expression \eqref{eq:G-real-NLO_g2g_2}, one arrives at
\begin{align}
g^{\textrm{n.r.}}\left({\textrm{x}}, \mu^2\right)\bigg|^{\ref{Fig:g2g_ladder}}
=&\, 
\frac{\alpha_s C_A}{\pi}\, 
\int \frac{d^D\Delta y}{2\pi\, \textrm{x}P^-}\, \langle P|{\mathcal T}\; {\cal F}^{-i}_{a}(\Delta y)\, {\cal F}^{-i}_a(0)|P\rangle_c\; 
\int \frac{d^{D-2}\l}{(2\pi)^{D-2}}\, e^{i\l \cdot \Delta \y}\,
 \int \frac{dl^+}{2\pi}\, e^{-i l^+  \Delta y^-}\,
  \int_{0}^{1}d \textrm{z}\; \textrm{z}^{-\epsilon}(1\!-\!\textrm{z})^{-\epsilon}\, 
 \nonumber\\
 &\, \times\,
\Gamma(1\!+\!\epsilon)\, \left(\frac{{\l^2\!-\! \frac{2\textrm{x}P^- l^+}{\textrm{z}}\!-\! i0^+}}{4\pi \mu^2}\right)^{-\epsilon}\, 
% \nonumber\\
% &\, \times\,
 \Bigg\{
\frac{(1\!-\!\epsilon)}{\epsilon}\, \frac{1}{(1\!-\! \textrm{z})}
\left[\textrm{z}\, e^{-i \frac{\textrm{x}}{\textrm{z}}P^-\, \Delta y^+} -e^{-i \textrm{x}P^-\, \Delta y^+}\right]
\nonumber\\
&\, 
+\Big[
\frac{(1\!-\!\epsilon)}{\epsilon}\, \frac{(1\!-\!\textrm{z})}{\textrm{z}}
+\frac{(1\!-\!\delta_s \epsilon)}{\epsilon}\, {\textrm{z}}{(1\!-\!\textrm{z})}
\Big]\,
e^{-i \frac{\textrm{x}}{\textrm{z}}P^-\, \Delta y^+}  \,
%\nonumber\\
%&\, 
+\frac{\l^2}{\textrm{z}\big({\l^2\!-\! \frac{2\textrm{x}P^- l^+}{\textrm{z}}\!-\! i0^+}\big)}\, 
e^{-i \textrm{x}P^-\, \Delta y^+}
\Bigg\}
+O(\alpha_s^2)
\label{eq:G-real-NLO_g2g_3}
\, .
\end{align}
Like in the $q\rightarrow g$ channel, the poles at $\epsilon=0$ correspond to UV divergences, and they are the only divergences in Eq.~\eqref{eq:G-real-NLO_g2g_3}. In particular, no further divergences would be obtained from the integrations over $\l$, $l^+$ and $\textrm{z}$ (provided that $\textrm{x}$ is non -zero). In particular, the main new feature in the expression \eqref{eq:G-real-NLO_g2g_3} compared to Eq.~\eqref{eq:G-real-NLO_q2g_3} is the potential divergence at $ \textrm{z}=1$ due to the  denominator $(1\!-\! \textrm{z})$ in the first term, which is turned into a removable singularity thanks to the ML prescription, as already mentioned.

In the $\overline{\textrm{MS}}$ scheme, the counterterm required to compensate the  UV divergence of the contribution \eqref{eq:G-real-NLO_g2g_3} is thus 
\begin{align}
g\left({\textrm{x}}, \mu^2\right)\bigg|^{\ref{Fig:g2g_ladder};\textrm{ c.t.}}
=&\, 
 -\frac{S_{\epsilon}}{\epsilon}\,
 \frac{\alpha_s C_A}{\pi}\, 
\int \frac{d^D\Delta y}{2\pi\, \textrm{x}P^-}\, \langle P|{\mathcal T}\; {\cal F}^{-i}_{a}(\Delta y)\, {\cal F}^{-i}_a(0)|P\rangle_c\; 
\int \frac{d^{D-2}\l}{(2\pi)^{D-2}}\, e^{i\l \cdot \Delta \y}\,
 \int \frac{dl^+}{2\pi}\, e^{-i l^+  \Delta y^-}\,
  \int_{0}^{1}d \textrm{z}\; 
 \nonumber\\
 &\, \times\,
 \Bigg\{
 \frac{1}{(1\!-\! \textrm{z})}
\left[\textrm{z}\, e^{-i \frac{\textrm{x}}{\textrm{z}}P^-\, \Delta y^+} -e^{-i \textrm{x}P^-\, \Delta y^+}\right]
+\Big[
 \frac{(1\!-\!\textrm{z})}{\textrm{z}}
+ {\textrm{z}}{(1\!-\!\textrm{z})}
\Big]\,
e^{-i \frac{\textrm{x}}{\textrm{z}}P^-\, \Delta y^+}  \,
\Bigg\}
\nonumber\\
=&\, 
 -\frac{S_{\epsilon}}{\epsilon}\,
 \frac{\alpha_s C_A}{\pi}\, 
\int \frac{d\Delta y^+}{2\pi\, \textrm{x}P^-}\, \langle P|{\mathcal T}\; {\cal F}^{-i}_{a}(\Delta y^+,0;0)\, {\cal F}^{-i}_a(0)|P\rangle_c\;
 \nonumber\\
 &\, \times\,
 \int_{0}^{1}d \textrm{z}\;   \Bigg\{
 \frac{1}{(1\!-\! \textrm{z})}
\left[\textrm{z}\, e^{-i \frac{\textrm{x}}{\textrm{z}}P^-\, \Delta y^+} -e^{-i \textrm{x}P^-\, \Delta y^+}\right]
+\Big[
 \frac{(1\!-\!\textrm{z})}{\textrm{z}}
+ {\textrm{z}}{(1\!-\!\textrm{z})}
\Big]\,
e^{-i \frac{\textrm{x}}{\textrm{z}}P^-\, \Delta y^+}  \,
\Bigg\}
\nonumber\\
=&\, 
 -\frac{S_{\epsilon}}{\epsilon}\,
 \frac{\alpha_s C_A}{\pi}\, 
  \int_{0}^{1}d \textrm{z}\;  
 \Bigg\{
 \frac{1}{(1\!-\! \textrm{z})}
\left[g^{\textrm{Bckgd}}\left(\frac{\textrm{x}}{\textrm{z}}, \mu^2\right) -g^{\textrm{Bckgd}}\left({\textrm{x}}, \mu^2\right)\right]
\nonumber\\
 &\, 
+\Big[
 \frac{(1\!-\!\textrm{z})}{\textrm{z}}
+ {\textrm{z}}{(1\!-\!\textrm{z})}
\Big]\,
\frac{1}{\textrm{z}}\, g^{\textrm{Bckgd}}\left(\frac{\textrm{x}}{\textrm{z}}, \mu^2\right) \,
\Bigg\}
\nonumber\\
=&\, 
 -\frac{S_{\epsilon}}{\epsilon}\,
 \frac{\alpha_s C_A}{\pi}\, 
  \int_{\textrm{x}}^{1}\frac{d \textrm{z}}{ \textrm{z}}\;  
 \Bigg\{
 \frac{ \textrm{z}}{(1\!-\! \textrm{z})_+}
+ \frac{(1\!-\!\textrm{z})}{\textrm{z}}
+ {\textrm{z}}{(1\!-\!\textrm{z})}
 \Bigg\}g^{\textrm{Bckgd}}\left(\frac{\textrm{x}}{\textrm{z}}, \mu^2\right) \,
\label{eq:G-real-NLO_g2g_ct}
\, ,
\end{align}
using the expression \eqref{eq:gluon-pdf_Bckgd} for the background gluon distribution. In the last step, we have used the fact that $g^{\textrm{Bckgd}}\left(\frac{\textrm{x}}{\textrm{z}}, \mu^2\right)$ vanishes for $\textrm{z}<\textrm{x}$, and the usual notation 
\begin{align}
  \int_{\textrm{x}}^{1} d \textrm{z}\;   \frac{1}{(1\!-\! \textrm{z})_+}\, f(\textrm{z}) 
  \equiv &\,
  \int_{\textrm{x}}^{1} d \textrm{z}\;   \frac{1}{(1\!-\! \textrm{z})}\, \big[f(\textrm{z})-f(1)\big] - f(1) \int_{0}^{\textrm{x}} d \textrm{z}\;   \frac{1}{(1\!-\! \textrm{z})}
  \label{def_plus_prescr}
  \, .
\end{align}
Comparing the counterterm \eqref{def_plus_prescr} with the expression \eqref{eq:q_pdf_renorm_1loop}, we can read-off 
\begin{align}
Z_{gg} (\textrm{z},g,\epsilon) Z_{3}(g,\epsilon)-\delta(\textrm{z}\!-\!1)
=&\,
 -\frac{S_{\epsilon}}{\epsilon}\,
 \frac{\alpha_s C_A}{\pi}\, 
 \left[\frac{ \textrm{z}}{(1\!-\! \textrm{z})_+}
+ \frac{(1\!-\!\textrm{z})}{\textrm{z}}
+ {\textrm{z}}{(1\!-\!\textrm{z})}
 \right]
 +O(\alpha_s^2)
\label{Eq:Zgg_Zg_1loop}
  \, .
\end{align}
since the diagram \ref{Fig:g2g_ladder} is the only non-zero real contribution to the $g\rightarrow g$ channel at order $\alpha_s$.

%%%%%%%%%%%%%%%%%%%%%%%%%%%%%%%%%%%%%%%%%%%%%%%%%%%%%%%%%%%%%%%%%%%%
%%%%%%%%%%%%%%%%%%%%%%%%%%%%%%%%%%%%%%%%%%%%%%%%%%%%%%%%%%%%%%%%%%%%
%%%%%%%%%%%%%%%%%%%%%%%%%%%%%%%%%%%%%%%%%%%%%%%%%%%%%%%%%%%%%%%%%%%%
%%%%%%%%%%%%%%%%%%%%%%%%%%%%%%%%%%%%%%%%%%%%%%%%%%%%%%%%%%%%%%%%%%%%
%%%%%%%%%%%%%%%%%%%%%%%%%%%%%%%%%%%%%%%%%%%%%%%%%%%%%%%%%%%%%%%%%%%%
%%%%%%%%%%%%%%%%%%%%%%%%%%%%%%%%%%%%%%%%%%%%%%%%%%%%%%%%%%%%%%%%%%%%

\section{Gluon field renormalization in the light-cone gauge}
\label{sec:virtual}

%%%%%%%%%%%%%%%%%%%%%%%%%%%%%%%%%%%%%%%%%%%%%%%%%%%%%%%%%%%%%%%%%%%%
%%%%%%%%%%%%%%%%%%%%%%%%%%%%%%%%%%%%%%%%%%%%%%%%%%%%%%%%%%%%%%%%%%%%
%%%%%%%%%%%%%%%%%%%%%%%%%%%%%%%%%%%%%%%%%%%%%%%%%%%%%%%%%%%%%%%%%%%%
%%%%%%%%%%%%%%%%%%%%%%%%%%%%%%%%%%%%%%%%%%%%%%%%%%%%%%%%%%%%%%%%%%%%
%%%%%%%%%%%%%%%%%%%%%%%%%%%%%%%%%%%%%%%%%%%%%%%%%%%%%%%%%%%%%%%%%%%%
%%%%%%%%%%%%%%%%%%%%%%%%%%%%%%%%%%%%%%%%%%%%%%%%%%%%%%%%%%%%%%%%%%%%

In this section, the calculation of the gluon vacuum polarization tensor at one loop in light-cone gauge (originally done in Ref.~\cite{Leibbrandt:1983pj}) is recalled, as well as the corresponding gluon field renormalization (originally done in Ref.~\cite{Bassetto:1987sw}).

%%%%%%%%%%%%%%%%%%%%%%%%%%%%%%%%%%%%%%%%%%%%%%%%%%%%%%%%%%%%%%%%%%%%%%%%%%%%%%%%%%%%%%
%%%%%%%%%%%%%%%%%%%%%%%%%%%%%%%%%%%%%%%%%%%%%%%%%%%%%%%%%%%%%%%%%%%%%%%%%%%%%%%%%%%%%%
\begin{figure}%[ht]
\subfloat[Gluon tadpole diagram \label{Fig:g_tad}]{%
       \includegraphics[width=0.3\textwidth]{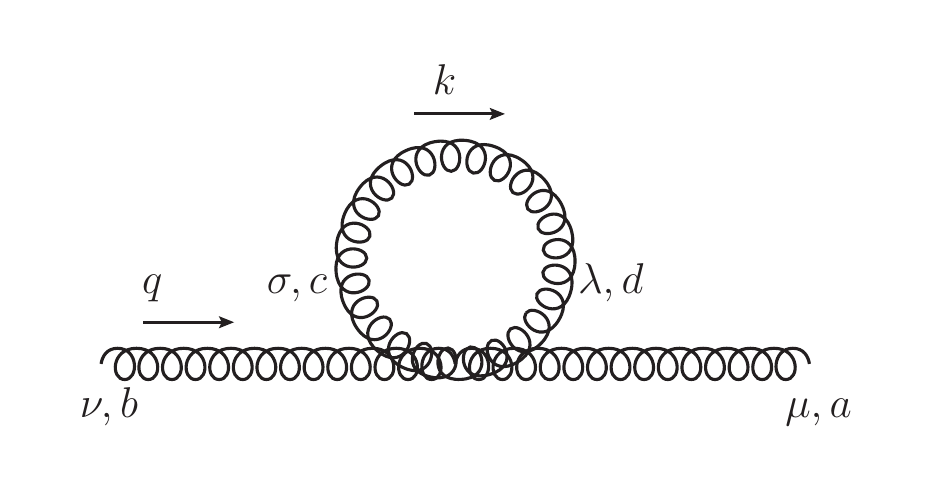}
     }
%\hfill
\subfloat[Gluon bubble diagram \label{Fig:g_bub}]{%
       \includegraphics[width=0.3\textwidth]{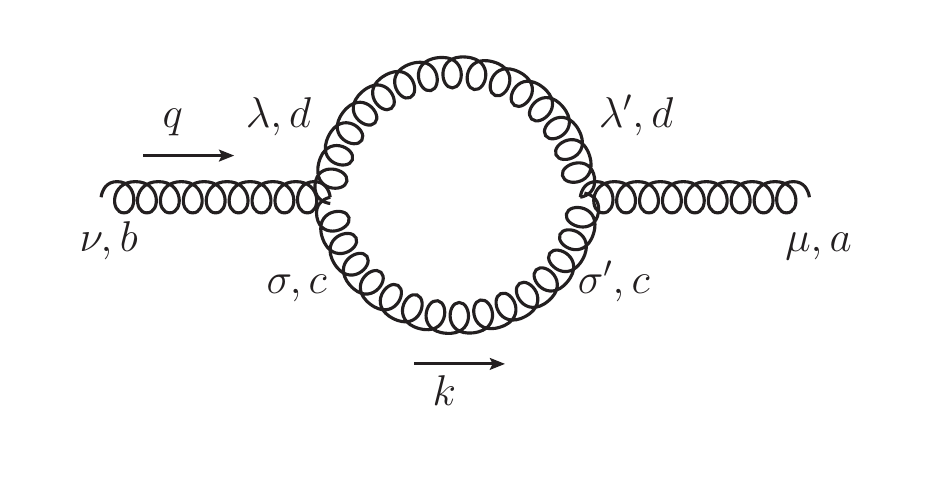}
     }
\subfloat[Quark bubble diagram\label{Fig:q_bub}]{%
       \includegraphics[width=0.3\textwidth]{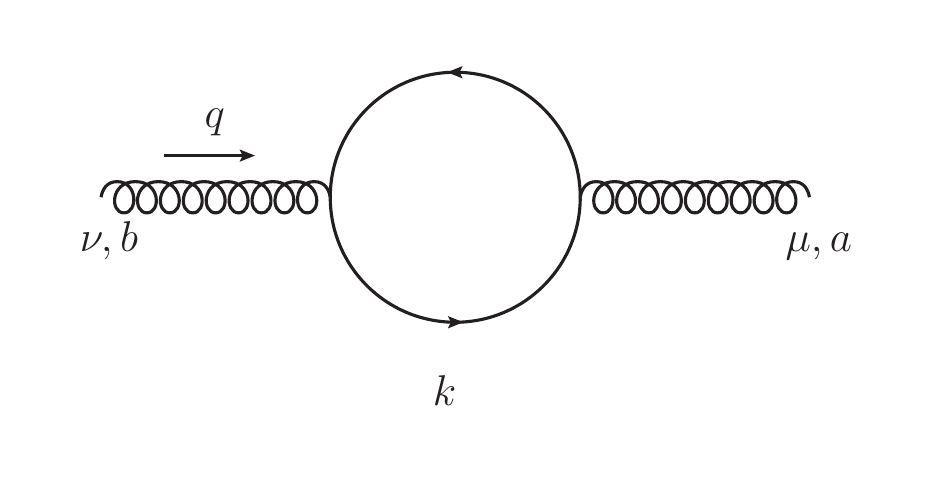}
     }
%\hfill
\caption{\label{Fig:glue_vac_pol_diags} One loop virtual diagrams for the QCD vacuum polarization tensor in the lightcone gauge. }
\end{figure}
%%%%%%%%%%%%%%%%%%%%%%%%%%%%%%%%%%%%%%%%%%%%%%%%%%%%%%%%%%%%%%%%%%%%%%%%%%%%%%%%%%%%%%
%%%%%%%%%%%%%%%%%%%%%%%%%%%%%%%%%%%%%%%%%%%%%%%%%%%%%%%%%%%%%%%%%%%%%%%%%%%%%%%%%%%%%%

% \begin{figure}
%\setbox1\hbox to 10cm{
%\includegraphics{tadpole_SE_diag.pdf}
%}
%\setbox2\hbox to 10cm{
%\includegraphics{bubble_SE_diag.pdf}
%}
%\begin{center}
%\resizebox*{12cm}{!}{\hspace{-5cm}
%\mbox{\box1 \hspace{8cm} \box2}}
%\caption{\label{Fig:glue_vac_pol_diags} The two diagrams contributing to the vacuum polarization tensor at one loop in the pure glue sector.}
%\end{center}
%\end{figure}

%%%%%%%%%%%%%%%%%%%%%%%%%%%%%%%%%%%%%%%%%%%%%%%%%%%%%%%%%%%%%%%%%%%%
%%%%%%%%%%%%%%%%%%%%%%%%%%%%%%%%%%%%%%%%%%%%%%%%%%%%%%%%%%%%%%%%%%%%
%%%%%%%%%%%%%%%%%%%%%%%%%%%%%%%%%%%%%%%%%%%%%%%%%%%%%%%%%%%%%%%%%%%%
%%%%%%%%%%%%%%%%%%%%%%%%%%%%%%%%%%%%%%%%%%%%%%%%%%%%%%%%%%%%%%%%%%%%

\subsection{Gluon vacuum polarization tensor at NLO in light-cone gauge\label{Sec:gluon_vac_pol_1_loop}}
 
 %%%%%%%%%%%%%%%%%%%%%%%%%%%%%%%%%%%%%%%%%%%%%%%%%%%%%%%%%%%%%%%%%%%%
%%%%%%%%%%%%%%%%%%%%%%%%%%%%%%%%%%%%%%%%%%%%%%%%%%%%%%%%%%%%%%%%%%%%
%%%%%%%%%%%%%%%%%%%%%%%%%%%%%%%%%%%%%%%%%%%%%%%%%%%%%%%%%%%%%%%%%%%%
%%%%%%%%%%%%%%%%%%%%%%%%%%%%%%%%%%%%%%%%%%%%%%%%%%%%%%%%%%%%%%%%%%%%
 
\subsubsection{Tadpole diagram}

The tadpole diagram (see Fig.~\ref{Fig:g_tad}) might a priori contribute to the gluon vacuum polarization tensor. It writes
\begin{align}
 i \Pi_{\mu \nu}^{ab}(q)\bigg|^{\ref{Fig:g_tad}}_{\textrm{tad.}}=&\, \frac{1}{2}\int \frac{d^Dk}{(2\pi)^D}\;  
 {V_{4g}}_{\mu \nu \sigma \lambda}^{abcd}\;
 \delta^{cd}\;  {\tilde G}_{0,F}^{\sigma \lambda}(k)\, ,
 \label{Eq:tadpole_1}
\end{align}
where the $1/2$ is the symmetry factor associated with the tadpole loop, ${V_{4g}}_{\mu \nu \sigma \lambda}^{abcd}\;$ is the QCD four-gluon vertex, and we use the gluon Feynman propagator in light-cone gauge~\eqref{eq:Feyn-prop_def} with the ML prescription~\eqref{eq:ML_prescription}.
Performing the numerator algebra, one arrives at
\begin{align}
 i \Pi_{\mu \nu}^{ab}(q)\bigg|^{\ref{Fig:g_tad}}_{\textrm{tad.}}
 =&\, \left[ {\mathbb {1}}\right]_{ab}\;  i\, g^2\, C_A\, \mu^{4\!-\!D} 
 \int \frac{d^Dk}{(2\pi)^D}\;  
 \frac{i}{\left(k^2+i0^+\right)}\: 
 \bigg[
 (1\!-\! 2\delta_s \epsilon) g_{\mu \nu}
 +\frac{\left(n_{\mu} k_{\nu}+k_{\mu} n_{\nu}\right)}{[n\!\cdot\!k]}
 \bigg]\,
 \, .
  \label{Eq:tadpole_2}
 \end{align}
In dimensional regularization, one has the vanishing of the standard scaleless integral
\begin{align}
 \mu^{4\!-\!D} 
 \int \frac{d^Dk}{(2\pi)^D}\;  
 \frac{i}{\left(k^2+i0^+\right)}   
 =&\, 0
 \label{Eq:int_A_0}
 \, , 
\end{align}
so that the first term in Eq.~\eqref{Eq:tadpole_2} vanishes. 
By contrast, the second term in Eq.~\eqref{Eq:tadpole_2} involves a vector integral with the extra denominator $[n\!\cdot\!k]$. Such integral in the ML prescription is a priori a linear combination of the two available vectors   
$n_{\nu}$ and $\bar n_{\nu}$, by Lorentz symmetry. Moreover, integrals with the ML prescription preserve homogeneity with respect to $n_{\nu}$ and with respect to $\bar n_{\nu}$. Hence, in this case, the only possible contribution is of the form 
\begin{align}
\mu^{4\!-\!D} 
 \int \frac{d^Dk}{(2\pi)^D}\;  
 \frac{i}{\left(k^2+i0^+\right)}\;   
 \frac{k_{\nu}}{[n\!\cdot\!k]} 
 =&\, C\, \frac{\bar n_{\nu}}{\bar n\cdot n}
 \, , \label{Eq:int_A_mu_ML_1}
\end{align}
with a constant coefficient $C$. Multiplying the relation~\eqref{Eq:int_A_mu_ML_1} with $n^{\nu}$, one finds
\begin{align}
C =&\,
n^{\nu}\:
\mu^{4\!-\!D} 
 \int \frac{d^Dk}{(2\pi)^D}\;  
 \frac{i}{\left(k^2+i0^+\right)}\;   
 \frac{k_{\nu}}{[n\!\cdot\!k]} 
=
\mu^{4\!-\!D} 
 \int \frac{d^Dk}{(2\pi)^D}\;  
 \frac{i}{\left(k^2+i0^+\right)}
 = 0\, ,\label{Eq:int_A_mu_ML_2}
\end{align}
thanks to Eq.~\eqref{Eq:int_A_0}. Hence, with the joint use of CDR and the ML prescription, the vector integral appearing in Eq.~\eqref{Eq:tadpole_2} vanishes identically as well.\footnote{This result can be checked explicitly by realizing that both poles in $n\!\cdot\!k$ always lie on the same side of the real axis in Eq.~\eqref{Eq:int_A_mu_ML_1}, thanks to the ML prescription.}  
Hence, the total contribution from the tadpole diagram on Fig.~\ref{Fig:g_tad} vanishes : 
\begin{align}
 i \Pi_{\mu \nu}^{ab}(q)\bigg|^{\ref{Fig:g_tad}}_{\textrm{tad.}}
 =&\, 
0 \label{Eq:tadpole_3}
\, .
\end{align}
%

%%%%%%%%%%%%%%%%%%%%%%%%%%%%%%%%%%%%%%%%%%%%%%%%%%%%%%%%%%%%%%%%%%%%%
%%%%%%%%%%%%%%%%%%%%%%%%%%%%%%%%%%%%%%%%%%%%%%%%%%%%%%%%%%%%%%%%%%%%%
%%%%%%%%%%%%%%%%%%%%%%%%%%%%%%%%%%%%%%%%%%%%%%%%%%%%%%%%%%%%%%%%%%%%%
%%%%%%%%%%%%%%%%%%%%%%%%%%%%%%%%%%%%%%%%%%%%%%%%%%%%%%%%%%%%%%%%%%%%%

\subsubsection{Bubble diagram contribution to the gluon vacuum polarization tensor}

The only one-loop contribution to the gluon vacuum polarization tensor in the pure glue sector is thus the so-called bubble diagram on Fig.~\ref{Fig:g_bub}. Its expression in light-cone gauge is given by 
\begin{align}
 i \Pi_{\mu \nu}^{ab}(q)\bigg|^{\ref{Fig:g_bub}}_{\textrm{g bub.}}=&\, \frac{1}{2}\int \frac{d^Dk}{(2\pi)^D}\;  
 {V_{3g}}_{\nu \sigma \lambda}^{bcd}(q,-k,k-q)\;
   {\tilde G}_{0,F}^{\sigma' \sigma}(k)\,
    {\tilde G}_{0,F}^{\lambda' \lambda}(q\!-\!k)\,
     {V_{3g}}_{\mu \sigma' \lambda'}^{acd}(-q,k,q-k)\;
 \label{Eq:vac_pol_tensor_1loop_1}
\end{align}
with the standard expression for the 3 gluon vertices in QCD, the symmetry factor $1/2$ due to the two internal gluon lines, and the Feynman propagators taken in the light-cone gauge with the ML prescription~\eqref{eq:ML_prescription}. The evaluation of the numerator of the integrand requires a significant amount of algebra. The obtained result can be simplified by performing the change of variables $k\mapsto q\!-\!k$ in some of the terms, and one eventually arrives at
\begin{align}
 i \Pi_{\mu \nu}^{ab}(q)\bigg|^{\ref{Fig:g_bub}}_{\textrm{g bub.}}
 =&\, 
 \left[ {\mathbb {1}}\right]_{ab}\;
 i\,g^2\, C_A\bigg\{
 -4(1\!-\!\delta_s \epsilon) {\cal B}_{\mu \nu}(q)
 +2(1\!-\!\delta_s \epsilon) \Big[q_{\mu}\, {\cal B}_{\nu}(q) + q_{\nu}\, {\cal B}_{\mu}(q)\Big]
 +\Big[(3\!+\!\delta_s \epsilon)q_{\mu}q_{\nu}  -4q^2 g_{\mu \nu}\Big]  {\cal B}_0(q)
 \nonumber\\
 &\, \;
 +2\left[q^2 n_{\mu} \!-\! (n\!\cdot\! q) q_{\mu}\right] {\mathbb B}_{\nu}(q) 
 +2\left[q^2 n_{\nu}\!-\! (n\!\cdot\! q) q_{\nu}\right] {\mathbb B}_{\mu}(q) 
 +\left[n_{\mu}q_{\nu} + q_{\mu}n_{\nu}-2 q^2 \frac{n_{\mu}n_{\nu}}{(n\!\cdot\! q)}\right] {\mathbb A}_0(q)
 \nonumber\\
 &\, \;
+q^2\left[-3 (n_{\mu}q_{\nu} + q_{\mu}n_{\nu})+4 (n\!\cdot\! q) g_{\mu \nu}+2 q^2 \frac{n_{\mu}n_{\nu}}{(n\!\cdot\! q)}\right] {\mathbb B}_0(q) 
  \bigg\}
  \, ,
 \label{Eq:vac_pol_tensor_1loop_2}
\end{align}
with the notations 
\begin{align}
{\cal B}_0(q)
\equiv &\, 
 \mu^{4\!-\!D} 
 \int \frac{d^Dk}{(2\pi)^D}\;  
 \frac{i}{\left(k^2+i0^+\right)\left((k\!-\!q)^2+i0^+\right)}   
 \label{Eq:def_B_0}
\\
{\cal B}_{\mu}(q)
\equiv &\, 
 \mu^{4\!-\!D} 
 \int \frac{d^Dk}{(2\pi)^D}\;  
 \frac{i\; k_{\mu}}{\left(k^2+i0^+\right)\left((k\!-\!q)^2+i0^+\right)}   
 \label{Eq:def_B_mu}
\\
{\cal B}_{\mu\nu}(q)
\equiv &\, 
 \mu^{4\!-\!D} 
 \int \frac{d^Dk}{(2\pi)^D}\;  
 \frac{i\; k_{\mu}k_{\nu}}{\left(k^2+i0^+\right)\left((k\!-\!q)^2+i0^+\right)}   
 \label{Eq:def_B_munu}
\end{align}
as well as
\begin{align}
{\mathbb A}_0(q)
\equiv &\, 
\mu^{4\!-\!D} 
\int \frac{d^Dk}{(2\pi)^D}\;  
\frac{i}{\left((k\!-\!q)^2+i0^+\right)}\;   
\frac{1}{[n\!\cdot\!k]} 
\label{Eq:int_A_0_ML_1}
\\
{\mathbb B}_0(q)
\equiv &\, 
\mu^{4\!-\!D} 
\int \frac{d^Dk}{(2\pi)^D}\;  
\frac{i}{\left(k^2+i0^+\right)\left((k\!-\!q)^2+i0^+\right)}\;   
\frac{1}{[n\!\cdot\!k]} 
\label{Eq:int_B_0_ML_1}
\\
{\mathbb B}_{\mu}(q)
\equiv &\, 
\mu^{4\!-\!D} 
\int \frac{d^Dk}{(2\pi)^D}\;  
\frac{i}{\left(k^2+i0^+\right)\left((k\!-\!q)^2+i0^+\right)}\;   
\frac{k_{\mu}}{[n\!\cdot\!k]} 
\label{Eq:int_B_mu_ML_1}
 \, ,
\end{align}
with the last denominator defined with the ML prescription~\eqref{eq:ML_prescription}.
In Eq.~\eqref{Eq:vac_pol_tensor_1loop_2}, we have already dropped the terms proportional to an integral which vanishes in CDR and with the use of the ML prescription, such as \eqref{Eq:int_A_0}, \eqref{Eq:int_A_mu_ML_1}, or 
\begin{align}
{\mathbb A}_0(0)
=&\,
\mu^{4\!-\!D} 
 \int \frac{d^Dk}{(2\pi)^D}\;  
 \frac{i}{\left(k^2+i0^+\right)}\;   
 \frac{1}{[n\!\cdot\!k]} 
 = 0
 \, , \label{Eq:int_A_0_ML_0}
\end{align}
since in that case both poles in terms of $n\!\cdot\!k$ are always on the same side of the real axis.

The integrals present in the expression \eqref{Eq:vac_pol_tensor_1loop_2} are not all independent. In particular, we can use the Passarino-Veltman reduction relations~\cite{Passarino:1978jh}
\begin{align}
{\cal B}_{\mu}(q) 
 =&\, 
 \frac{q_{\mu}}{2}\;
 {\cal B}_0(q)
 \label{Eq:PV_red_B_mu}
\\
{\cal B}_{\mu\nu}(q)
= &\, 
\frac{1}{(3\!-\!2\epsilon)}
\left[\left(1\!-\!\frac{\epsilon}{2}\right)  q_{\mu}q_{\nu}
-\frac{q^{2}}{4}\, g_{\mu\nu}
\right]
  {\cal B}_0(q)
 \label{Eq:PV_red_B_munu}
\end{align}
to eliminate ${\cal B}_{\mu}(q)$ and ${\cal B}_{\mu\nu}(q)$ from Eq.~\eqref{Eq:vac_pol_tensor_1loop_2}.
From their integrands, it is clear that the integrals \eqref{Eq:int_A_0_ML_1}, \eqref{Eq:int_B_0_ML_1} and \eqref{Eq:int_B_mu_ML_1} satisfy the relation
\begin{align}
q^2\, {\mathbb B}_0(q) -2q^{\mu}\,  {\mathbb B}_{\mu}(q)+{\mathbb A}_0(q) = {\mathbb A}_0(0)=0 
 \, , \label{Eq:int_B_mu_ML_PVred}
\end{align}
which can be used to eliminate for example ${\mathbb A}_0(q)$ in favor of ${\mathbb B}_0(q)$ and ${\mathbb B}_{\mu}(q)$.
Using these relations between integrals, one reduces the expression \eqref{Eq:vac_pol_tensor_1loop_2} to
\begin{align}
 i \Pi_{\mu \nu}^{ab}(q)\bigg|^{\ref{Fig:g_bub}}_{\textrm{g bub.}}
 =&\, 
 \left[ {\mathbb {1}}\right]_{ab}\;
 i\,g^2\, C_A\bigg\{ \Big[q_{\mu}q_{\nu}\!-\!q^2 g_{\mu\nu}\Big]\,
 \frac{(11\!-\!8\epsilon\!+\!\delta_s \epsilon)}{(3\!-\!2\epsilon)}\, {\cal B}_0(q)
 \nonumber\\
 &\, \;
  -\frac{2}{ (n\!\cdot\! q)} 
 \bigg[(n\!\cdot\! q)g_{\mu\sigma} \!-\! n_{\mu} q_{\sigma} \bigg]
 \bigg[(n\!\cdot\! q)g_{\rho\nu} \!-\!  q_{\rho} n_{\nu} \bigg]
  \bigg[q^{\sigma}\,  {\mathbb B}^{\rho}(q)   +q^{\rho}\, {\mathbb B}^{\sigma}(q)\bigg]
 \nonumber\\
 &\, \;
+4q^2\left[ (n\!\cdot\! q) g_{\mu \nu}-n_{\mu}q_{\nu} - q_{\mu}n_{\nu}+ q^2 \frac{n_{\mu}n_{\nu}}{(n\!\cdot\! q)}\right] {\mathbb B}_0(q) 
  \bigg\}
  \, .
 \label{Eq:vac_pol_tensor_1loop_3}
\end{align}
At this stage, the contribution \eqref{Eq:vac_pol_tensor_1loop_3} to the vacuum polarization tensor is manifestly transverse
\begin{align}
q^{\mu} i \Pi_{\mu \nu}^{ab}(q)\bigg|^{\ref{Fig:g_bub}}_{\textrm{g bub.}}
 =&\, 0
\label{Eq:transversity_of_vac_pol}
 \, .
\end{align}
That property is required for the full one-loop result in light-cone gauge from the Ward identity~\cite{Leibbrandt:1983pj,Bassetto:1987sw}.

The integral ${\cal B}_0(q)$, defined in Eq.~\eqref{Eq:def_B_0}, can be calculated in the standard way, and one finds
\begin{align}
 {\cal B}_0(q)
  = &\,
   -\frac{1}{(4\pi)^2}\, \frac{\Gamma(1\!+\!\epsilon)}{\epsilon}\, \left(\frac{-q^2\!-\!i0^+}{4\pi \mu^2}\right)^{-\epsilon}\,
 \frac{1}{(1\!-\!2\epsilon)}\, \frac{\Gamma(1\!-\!\epsilon)\Gamma(1\!-\!\epsilon)}{\Gamma(1\!-\!2\epsilon)} 
  \nonumber\\
  = &\,
   -\frac{1}{(4\pi)^2}\, \Bigg\{
    \frac{S_{\epsilon}}{\epsilon} -\log\left(\frac{-q^2\!-\!i0^+}{\mu^2}\right)+2 +O(\epsilon)
   \Bigg\}
 \label{Eq:B_0_scal_int}
 \, ,
\end{align}
with a UV pole at $\epsilon=0$.
Moreover, let us note that 
\begin{align}
n_{\mu}\,  {\mathbb B}^{\mu}(q) =&\,    {\cal B}_0(q)
 \, , \label{Eq:rel_int_B_min_ML_vs_B0_PV}
\end{align}
which is then UV divergent as well. By contrast, the other components of ${\mathbb B}^{\mu}(q)$ and the integral ${\mathbb B}_0(q)$ are finite at $D=4$ thanks to the ML prescription~\eqref{eq:ML_prescription}. Their calculation and expressions (that we could not find in the literature), are provided in appendix \ref{sec:ML_int} for completeness. Hence, one finds in particular 
\begin{align}
 {\mathbb B}^{\mu}(q) =&\,    {\cal B}_0(q)\, \frac{\bar n^{\mu}\, }{(\bar n\!\cdot\! n)}  + \textrm{finite}
 \label{Eq:rel_int_B_min_ML_vs_B0_PV_2}
\end{align}
thanks to the ML prescription, in agreement with the result of Ref.~\cite{Leibbrandt:1983pj}. Note that thanks to the ML prescription, apart from the abovementioned  UV divergences, there is no other divergence of any kind in any of the integrals encountered in the calculation of the gluon loop diagram from Fig.~\eqref{Fig:g_bub}. By contrast, other prescriptions like the Cauchy principal value or the retarded prescription for the light-cone denominators would typically produce spurious IR divergences.

%%%%%%%%%%%%%%%%%%%%%%%%%%%%%%%%%%%%%%%%%%%%%%%%%%%%%%%%%%%%%%%%%%%%%
%%%%%%%%%%%%%%%%%%%%%%%%%%%%%%%%%%%%%%%%%%%%%%%%%%%%%%%%%%%%%%%%%%%%%
%%%%%%%%%%%%%%%%%%%%%%%%%%%%%%%%%%%%%%%%%%%%%%%%%%%%%%%%%%%%%%%%%%%%%
%%%%%%%%%%%%%%%%%%%%%%%%%%%%%%%%%%%%%%%%%%%%%%%%%%%%%%%%%%%%%%%%%%%%%

\subsubsection{Quark loop contribution to the gluon vacuum polarization tensor}

The remaining one-loop diagram for the gluon vacuum polarization tensor is the quark loop diagram, see Fig.~\eqref{Fig:q_bub}. Since the loop involves only the quark propagator, it does not depend on the gauge choice, and its calculation is standard. Including the $-1$ factor for the closed fermion loop, it can be written, for each light quark flavor (considered massless), as
\begin{align}
 i \Pi_{\mu \nu}^{ab}(q)\bigg|^{\ref{Fig:q_bub}}_{\textrm{q bub.}}=&\, -\int \frac{d^Dk}{(2\pi)^D}\;  {\textrm{tr}}\bigg\{
(-ig \mu^{\epsilon} t^a \gamma_{\mu}) 
\frac{i \slk}{\big(k^2\!+\!i0^+\big)}
(-ig \mu^{\epsilon} t^b \gamma_{\nu})
\frac{i (\slk\!-\! \slq)}{\big((k\!-\!q)^2\!+\!i0^+\big)}
 \bigg\}
 \nonumber\\
 =&\,
  \left[ {\mathbb {1}}\right]_{ab}\;
 i\,g^2\,
 \bigg\{
 4 {\cal B}_{\mu \nu}(q)   -2q_{\mu}\, {\cal B}_{\nu}(q)  -2q_{\nu}\, {\cal B}_{\mu}(q)  +q^2 g_{\mu \nu}  {\cal B}_0(q)
 \bigg\}
\, ,
 \label{Eq:vac_pol_tensor_q_loop_1}
\end{align}
with the notations \eqref{Eq:def_B_0}, \eqref{Eq:def_B_mu} and \eqref{Eq:def_B_munu}. Then, using the tensor reduction relations \eqref{Eq:PV_red_B_mu} and \eqref{Eq:PV_red_B_munu}, one obtains
\begin{align}
 i \Pi_{\mu \nu}^{ab}(q)\bigg|^{\ref{Fig:q_bub}}_{\textrm{q bub.}}
 =&\,
  \left[ {\mathbb {1}}\right]_{ab}\;
 (-i)\,g^2\,
 \Big[q_{\mu}q_{\nu}\!-\!q^2 g_{\mu\nu}\Big]\,
 \frac{2(1\!-\!\epsilon)}{(3\!-\!2\epsilon)}\, {\cal B}_0(q)
\, ,
 \label{Eq:vac_pol_tensor_q_loop_2}
\end{align}
which is manifestly transverse, and which has a logarithmic UV divergence, see Eq.~\eqref{Eq:B_0_scal_int}.

%%%%%%%%%%%%%%%%%%%%%%%%%%%%%%%%%%%%%%%%%%%%%%%%%%%%%%%%%%%%%%%%%%%%%
%%%%%%%%%%%%%%%%%%%%%%%%%%%%%%%%%%%%%%%%%%%%%%%%%%%%%%%%%%%%%%%%%%%%%
%%%%%%%%%%%%%%%%%%%%%%%%%%%%%%%%%%%%%%%%%%%%%%%%%%%%%%%%%%%%%%%%%%%%%
%%%%%%%%%%%%%%%%%%%%%%%%%%%%%%%%%%%%%%%%%%%%%%%%%%%%%%%%%%%%%%%%%%%%%

\subsubsection{Total UV divergence of the gluon vacuum polarization tensor at NLO}

In summary, out of the three possible diagrams in light-cone gauge for the gluon vacuum polarization tensor at NLO, one finds that, when using the ML prescription, the gluon tadpole diagram from Fig.~\ref{Fig:g_tad} vanishes, whereas the other gluon loop diagram, from Fig.~\ref{Fig:g_bub}, and the quark loop diagram, from  Fig.~\eqref{Fig:q_bub}, are UV divergent but IR finite. Focusing on the UV divergent piece, the total one-loop result obtained from Eqs.~\eqref{Eq:tadpole_3}, \eqref{Eq:vac_pol_tensor_1loop_3} and \eqref{Eq:vac_pol_tensor_q_loop_2} is of the form
\begin{align}
 i \Pi_{\mu \nu}^{ab}(q)\bigg|^{1\textrm{ loop}}
 =&\,
  \left[ {\mathbb {1}}\right]_{ab}\; i\,g^2\, {\cal B}_0(q)
  \bigg\{
 \left[\frac{11 C_A}{3}-\frac{2 N_f}{3}\right] \Big[q_{\mu}q_{\nu}\!-\!q^2 g_{\mu\nu}\Big]\,
 \nonumber\\
 &\,
  -\frac{2C_A}{ (n\!\cdot\! q)(\bar n\!\cdot\! n)} 
 \bigg[(n\!\cdot\! q)g_{\mu\sigma} \!-\! n_{\mu} q_{\sigma} \bigg]
 \bigg[(n\!\cdot\! q)g_{\rho\nu} \!-\!  q_{\rho} n_{\nu} \bigg]
  \Big[q^{\sigma}\,  \bar n^{\rho}  +q^{\rho}\, \bar n^{\sigma}  \Big]
  \bigg\}
  + \textrm{finite}
\, ,
 \label{Eq:vac_pol_tensor_1_loop_tot}
\end{align}
with $N_f$ the number of active light quark flavors, and $ {\cal B}_0(q)$ given by Eq.~\eqref{Eq:B_0_scal_int}.
 The $1/\epsilon$ UV pole contribution found in the expression \eqref{Eq:vac_pol_tensor_1_loop_tot} is in agreement with the result from Ref.~\cite{Leibbrandt:1983pj}. 

Then, attaching to the 1-loop vacuum polarization tensor \eqref{Eq:vac_pol_tensor_1_loop_tot} two gluon fields in momentum space (taken in light-cone gauge) as external legs, one obtains
\begin{align}
 i \Pi_{\mu \nu}^{ab}(q)\bigg|^{1\textrm{ loop}}\: \tilde{A}^{\mu}_a(-q)\,  \tilde{A}^{\nu}_b(q)
 =&\,
 i\,g^2\, {\cal B}_0(q)
  \bigg\{
 \left[\frac{11 C_A}{3}-\frac{2 N_f}{3}\right] \Big[q_{\mu}q_{\nu}\!-\!q^2 g_{\mu\nu}\Big]\,
 \nonumber\\
 &\,
  -2C_A\, \frac{(n\!\cdot\! q)}{ (\bar n\!\cdot\! n)} 
  \Big[q_{\mu}\,  \bar n_{\nu}  +q_{\nu}\, \bar n_{\mu} \Big]
  \bigg\}\tilde{A}^{\mu}_a(-q)\,  \tilde{A}^{\nu}_a(q) 
  + \textrm{finite}
\, .
 \label{Eq:vac_pol_tensor_1_loop_tot_with_fields}
\end{align}
On the one hand, the divergent part of the  vacuum polarization tensor \eqref{Eq:vac_pol_tensor_1_loop_tot} contains some non-local part, due to the $(n\!\cdot\! q)$ denominator in the second line. On the other hand, that nonlocality drops when contracting that expression with gluon fields (or gluon propagators) in light-cone gauge, as can be seen from Eq.~\eqref{Eq:vac_pol_tensor_1_loop_tot_with_fields}.

%%%%%%%%%%%%%%%%%%%%%%%%%%%%%%%%%%%%%%%%%%%%%%%%%%%%%%%%%%%%%%%%%%%%%
%%%%%%%%%%%%%%%%%%%%%%%%%%%%%%%%%%%%%%%%%%%%%%%%%%%%%%%%%%%%%%%%%%%%%
%%%%%%%%%%%%%%%%%%%%%%%%%%%%%%%%%%%%%%%%%%%%%%%%%%%%%%%%%%%%%%%%%%%%%
%%%%%%%%%%%%%%%%%%%%%%%%%%%%%%%%%%%%%%%%%%%%%%%%%%%%%%%%%%%%%%%%%%%%%

\subsection{Gluon field renormalization in light-cone gauge}

An in-depth study of the all-order renormalizability of pure Yang-Mills theory and QCD in light-cone gauge (with the ML prescription \eqref{eq:ML_prescription}) was performed in Ref.~\cite{Bassetto:1987sw}, with the gauge condition $n^{\mu} A_{\mu}^a(x)=0$ imposed as a constraint via a Lagrange multiplier $\Lambda^a(x)$.
It was shown that, although 1PI vertex functions have UV divergences which are nonpolynomial in the external momenta, they can be made finite order by-order by introduction of a specific non-local counterterm and a small number of local counterterms. Moreover, the non-polynomiality (or non-locality) of the UV divergences and counterterm was shown to disappear at the level of Green's function and thus of physical observables.  

In the case of pure Yang-Mills theory, the full renormalized Lagrangian, including the local and non-local counterterms, can be obtained~\cite{Bassetto:1987sw} from the local bare Lagrangian 
\begin{align}
{\cal L} 
=&\,
-\frac{1}{4}\, F^{(0)\mu\nu}_{a}(x)\, F^{(0)a}_{\mu\nu}(x)
-\Lambda^{(0)a}(x)\,  n^{\mu} A_{\mu}^{(0)a}(x)
 \label{Eq:bare_Lagr_pure_YM}
\end{align}
thanks to the transformation
\begin{align}
A_{\mu}^{(0)a} =&\, {Z_3}^{\frac{1}{2}}\,
\left[A_{\mu}^{a}-\left(1\!-\!  {\tilde{Z}_3}^{-1}\right)n_{\mu}\, \Omega^{a}
\right]
 \label{Eq:bare_renorm_transfo_gluon}\\
\Lambda^{(0)a}=&\, {Z_3}^{-\frac{1}{2}}\, \Lambda^{a}
 \label{Eq:bare_renorm_transfo_LagMult}\\
g_0 =&\, \mu^{\epsilon}\,  {Z_3}^{-\frac{1}{2}}\, g(\mu^2)
\label{Eq:bare_renorm_transfo_g}
\end{align}
between the bare and renormalized gauge field, Lagrange multiplier and coupling. At this stage, the renormalization of the gluon field in light-cone gauge is not only multiplicative, but also involve an additive part, involving the non-local quantity
\begin{align}
\Omega \equiv &\, \frac{n^{\mu} \bar n^{\nu}}{(\bar n\!\cdot\! n)}\, \left(n^{\rho} D_{\rho}\right)^{-1} F_{\mu\nu}
\, .
\label{Eq:def_Omega}
\end{align}
In Eq.~\eqref{Eq:def_Omega}, the inverse covariant derivative can be interpreted in terms of a light-like gauge link.
In the case of full QCD in light-cone gauge, the renormalization of the gluon field is still of the form \eqref{Eq:bare_renorm_transfo_gluon}, with the presence of the quarks only affecting the value of the renormalization constant $Z_3$ \cite{Bassetto:1987sw,Bassetto:1985dr}.

Interestingly, the shift involving $\Omega$ in Eq.~\eqref{Eq:bare_renorm_transfo_gluon} affects only the component ${\bar n}^{\mu} A_{\mu}^{a}$ of the gauge field.
By contrast, one has
\begin{align}
n^{\mu} A_{\mu}^{(0)a} =&\, {Z_3}^{\frac{1}{2}}\,
n^{\mu} A_{\mu}^{a}
\, ,
\end{align}
so that the gauge condition $n^{\mu} A_{\mu}^{a} =0$ is not spoiled by renormalization, and the transverse components are multiplicatively renormalized as well,
\begin{align}
 A_{i}^{(0)a} =&\, {Z_3}^{\frac{1}{2}}\,
 A_{i}^{a}
 \label{Eq:Aperp_renorm}
\, .
\end{align}

Then, if one imposes the gauge condition $n^{\mu} A_{\mu}^{a} =0$ strongly, one finds the simplifications
\begin{align}
n^{\rho} D_{\rho}  \to &\, n^{\rho} {\partial}_{\rho}
\\
n^{\mu} F_{\mu\nu}  \to &\, n^{\mu} {\partial}_{\mu} A_{\nu}
\, ,
\end{align}
and thus
\begin{align}
\Omega^a 
 \to &\, 
 \frac{\bar n^{\nu}}{(\bar n\!\cdot\! n)}\,   \left( n^{\rho} {\partial}_{\rho}\right)^{-1}     n^{\mu} {\partial}_{\mu} A_{\nu}^a
\to
 \frac{\bar n^{\nu}}{(\bar n\!\cdot\! n)}\,   A_{\nu}^a 
\, , 
\end{align}
so that the transformation \eqref{Eq:bare_renorm_transfo_gluon} reduces to
\begin{align}
A_{\mu}^{(0)a} =&\, {Z_3}^{\frac{1}{2}}\,
\left[{g_{\mu}}^{\nu}-\left(1\!-\!  {\tilde{Z}_3}^{-1}\right)\,  \frac{n_{\mu} \bar n^{\nu}}{(\bar n\!\cdot\! n)}
\right]\, A_{\nu}^{a}
 \label{Eq:bare_renorm_transfo_gluon_OS}
 \, .
\end{align}

In order to extract the one-loop expressions for the renormalization constants $Z_3$ and $\tilde{Z}_3$ from the calculation of the gluon polarization tensor in the previous subsection, one can consider the tree-level two points function for gluons $i(q_{\mu}q_{\nu}\!-\!q^2 g_{\mu\nu})$ contracted with two bare gluon fields in momentum space as
\begin{align}
&\,  i (q_{\mu}q_{\nu}\!-\!q^2 g_{\mu\nu}) \left[{\mathbb {1}}\right]_{ab}\: {\tilde{A}}^{(0)\mu}_a(-q)\,  {\tilde{A}}^{(0)\nu}_b(q)
\nonumber\\
  =&\,
  i Z_3\bigg\{(q_{\mu}q_{\nu}\!-\!q^2 g_{\mu\nu})
  -\left(1\!-\!  {\tilde{Z}_3}^{-1}\right)\, \frac{(q\!\cdot\! n)}{(\bar n\!\cdot\! n)}\, \left[q_{\mu} {\bar n}_{\nu}\!+\!{\bar n}_{\mu} q_{\nu}\right] 
  +\left(1\!-\!  {\tilde{Z}_3}^{-1}\right)^2 \frac{(q\!\cdot\! n)^2}{(\bar n\!\cdot\! n)^2}\,{\bar n}_{\mu}  {\bar n}_{\nu}
  \bigg\}\tilde{A}^{\mu}_a(-q)\,  \tilde{A}^{\nu}_a(q) 
  \nonumber\\
  =&\,
  i \bigg\{Z_3 (q_{\mu}q_{\nu}\!-\!q^2 g_{\mu\nu})
  -\left({\tilde{Z}_3}\!-\!  1\right)\, \frac{(q\!\cdot\! n)}{(\bar n\!\cdot\! n)}\, \left[q_{\mu} {\bar n}_{\nu}\!+\!{\bar n}_{\mu} q_{\nu}\right] 
 +O(g^4) \bigg\}\tilde{A}^{\mu}_a(-q)\,  \tilde{A}^{\nu}_a(q) 
 \label{Eq:tree_plus_ct}
\end{align}
using Eq.~\eqref{Eq:bare_renorm_transfo_gluon_OS} and then $Z_3=1+O(g^2)$ and $\tilde{Z}_3=1+O(g^2)$. Requiring the UV divergences to cancel in the sum of Eqs.~\eqref{Eq:vac_pol_tensor_1_loop_tot_with_fields} and \eqref{Eq:tree_plus_ct}, one obtains the gluon renormalization constants in light-cone gauge in the $\overline{\textrm{MS}}$ scheme
\begin{align}
Z_3 
=&\, 
1+ \frac{g^2}{(4\pi)^2}\,  \left[\frac{11 C_A}{3}\!-\!\frac{2 N_f}{3}\right]  \frac{S_{\epsilon}}{\epsilon} +O(g^4)
=1+ \frac{\alpha_s}{12\pi}\,  \big[11 C_A\!-\!2 N_f\big] \frac{S_{\epsilon}}{\epsilon} +O(\alpha_s^2)
\label{Eq:Z_3_one_loop}
\\
{\tilde{Z}}_3
=&\, 
1+ \frac{2 g^2 C_A}{(4\pi)^2}\,   \frac{S_{\epsilon}}{\epsilon} +O(g^4)
=1+ \frac{\alpha_s C_A}{2\pi}\,  \frac{S_{\epsilon}}{\epsilon} +O(\alpha_s^2)
\label{Eq:tilde_Z_3_one_loop}
\, ,
\end{align}
which are indeed consistent\footnote{Note that in Ref.~\cite{Bassetto:1985dr}, the results for the renormalization constants were given in the specific case of the $SU(3)$ gauge group, with in particular $C_A=N_c=3$.} with the results of Refs.~\cite{Bassetto:1987sw,Bassetto:1985dr}. 

In the case of the gluon distribution considered in this study, only the field strength components $n_{\mu} F_a^{\mu i}=n^{\mu} \partial_{\mu} A_a^{ i}$ appears in the operator definition \eqref{eq:gluon-def_T-ord} in ligh-cone gauge. The renormalization of these field strength components directly follow from Eq.~\eqref{Eq:Aperp_renorm}. Hence the relation \eqref{Eq:fully_vs_partially_bare_pdfs} between the fully bare parton distribution and the parton distribution obtained from the operator definition \eqref{eq:gluon-def_T-ord} written in terms of renormalized fields becomes  
\begin{align}
g^{(0)}\left({\textrm{x}}\right)
=&\,
Z_{3}(g,\epsilon)\;
g^{\textrm{n.r.}}\left({\textrm{x}}, \mu^2\right)
\label{Eq:fully_vs_partially_bare_pdfs_g}
\end{align}
in the case of the gluon distribution in light-cone gauge,
with the standard gluon renormalization constant $Z_3$ given in Eq.~\eqref{Eq:Z_3_one_loop}. 

%%%%%%%%%%%%%%%%%%%%%%%%%%%%%%%%%%%%%%%%%%%%%%%%%%%%%%%%%%%%%%%%%%%%%
%%%%%%%%%%%%%%%%%%%%%%%%%%%%%%%%%%%%%%%%%%%%%%%%%%%%%%%%%%%%%%%%%%%%%
%%%%%%%%%%%%%%%%%%%%%%%%%%%%%%%%%%%%%%%%%%%%%%%%%%%%%%%%%%%%%%%%%%%%%
%%%%%%%%%%%%%%%%%%%%%%%%%%%%%%%%%%%%%%%%%%%%%%%%%%%%%%%%%%%%%%%%%%%%%

%%%%%%%%%%%%%%%%%%%%%%%%%%%%%%%%%%%%%%%%%%%%%%%%%%%%%%%%%%%%%%%%%%%%
%%%%%%%%%%%%%%%%%%%%%%%%%%%%%%%%%%%%%%%%%%%%%%%%%%%%%%%%%%%%%%%%%%%%
%%%%%%%%%%%%%%%%%%%%%%%%%%%%%%%%%%%%%%%%%%%%%%%%%%%%%%%%%%%%%%%%%%%%
%%%%%%%%%%%%%%%%%%%%%%%%%%%%%%%%%%%%%%%%%%%%%%%%%%%%%%%%%%%%%%%%%%%%
%%%%%%%%%%%%%%%%%%%%%%%%%%%%%%%%%%%%%%%%%%%%%%%%%%%%%%%%%%%%%%%%%%%%
%%%%%%%%%%%%%%%%%%%%%%%%%%%%%%%%%%%%%%%%%%%%%%%%%%%%%%%%%%%%%%%%%%%%

 \section{Extracting the DGLAP evolution for the gluon distribution}
 
\label{sec:dglap}	
%%%%%%%%%%%%%%%%%%%%%%%%%%%%%%%%%%%%%%%%%%%%%%%%%%%%%%%%%%%%%%%%%%%%
%%%%%%%%%%%%%%%%%%%%%%%%%%%%%%%%%%%%%%%%%%%%%%%%%%%%%%%%%%%%%%%%%%%%
%%%%%%%%%%%%%%%%%%%%%%%%%%%%%%%%%%%%%%%%%%%%%%%%%%%%%%%%%%%%%%%%%%%%
%%%%%%%%%%%%%%%%%%%%%%%%%%%%%%%%%%%%%%%%%%%%%%%%%%%%%%%%%%%%%%%%%%%%
%%%%%%%%%%%%%%%%%%%%%%%%%%%%%%%%%%%%%%%%%%%%%%%%%%%%%%%%%%%%%%%%%%%%
%%%%%%%%%%%%%%%%%%%%%%%%%%%%%%%%%%%%%%%%%%%%%%%%%%%%%%%%%%%%%%%%%%%%

After having presented our method in Sec.~\ref{sec:general}, we have calculated the real corrections to the gluon distribution in Sec.~\ref{sec:real}, and the gluon field renormalization in Sec.~\ref{sec:virtual}.  It remains now to combine the results from the various sections in order to obtain the splitting functions for the DGLAP evolution of the gluon distribution at LO.
For the $g\rightarrow g$ channel, we have found in Sec.~\ref{sec:real} the result \eqref{Eq:Zgg_Zg_1loop}, which can be written as
\begin{align}
Z_{gg} 
=&\,
\frac{1}{Z_{3}}\bigg\{
\delta(\textrm{z}\!-\!1) 
-\frac{S_{\epsilon}}{\epsilon}\,
 \frac{\alpha_s C_A}{\pi}\, 
 \left[\frac{ \textrm{z}}{(1\!-\! \textrm{z})_+}
+ \frac{(1\!-\!\textrm{z})}{\textrm{z}}
+ {\textrm{z}}{(1\!-\!\textrm{z})}
 \right]
 +O(\alpha_s^2)
 \bigg\}
\label{Eq:Zgg_Zg_1loop_2}
  \, .
\end{align}
Then, using the expression \eqref{Eq:Z_3_one_loop} for $Z_3$, one finds
\begin{align}
Z_{gg} 
=&\,
\delta(\textrm{z}\!-\!1) 
-\frac{S_{\epsilon}}{\epsilon}\,  \frac{\alpha_s }{\pi}\, 
\bigg\{
  C_A\left[\frac{ \textrm{z}}{(1\!-\! \textrm{z})_+}
+ \frac{(1\!-\!\textrm{z})}{\textrm{z}}
+ {\textrm{z}}{(1\!-\!\textrm{z})} \right]
+\frac{\big[11C_A\!-\! 2N_f\big]}{12}\, 
\delta(\textrm{z}\!-\!1) 
 \bigg\}
 +O(\alpha_s^2)
\label{Eq:Zgg_1loop}
  \, .
\end{align}
Concerning the $q\rightarrow g$ and $\bar q\rightarrow g$ channels, we have found in Sec.~\ref{sec:real}
\begin{align}
Z_{g{\bar q}}
=&\, 
Z_{gq} 
=-\frac{S_{\epsilon}}{\epsilon}\,\frac{ \alpha_s C_F}{2\pi}\,  
\frac{\big[1+(1\!-\!\textrm{z})^2\big]}{\textrm{z}}
+O\left(\alpha_s^2\right)
\label{Eq:Zgq_1loop_bis}
\, ,
\end{align}
The renormalization factors \eqref{Eq:Zgg_1loop} and \eqref{Eq:Zgq_1loop_bis} are expressed as perturbative series in the renormalized coupling $\alpha_s=g^2/(4\pi)$, at finite value of the dimensional regulator $\epsilon>0$. These factors depend on the scale $\mu$ only through the renormalized coupling $g(\mu^2)$.

From the renormalization relation for the coupling \eqref{Eq:bare_renorm_transfo_g}, one has
\begin{align}
g_0 =&\, \mu^{\epsilon}\,  g(\mu^2)\, \Big[1+O(g^2)\Big]
\label{Eq:bare_renorm_transfo_g_2}
\end{align}
and thus
\begin{align}
\alpha_s^{(0)} \equiv&\, \frac{g_0^2}{4\pi} = \mu^{2\epsilon}\,  \alpha_s(\mu^2)\, \Big[1+O(\alpha_s)\Big]
\label{Eq:bare_renorm_transfo_alpha_s}
\, .
\end{align}
Since the bare couplings $g_0$ and $\alpha_s^{(0)}$ are by definition independent of the scale $\mu$, one finds from Eq.~\eqref{Eq:bare_renorm_transfo_alpha_s} the dependence on $\mu$ of the renormalized coupling at finite $\epsilon$ as
\begin{align}
\mu^2\frac{d}{d\mu^2}\alpha_s(\mu^2) =&\, -\epsilon\,  \alpha_s(\mu^2)+O(\alpha_s^2)
\label{Eq:alpha_s_evol}
\, .
\end{align}

The splitting functions are given via Eq.~\eqref{eq:def_split_funct_1L} in term of the evolution in $\mu$ of the renormalization factors \eqref{Eq:Zgg_1loop} and \eqref{Eq:Zgq_1loop_bis}. We thus find
\begin{align}
 P_{gg} (\textrm{z},g,\epsilon) 
 =&\, 
S_{\epsilon}\,  \frac{\alpha_s }{\pi}\, 
\bigg\{
  C_A\left[\frac{ \textrm{z}}{(1\!-\! \textrm{z})_+}
+ \frac{(1\!-\!\textrm{z})}{\textrm{z}}
+ {\textrm{z}}{(1\!-\!\textrm{z})} \right]
+\frac{\big[11C_A\!-\! 2N_f\big]}{12}\, 
\delta(\textrm{z}\!-\!1) 
 \bigg\}
 +O(\alpha_s^2)
 \nonumber\\
  =&\, 
  \frac{\alpha_s }{\pi}\, 
\bigg\{
  C_A\left[\frac{ \textrm{z}}{(1\!-\! \textrm{z})_+}
+ \frac{(1\!-\!\textrm{z})}{\textrm{z}}
+ {\textrm{z}}{(1\!-\!\textrm{z})} \right]
+\frac{\big[11C_A\!-\! 2N_f\big]}{12}\, 
\delta(\textrm{z}\!-\!1) 
 \bigg\}
+O(\epsilon)
 +O(\alpha_s^2)
\, ,
\label{Eq:Pgg_result}
\end{align}
and
\begin{align}
P_{gq} (\textrm{z},g,\epsilon)  =&\, P_{g\bar q} (\textrm{z},g,\epsilon) 
=
S_{\epsilon}\,\frac{ \alpha_s C_F}{2\pi}\,  
\frac{\big[1+(1\!-\!\textrm{z})^2\big]}{\textrm{z}}
+O\left(\alpha_s^2\right)
=
\frac{ \alpha_s C_F}{2\pi}\,  
\frac{\big[1+(1\!-\!\textrm{z})^2\big]}{\textrm{z}}
+O(\epsilon)
+O\left(\alpha_s^2\right)
\,.
\label{Eq:Pgq_result}
\end{align}
The $\epsilon\rightarrow 0$ can indeed be taken safely at this stage.
The standard LO DGLAP evolution of the gluon distribution is thus recovered as
\begin{align}
\mu^2 \frac{d}{d\mu^2} g(\textrm{x},\mu^2)
=&\,
\int_{\textrm{x}}^1 \frac{d\textrm{z}}{\textrm{z}}\; P_{gg} (\textrm{z},g,0)\;
g\left(\frac{\textrm{x}}{\textrm{z}},\mu^2\right)
+
\int_{\textrm{x}}^1 \frac{d\textrm{z}}{\textrm{z}}\; P_{gq} (\textrm{z},g,0)\;
\sum_{f}\bigg[{q}_f\left(\frac{\textrm{x}}{\textrm{z}},\mu^2\right)+\bar{q}_f\left(\frac{\textrm{x}}{\textrm{z}},\mu^2\right)
\bigg]
\label{eq:DGALP_final}
\, ,
\end{align}
with the splitting functions \eqref{Eq:Pgg_result} and \eqref{Eq:Pgq_result}.

Finally, let us comment on the case of the renormalization and evolution of quark distribution. It can be studied along the same lines as the (more complicated) gluon distribution case that we have considered in the present work. On the one hand, the calculation of the $g\rightarrow q$ channel is similar as the one of the $q\rightarrow g$ channel from Sec.~\eqref{sec:q2g}. The main technical difference, is that the $g\rightarrow q$ channel requires using the relation \eqref{gauge_field_to_field_strength}, that we needed only in the $g\rightarrow g$ channel. On the other hand, the real NLO contribution to the $q\rightarrow q$ channel is similar to (but algebraically simpler than)  its $g\rightarrow g$ channel analog, from Sec.~\ref{sec:g2g}. In particular it contains a gluon propagator as the rung in the ladder diagram, and spurious soft divergences associated with that propagator are avoided thanks to the use of the ML prescription, in the same way as in Sec.~\ref{sec:g2g}. The renormalization of the quark field in light-cone gauge has also been presented in Refs.~\cite{Bassetto:1987sw,Bassetto:1985dr}, based on the one-loop corrections to the quark self-energy (and quark-gluon vertex) from Ref.~\cite{Leibbrandt:1983zd}, which can be checked along the same line as in Sec.~\ref{Sec:gluon_vac_pol_1_loop}.

%%%%%%%%%%%%%%%%%%%%%%%%%%%%%%%%%%%%%%%%%%%%%%%%%%%%%%%%%%%%%%%%%%%%%
%%%%%%%%%%%%%%%%%%%%%%%%%%%%%%%%%%%%%%%%%%%%%%%%%%%%%%%%%%%%%%%%%%%%%
%%%%%%%%%%%%%%%%%%%%%%%%%%%%%%%%%%%%%%%%%%%%%%%%%%%%%%%%%%%%%%%%%%%%%
%%%%%%%%%%%%%%%%%%%%%%%%%%%%%%%%%%%%%%%%%%%%%%%%%%%%%%%%%%%%%%%%%%%%%
%%%%%%%%%%%%%%%%%%%%%%%%%%%%%%%%%%%%%%%%%%%%%%%%%%%%%%%%%%%%%%%%%%%%%
%%%%%%%%%%%%%%%%%%%%%%%%%%%%%%%%%%%%%%%%%%%%%%%%%%%%%%%%%%%%%%%%%%%%%
%%%%%%%%%%%%%%%%%%%%%%%%%%%%%%%%%%%%%%%%%%%%%%%%%%%%%%%%%%%%%%%%%%%%%
%%%%%%%%%%%%%%%%%%%%%%%%%%%%%%%%%%%%%%%%%%%%%%%%%%%%%%%%%%%%%%%%%%%%%
%%%%%%%%%%%%%%%%%%%%%%%%%%%%%%%%%%%%%%%%%%%%%%%%%%%%%%%%%%%%%%%%%%%%%
%%%%%%%%%%%%%%%%%%%%%%%%%%%%%%%%%%%%%%%%%%%%%%%%%%%%%%%%%%%%%%%%%%%%%
%%%%%%%%%%%%%%%%%%%%%%%%%%%%%%%%%%%%%%%%%%%%%%%%%%%%%%%%%%%%%%%%%%%%%
%%%%%%%%%%%%%%%%%%%%%%%%%%%%%%%%%%%%%%%%%%%%%%%%%%%%%%%%%%%%%%%%%%%%%

\section{Conclusion}

We have derived the DGLAP evolution equation for the gluon distribution function using the
background field method, also used in the Color Glass Condensate formalism. Starting with the operator 
definition of the gluon distribution function in "target" light cone gauge ($A^- = 0$ for a 
left moving proton) in the presence of a 
background field the DGLAP evolution is cast as the standard UV renormalization of a 
composite operator constructed from quantum fields, decomposed into background field and fluctuations around it. One loop corrections generate UV divergences for the transverse components 
of this operator and are regulated and renormalized via standard techniques using dimensional 
regularization in $\overline{MS}$ scheme. 

While the results presented here are well-known, there are several aspects of this calculation 
which will be very useful in the pursuit of a unified framework for QCD evolution. 
With the method developed in this paper we make the connections between the collinear factorization and the CGC effective theory more clear. Having understood how the scale evolution of gluon distribution function arise in the background field formalism in the target light cone gauge,  
we will repeat the calculation in the "projectile" light cone gauge ($A^+ = 0$ for a left moving proton) where CGC formalism is most conveniently applied. This will require to fully take into account the gauge link between the field strength tensors in the UV renormalization of the composite operator in the background field formalism.
Moreover, this setup and the methods developed here can also adapted to the case of
transverse momentum distribution functions (TMDs). We are planning to study the evolution of the TMDs in the presence of the background 
fields to further investigate the relation between CGC formalism and TMD framework.

%\section*{Acknowledgements}
\acknowledgements{ TA is supported in part by the National Science Centre (Poland) under the research Grant No. 2023/50/E/ST2/00133
(SONATA BIS 13). GB is supported in part by the National Science Centre (Poland) under the research grant no 2020/38/E/ST2/00122 (SONATA BIS 10). This material is based upon work supported by the U.S. Department of Energy, Office of Science, Office of Nuclear Physics, within the framework of the Saturated Glue (SURGE) Topical Theory Collaboration.
JJM is supported by ULAM program of NAWA No. BPN/ULM/2023/1/00073/U/00001 and by the US DOE Office of Nuclear Physics through Grant No. DE-SC0002307.}

\appendix

\section{Calculation of ML integrals\label{sec:ML_int}} 

This appendix is devoted to the calculation of the integrals ${\mathbb A}_0(q)$, ${\mathbb B}_0(q)$, and ${\mathbb B}^{\mu}(q)$ defined in Eqs.~\eqref{Eq:int_A_0_ML_1}, \eqref{Eq:int_B_0_ML_1} and \eqref{Eq:int_B_mu_ML_1} with the ML prescription \eqref{eq:ML_prescription}.
Like in the rest of this study, we choose $n^{\mu}= g^{\mu-}$ and  $\bar n^{\mu} = g^{\mu+}$. The relation \eqref{Eq:rel_int_B_min_ML_vs_B0_PV} becomes ${\mathbb B}^{-}(q)={\cal B}_{0}(q)$, so that we still have the components  ${\mathbb B}^{+}(q)$ and ${\mathbb B}^{j}(q)$ to calculate on top of the scalar integrals ${\mathbb A}_0(q)$ and ${\mathbb B}_0(q)$. 
Thus, we consider
\begin{align}
\left[
\begin{array}{c}
{\mathbb B}_{0}(q)\\
{\mathbb B}^{j}(q)\\
{\mathbb B}^{+}(q)
\end{array}
\right]
\equiv &\, 
\mu^{2\epsilon} 
\int \frac{d^{4\!-\!2\epsilon}k}{(2\pi)^{4\!-\!2\epsilon} }\;  
\frac{i}{\left(k^2+i0^+\right)\left((k\!-\!q)^2+i0^+\right)}\;   
\frac{1}{[k^-]} 
\left[
\begin{array}{c}
1\\
\k^{j}\\
k^{+}
\end{array}
\right]
\nonumber\\
=&\,
\mu^{2\epsilon} 
\int \frac{d^{2\!-\!2\epsilon}\k}{(2\pi)^{2\!-\!2\epsilon} }\;  
\int \frac{dk^+}{2\pi}\;  
\left[
\begin{array}{c}
1\\
\k^{j}\\
k^{+}
\end{array}
\right]
\int \frac{dk^-}{2\pi}\;  
\frac{i}{\left(k^2+i0^+\right)\left((k\!-\!q)^2+i0^+\right)}\;   
\frac{2k^+}{\left(2k^+k^-+i0^+\right)} 
\label{Eq:finite_ML_ints_1}
 \, .
\end{align}
If $k^+$ and $k^+\!-\!q^+$ are of the same sign, the three poles in $k^-$ are on the same side of the real axis. The integration contour can then be closed on the opposite side, without encircling any sigularity and leading to a vanishing integral. 
Hence, the integration in $k^-$ can give a non-zero result only in the regime $0<k^+<q^+$ and in the regime $q^+<k^+<0$, which is
\begin{align}
&\int \frac{dk^-}{2\pi}\;  
\frac{i}{\left(k^2+i0^+\right)\left((k\!-\!q)^2+i0^+\right)}\;   
\frac{2k^+}{\left(2k^+k^-+i0^+\right)} 
\nonumber\\
=&\,
\frac{k^+}{(q^+\!-\!k^+)}\, \frac{\Big[\theta(k^+)\, \theta(q^+\!-\!k^+)
-\theta(-k^+)\, \theta(k^+\!-\!q^+)\Big]}{\left[
2k^+\left(q^-\!-\!\frac{(\k\!-\!\q)^2}{2(q^+\!-\!k^+)}\right)
-\k^2+i0^+
\right]
\left[
2k^+\left(q^-\!-\!\frac{(\k\!-\!\q)^2}{2(q^+\!-\!k^+)}\right)
+i0^+
\right]}
\label{Eq:finite_ML_ints_2}
 \, .
\end{align}
Then the $k^+$ integral of the contribution from the regime $0<k^+<q^+$ is of the form
\begin{align}
\int \frac{dk^+}{2\pi}\;  \theta(k^+)\, \theta(q^+\!-\!k^+)\; f(k^+)
=&\, \theta(q^+)\,\int_0^{q^+} \frac{dk^+}{2\pi}\; f(k^+)
= \theta(q^+)\, \frac{q^+}{2\pi} \int_0^{1} d\xi \; f(\xi q^+)
\label{Eq:finite_ML_ints_3}
 \, ,
\end{align}
whereas the contribution from the regime $q^+<k^+<0$  is of the form
\begin{align}
\int \frac{dk^+}{2\pi}\; (-1)\,  \theta(-k^+)\, \theta(k^+\!-\!q^+)\; f(k^+)
=&\, - \theta(-q^+)\,\int_0^{|q^+|} \frac{d|k^+|}{2\pi}\; f(-|k^+|)
=  - \theta(-q^+)\, \frac{|q^+|}{2\pi} \int_0^{1} d\xi \; f(-\xi |q^+|)
\nonumber\\
=&\,
 \theta(-q^+)\, \frac{q^+}{2\pi} \int_0^{1} d\xi \; f(\xi q^+)
\label{Eq:finite_ML_ints_4}
 \, ,
\end{align}
so that the total can be written as
\begin{align}
\int \frac{dk^+}{2\pi}\;  \Big[\theta(k^+)\, \theta(q^+\!-\!k^+)
-\theta(-k^+)\, \theta(k^+\!-\!q^+)\Big]\; f(k^+)
=&\,  \frac{q^+}{2\pi} \int_0^{1} d\xi \; f(\xi q^+)
\label{Eq:finite_ML_ints_5}
 \, .
\end{align}
Hence, at this stage one has
\begin{align}
\left[
\begin{array}{c}
{\mathbb B}_{0}(q)\\
{\mathbb B}^{j}(q)\\
{\mathbb B}^{+}(q)
\end{array}
\right]
=&\, \frac{q^+}{2\pi} \int_0^{1} d\xi \;
\mu^{2\epsilon} 
\int \frac{d^{2\!-\!2\epsilon}\k}{(2\pi)^{2\!-\!2\epsilon} }\;  
\left[
\begin{array}{c}
1\\
\k^{j}\\
\xi q^{+}
\end{array}
\right]
\frac{\xi}{(1\!-\!\xi)}\, \frac{1}{\left[2\xi q^+q^- \!-\! \frac{\xi (\k\!-\!\q)^2}{(1\!-\!\xi)}
-\k^2+i0^+
\right]
\left[
2\xi q^+q^- \!-\! \frac{\xi (\k\!-\!\q)^2}{(1\!-\!\xi)}
+i0^+
\right]}
\nonumber\\
=&\, \frac{q^+}{2\pi} \int_0^{1} d\xi \; (1\!-\!\xi)\,
\mu^{2\epsilon} 
\int \frac{d^{2\!-\!2\epsilon}\k}{(2\pi)^{2\!-\!2\epsilon} }\;  
\left[
\begin{array}{c}
1\\
\k^{j}\\
\xi q^{+}
\end{array}
\right]
 \frac{1}{\left[\k^2\!-\!2\xi \k\cdot\q + \xi \q^2
+\xi (1\!-\!\xi)\big(-2 q^+q^-\!-\! i0^+\big)
\right]}
\nonumber\\
&\, \times\,
 \frac{1}{
\left[ (\k\!-\!\q)^2 + (1\!-\!\xi)\big(-2 q^+q^-\!-\! i0^+\big)
\right]}
\label{Eq:finite_ML_ints_6}
 \, .
\end{align}
Combining the two denominators with a Feynman parameter $y$ and integrating over $\k$ leads to
\begin{align}
\left[
\begin{array}{c}
{\mathbb B}_{0}(q)\\
{\mathbb B}^{j}(q)\\
{\mathbb B}^{+}(q)
\end{array}
\right]
=&\, 
\frac{q^+}{2\pi}\, \frac{\Gamma(1\!+\!\epsilon)}{4\pi}\, \left({4\pi \mu^2}\right)^{\epsilon}\, 
\int_0^{1} d\xi \; (1\!-\!\xi)^{-\epsilon}\,\, \int_0^{1} dy
\left[
\begin{array}{c}
1\\
\big(1\!-\!y\!+\!\xi y\big)\q^{j}\\
\xi q^{+}
\end{array}
\right]\,
\nonumber\\
&\, \times\, 
\big(1\!-\!y\!+\!\xi y\big)^{-1-\epsilon}\, 
\big(-2 q^+q^-\!+\!y \q^2 \!-\! i0^+\big)^{-1-\epsilon}\, 
%\nonumber\\
%=&\, 
%\frac{2q^+}{(4\pi)^2}\, \Gamma(1\!+\!\epsilon)\, \left({4\pi \mu^2}\right)^{\epsilon}\, 
%\int_0^{1} dy\, \big(-2 q^+q^-\!+\!y \q^2 \!-\! i0^+\big)^{-1-\epsilon}\, 
%\nonumber\\
%&\, \times\,
%\int_0^{1} d\xi \, (1\!-\!\xi)^{-\epsilon}\, \big(1\!-\!y(1\!-\!\xi)\big)^{-1-\epsilon}\, 
%\left[
%\begin{array}{c}
%1\\
%\big(1\!-\!y(1\!-\!\xi)\big)\q^{j}\\
%\xi q^{+}
%\end{array}
%\right]\,
\label{Eq:finite_ML_ints_7}
 \, .
\end{align}
Taking the $\epsilon=0$ limit and integrating over $\xi$, one finds
\begin{align}
\left[
\begin{array}{c}
{\mathbb B}_{0}(q)\\
{\mathbb B}^{j}(q)\\
{\mathbb B}^{+}(q)
\end{array}
\right]
=&\, 
\frac{2q^+}{(4\pi)^2}\, 
\int_0^{1} dy\, \frac{1}{\big(-2 q^+q^-\!+\!y \q^2 \!-\! i0^+\big)}\, 
\left[
\begin{array}{c}
-\frac{1}{y}\, \log(1\!-\!y) \\
\q^{j}\\
\left(\frac{1}{y} + \frac{(1\!-\!y)}{y^2}\,  \log(1\!-\!y)\right) q^{+}
\end{array}
\right]\,
+O(\epsilon)
\label{Eq:finite_ML_ints_8}
 \, .
\end{align}
Finally, one arrives at the results
\begin{align}
{\mathbb B}_0(q)
= &\, 
\frac{2q^+}{(4\pi)^2}\, \frac{1}{\big(2 q^+q^- \!+\! i0^+\big)}
\left\{\textrm{Li}_2\!\left(\frac{\q^2}{-q^2 \!-\! i0^+}\right) -\frac{\pi^2}{6}
\right\}
+O(\epsilon)
%\label{Eq:int_B_0_ML_result}
\nonumber\\
{\mathbb B}^j(q)
= &\, 
\frac{2q^+}{(4\pi)^2}\, \frac{\q^j}{\q^2}\,\log\left(\frac{-q^2 \!-\! i0^+}{-2 q^+q^- \!-\! i0^+}\right)
+O(\epsilon)
%\label{Eq:int_B_j_ML_result}
\nonumber\\
{\mathbb B}^+(q)
= &\, 
\frac{2(q^+)^2}{(4\pi)^2}\,  \frac{1}{\big(2 q^+q^- \!+\! i0^+\big)}\,
\left\{
\frac{q^2}{\big(2 q^+q^- \!+\! i0^+\big)}\left[\textrm{Li}_2\!\left(\frac{\q^2}{-q^2 \!-\! i0^+}\right) -\frac{\pi^2}{6}\right]
+\log\left(\frac{-q^2 \!-\! i0^+}{-2 q^+q^- \!-\! i0^+}\right)+1
\right\}
+O(\epsilon)
%\label{Eq:int_B_plus_ML_result}
 \, .
\end{align}
Note that these three integrals are finite at $D=4$, by contrast to ${\mathbb B}^-(q)={\cal B}_0(q)$ which has a logarithmic UV divergence.

The integral ${\mathbb A}_0(q)$ defined in Eq.~\eqref{Eq:int_A_0_ML_1} can be calculated in a similar fashion, and one obtains 
\begin{align}
{\mathbb A}_0(q)
= &\, 
\mu^{2\epsilon} 
\int \frac{d^{4\!-\!2\epsilon}k}{(2\pi)^{4\!-\!2\epsilon} }\;  
\frac{i}{\left((k\!-\!q)^2+i0^+\right)}\;     
\frac{2k^+}{\left(2k^+k^-+i0^+\right)} 
\nonumber\\
= &\, 
-\frac{2q^+}{(4\pi)^2}\, \frac{\Gamma(1\!+\!\epsilon)}{\epsilon(1\!-\!\epsilon)}\, 
\left(\frac{-2 q^+q^- \!-\! i0^+}{4\pi \mu^2}\right)^{-\epsilon}
\nonumber\\
= &\, 
-\frac{2q^+}{(4\pi)^2}\, 
\Bigg\{\frac{S_{\epsilon}}{\epsilon}
- \log\left(\frac{-2 q^+q^- \!-\! i0^+}{\mu^2 }\right)
+1 
\Bigg\}+O(\epsilon)
\label{Eq:int_A_0_ML_result}
 \, .
\end{align}
The pole at $\epsilon=0$ corresponds to a logarithmic UV divergence.

These results can also be written without using the light-cone coordinates, but only the light cone vectors $n^{\mu}$ and $\bar n^{\mu}$, as
\begin{align}
{\mathbb A}_0(q)
= &\, 
-\frac{1}{(4\pi)^2}\, \frac{2(\bar n\!\cdot\!q)}{(\bar n\!\cdot\!n)}\, \frac{\Gamma(1\!+\!\epsilon)}{\epsilon(1\!-\!\epsilon)}\, 
\left(\frac{-2 (\bar n\!\cdot\!q)(n\!\cdot\!q) \!-\! i0^+}{4\pi \mu^2(\bar n\!\cdot\!n)}\right)^{-\epsilon}
\label{Eq:int_A_0_ML_result_covar}
\\
{\mathbb B}_0(q)
= &\, 
\frac{1}{(4\pi)^2}\, \frac{2(\bar n\!\cdot\!q)}{\big[2 (\bar n\!\cdot\!q)(n\!\cdot\!q) \!+\! i0^+\big]}
\left\{\textrm{Li}_2\!\left(\frac{-\hat{q}^2}{-q^2 \!-\! i0^+}\right) -\frac{\pi^2}{6}
\right\}
+O(\epsilon)
\label{Eq:int_B_0_ML_result_covar}
\\
{\mathbb B}^{\mu}(q)
= &\, 
 {\cal B}_0(q)\, \frac{\bar n^{\mu}\, }{(\bar n\!\cdot\! n)}
 -\frac{1}{(4\pi)^2}\, \frac{2(\bar n\!\cdot\!q)}{(\bar n\!\cdot\!n)}\, 
 \log\left(\frac{(-q^2 \!-\! i0^+)(\bar n\!\cdot\!n)}{-2(\bar n\!\cdot\!q)(n\!\cdot\!q) \!-\! i0^+}\right)
 \frac{\hat{q}^{\mu}}{\hat{q}^2}
 \nonumber\\
 &\,
 +\frac{1}{(4\pi)^2}\, \frac{2(\bar n\!\cdot\!q)^2}{\big[2 (\bar n\!\cdot\!q)(n\!\cdot\!q) \!+\! i0^+\big]}
 \bigg\{
 \frac{q^2\, (\bar n\!\cdot\!n)}{\big[2 (\bar n\!\cdot\!q)(n\!\cdot\!q) \!+\! i0^+\big]}\,
 \left[\textrm{Li}_2\!\left(\frac{-\hat{q}^2}{-q^2 \!-\! i0^+}\right) -\frac{\pi^2}{6}
\right]
 \nonumber \\
 &\, \hspace{5cm}
 +\log\left(\frac{(-q^2 \!-\! i0^+)(\bar n\!\cdot\!n)}{-2(\bar n\!\cdot\!q)(n\!\cdot\!q) \!-\! i0^+}\right)
 +1\bigg\}
 \, \frac{ n^{\mu} }{(\bar n\!\cdot\! n)}+O(\epsilon)
 \, ,
\end{align}
where
\begin{align} 
\hat{q}^{\mu}
\equiv &\,
{q}^{\mu}-\frac{(\bar n\!\cdot\!q)}{(\bar n\!\cdot\!n)}\, n^{\mu} -\frac{( n\!\cdot\!q)}{(\bar n\!\cdot\!n)}\, {\bar n}^{\mu}
\, ,
\end{align}
so that in particular $\hat{q}^2 =-\q^2$ in lightcone coordinates.

\vskip 1cm
\vskip 1cm
%\bibliographystyle{elsarticle-num} 
%\bibliography{mybib}

%\bibliographystyle{apsrev}
%\bibliographystyle{elsarticle-num-names} 
\bibliography{mybib}

\end{document}